\documentclass[a4paper,11pt]{article}

\pdfoutput=1 

\usepackage{jheppub} 

\usepackage{amsfonts}
\usepackage{amssymb}
\usepackage{graphicx}
\usepackage{epstopdf}%
\providecommand{\U}[1]{\protect\rule{.1in}{.1in}}

\newcommand{\Dslash}[1] {
\setbox0=\hbox{$#1$}     \dimen0=\wd0        \setbox1=\hbox{/} \dimen1=\wd1
\ifdim\dimen0>\dimen1         \rlap{\hbox to \dimen0{\hfil/\hfil}}       #1
\else           \rlap{\hbox to \dimen1{\hfil$#1$\hfil}}       /         \fi
}

\newcommand{\ba}{\begin{array}}
\newcommand{\ea}{\end{array}}

\newcommand{\be}[1]{
\begin{eqnarray}\label{#1}}
\newcommand{\ee}{\end{eqnarray}}

\newcommand{\bea}{\begin{eqnarray}}
\newcommand{\eea}{\end{eqnarray}}

\newcommand{\ns}{\Dslash{n}}
\newcommand {\nbs}{\Dslash{\bar n}}
\newcommand{\nbn}{\frac{\nbs\ns}{4}}
\newcommand{\nnb}{\frac{\ns\nbs}{4}}

\title{\boldmath  Two-photon exchange corrections to elastic electron-proton scattering at large momentum transfer  within the SCET approach}

\author[1]{
N. Kivel\note{ On leave of absence from St.~Petersburg Nuclear Physics Institute,
188350, Gatchina, Russia} 
}
\author{ and M. Vanderhaeghen}

\affiliation{Helmholtz Institut Mainz, Johannes Gutenberg-Universit\"at, D-55099 Mainz, Germany
\\
Institut f\"ur Kernphysik, Johannes Gutenberg-Universit\"at, D-55099 Mainz, Germany
}

\abstract{
We calculate  the two-photon exchange (TPE) corrections  in the region  where the kinematical variables describing the elastic $ep$ scattering  
are moderately large momentum scales relative to the soft hadronic scale. 
For such kinematics we use  the  QCD factorization  approach  formulated  
in the framework of the  soft-collinear effective theory (SCET).  Such  technique allows us to develop a description  for  the soft-spectator scattering contribution which is found to be important   in the region of moderately  large scales.  

Together with the hard-spectator contribution we present the complete factorization formulas for the TPE amplitudes at the leading power and leading logarithmic
accuracy.  The momentum region where both photons are hard is described by  only  one new 
nonperturbative SCET form factor.   It turns out that the same form factor also arises  for  wide-angle Compton scattering 
which is also described in the framework  of the  SCET approach.  This allows us to  estimate the soft-spectator contribution associated 
with the hard photons in a model independent  way.  

The main unknown in our  description of the  TPE contribution is related with the  configuration where one photon is soft. The  nonperturbative
dynamics in this case is described by two unknown SCET amplitudes.  We use a simple model in order to estimate their contribution.   
  
The formalism is then applied to a  phenomenological analysis  of existing data for the  reduced cross section as well as for the 
 transverse and longitudinal  polarization observables.  
}

\begin{document} 
\maketitle

\section{Introduction}
The electromagnetic form factors of the nucleon have been explored extensively during the past 
50 years with ever increasing accuracy. 
The tool to extract the electromagnetic form factors is provided by the one-photon ($1 \gamma$) exchange approximation to elastic electron-nucleon scattering. 
Precision measurements of the proton electric to magnetic form factor ratio at larger $Q^2$ using polarization experiments~
\cite{Jones00,Punjabi:2005wq,Gayou02,Puckett:2010ac,Puckett:2011xg,Meziane:2010xc} have 
revealed significant discrepancies in recent years  with unpolarized experiments using the Rosenbluth technique~\cite{Andivahis:1994rq, 
Christy:2004rc, Qattan:2005zd}, when analyzing both within the $1 \gamma$-exchange framework. 
This discrepancy between polarized and unpolarized measurements 
has generated a lot of activity, both theoretical and experimental, over 
 the past decade. The most plausible solution of
this problem is the correct calculation of the radiative corrections (RC) to
elastic lepton-proton scattering. The well-known calculations  in Ref.~\cite{Tsai:1961zz}  were performed in  the so-called soft photon approximation 
which allows one  to compute  the two-photon exchange (TPE) diagrams.  This calculation has been recently  reconsidered and improved   
in a series of works \cite{Maximon:2000hm, Blunden:2003sp, Blunden:2005ew, Kondratyuk:2005kk, Borisyuk:2006fh, Borisyuk:2008es, Borisyuk:2012he} 
within a hadronic  framework. 
A more detailed  review of the subject can be fond  in Refs.\cite{Perdrisat:2006hj, Carlson:2007sp, Arrington:2011dn}.  
However with  increasing  energy, 
calculations using  hadronic degrees of freedom become less and less reliable.        
 At large  energy $s$ and momentum transfer $t\equiv-Q^{2}$ one has to properly  take
into account  the interactions of multiple hard photons with constituents
inside the proton. Such dynamics was not  considered  in the pioneering papers \cite{Tsai:1961zz} 
 which have been published before the formulation of the underlying microscopic theory of the strong interactions QCD.

A consistent description of the hard and soft QCD dynamics can be carried out
using the factorization approach which was applied for the analysis of many
exclusive reactions, see e.g. \cite{Efremov:1978rn, Lepage:1979zb, Chernyak:1983ej}. However even this consideration may be sometimes
complicated and challenging. The well known example is the description of the
nucleon form factors (FFs) at large momentum transfers. For  many years
a proper theoretical framework for this regime has been
 the subject of theoretical debates. There are two
different points of view on the description of QCD dynamics at large $Q^{2}$.
 They are related with the different mechanisms of the underlying  scattering which we  describe  as a 
  hard and soft spectator scattering.  The hard spectator scattering
mechanism was studied long time ago, see e.g.  \cite{Efremov:1978rn, Chernyak:1983ej, Lepage:1979zb} and
references therein,  and results in  the well known  factorization formula for
the FF $F_{1}$ . On the other hand the soft spectator scattering has only been
estimated  using various phenomenological approaches such as light-cone wave
functions and QCD sum rules  
\cite{Isgur:1984jm, Isgur:1988iw, Isgur:1989cy, Ioffe:1982qb, Nesterenko:1982gc, Braun:2001tj, Braun:2006hz}. 
Such estimates show that  at moderate values of
$Q^{2}\leq 10$~GeV$^{2}$  a dominant contribution to the nucleon FFs  
originates from the soft overlap mechanism.  

Recently we suggested the factorization formula for the nucleon FFs \cite{Kivel:2010ns} which systematically
includes both  contributions. We used the soft collinear
effective theory (SCET) framework in order to describe the factorization of the
soft spectator scattering contribution.  In this approach the factorization of the soft spectator scattering contribution 
at moderate values of $Q^{2}$ can be defined  in terms of SCET FFs  which  can be rigorously defined 
 in the intermediate  effective theory SCET-I.   If   the hard-collinear virtualities  are relatively small then the further 
 factorization can not be performed and  the SCET FFs  must be considered as nonperturbative functions.   

\begin{figure}[h]%
\centering
\includegraphics[
width=5.5in
]%
{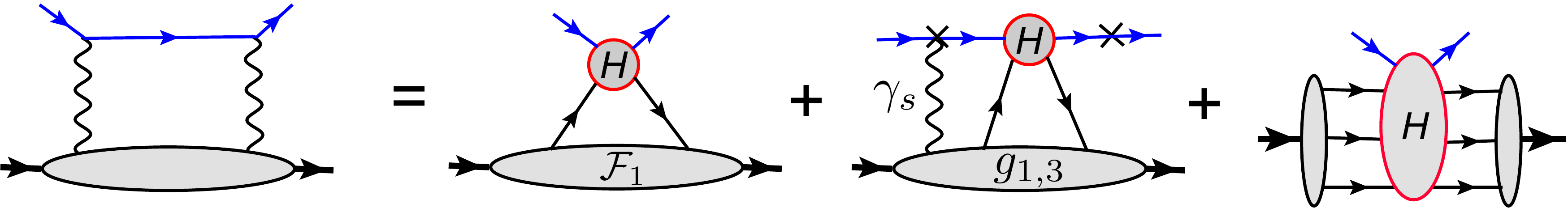}%
\caption{The factorization of the TPE contribution in elastic $ep$ scattering. The crossed box diagram on the $lhs$ is not shown for simplicity.  
The possible attachments of the soft photon
 $\gamma_{s}$ to the lepton lines on the $rhs$  are shown by crosses. }%
\label{tpe}%
\end{figure}

In the present work we follow the  same approach in order to describe the factorization of the short and long
distances  for the two-photon exchange (TPE) contribution shown  on the $lhs$ in Fig.\ref{tpe}. 
This allows us to perform an unambiguous and  consistent separation of the different regions in the QED loops  in Fig.\ref{tpe} associated with  hard and
soft configurations of the photons. We restrict our considerations to  the region where the Mandelstam variables
are much larger then the typical QCD scale  $s\sim -t\sim -u\gg \Lambda^{2}$, 
where $\Lambda \sim 0.5$~GeV is a soft hadronic scale.   
In  full analogy with the nucleon FFs,  the  leading power behavior of the TPE amplitudes  can be  described by  two different configurations 
 associated with the soft and hard spectator contributions.   Schematically, the structure of the leading contributions is shown in  Fig.\ref{tpe}.
 The first two graphs on the $rhs$ describe the soft spectator contribution and the third one corresponds to the hard spectator configuration.    
  The hard subprocesses, which include  one or two hard photons are shown as blobs with the symbol $H$.  The hard spectator configuration
  has already been  studied in Ref.\cite{Borisyuk:2008db, Kivel:2009eg}.   The analysis of the soft spectator terms is new. 
The soft  QCD dynamics for the corresponding diagram   is described by a SCET FF denoted by  $\mathcal{F}_{1}$ and  amplitudes $g_{1,3}$  which are defined in SCET-I.  
At leading order in the QCD coupling, we obtain  these  new SCET amplitudes  which do not appear in the factorization formulas for the nucleon FFs. 
 It turns out that  $\mathcal{F}_{1}$ can be  fixed from the wide-angle Compton scattering  using the universality of its definition in SCET-I.  
 The  amplitudes  $g_{1,3}$  involve matrix element which cannot be related to known objects, and at present can only be  estimated  within some model approach.  As a first step  we consider  an estimate using the effective theory 
 with hadronic degrees of freedom.  In this way  the  TPE contribution is completely defined.    We then perform 
  phenomenological  studies of the TPE effects and compare our results with existing experimental data.

Our paper is organized as follows. Section~\ref{sec-epscat}  is devoted to the general properties of  elastic $ep$ scattering. We specify  notations and
  kinematics,  discuss the  general properties of the amplitudes,  as well as the structure  of the reduced cross section and asymmetries. In  section~\ref{sec-scet}
 we derive the leading order SCET-I factorization formula for the soft spectator scattering contribution in the TPE amplitudes. In the next section we  perform the 
 matching and compute the one-loop,  leading order in $\alpha_{s}$ hard coefficient functions.  
 Section~\ref{sec-wacs}  is devoted to the extraction of the second unknown FF $\mathcal{F}_{1}$  from the data for wide-angle  Compton scattering.  
 In  section~\ref{sec-g1}  we  discuss the  the
  SCET amplitudes $g_{1,3}$ which describe the TPE contribution when  one of the photons is soft.   
  We  use a simple  hadronic model in order to estimate  this contribution.  
  
 In section~\ref{sec-phnm}  we use the obtained result for the phenomenological analysis  and estimate the effect of the  TPE contribution for different observables. 
 The summary of our obtained results is presented in section~\ref{sec-cnc}. In Appendices A-E we present  more details  of  some calculations.

\section{Elastic lepton-nucleon scattering at large $Q^{2}$}
\label{sec-epscat}

We start by briefly reviewing the main definitions and some results for the process
$e(k)+p(p)\rightarrow e(k^{\prime})+p(p^{\prime})$. In order to describe the 
electron-nucleon elastic scattering 
process we introduce the following notations%
\begin{equation}
P=\frac{1}{2}(p+p^{\prime}),~~K=\frac{1}{2}(k+k^{\prime}),~q=p^{\prime
}-p=k-k^{\prime},
\end{equation}
and define the Mandelstam variables%
\begin{equation}
s=(p+k)^{2},~t=q^{2}=-Q^{2},~u=(p-k^{\prime})^{2},~~\tau=\frac{Q^{2}}{4m^{2}%
}~,
\end{equation}
where $m$ is the nucleon mass. For further use, we  introduce two more convenient
variables
\begin{equation}
\varepsilon=\left(  1+2(1+\tau)\tan^{2}\frac{\theta}{2}\right)  ^{-1}%
=\frac{(s-u)^{2}+t(4m^{2}-t)}{(s-u)^{2}-t(4m^{2}-t)},~~\ 0<\varepsilon<1,
\end{equation}%
\begin{equation}
\nu=(K\cdot P)=\frac{s-u}{4}, \label{nu}%
\end{equation}
where $\theta$ is the electron Lab scattering angle. One can choose any two
independent variables for the description of the physical amplitudes of the
process. It is customary  to use the variables $Q^{2}$ and $\varepsilon$
 for a description of the cross sections and related observables.

Then the general parametrization of the \ $ep-$scattering amplitude reads \cite{Guichon:2003qm}
\begin{equation}
\left\langle p^{\prime},k^{\prime}~out\right\vert in\left.  k,p\right\rangle
=i(2\pi)^{4}\delta(p+k-p-k)~A_{ep},
\end{equation}
where%
\begin{equation}
A_{ep}=\frac{e^{2}}{Q^{2}}\bar{u}(k^{\prime})\gamma^{\mu}u(k)~\bar
{N}(p^{\prime})\left[  \gamma^{\mu}\tilde{G}_{M}(\varepsilon,Q^{2})-\frac{P^{\mu}%
}{m}\tilde{F}_{2}(\varepsilon,Q^{2})+\frac{P^{\mu}}{m^{2}}\Dslash{K}~\tilde{F}%
_{3}(\varepsilon,Q^{2})\right]  N(p).
 \label{def:A}%
\end{equation}
In the one-photon exchange approximation, this amplitude is given by the well known formula
\begin{equation}
A_{ep}^{\gamma}=\frac{e^{2}}{Q^{2}}\bar{u}(k^{\prime})\gamma^{\mu}u(k)~\bar
{N}(p^{\prime})\left[  \gamma^{\mu}G_{M}(Q^{2})-\frac{P^{\mu}}{m}F_{2}(Q^{2})\right]
N(p). 
\label{def:Ag}%
\end{equation}
The following difference can be considered as definition of the TPE corrections arising from the QED next-to-leading corrections  
\begin{align}
A_{ep}-A_{ep}^{\gamma}&  \equiv A_{ep}^{\gamma\gamma}=\frac{e^{2}}{Q^{2}}\bar{u}(k^{\prime
})\gamma^{\mu}u(k)~\label{def:Agg}\\
&  \bar{N}(p^{\prime})\left[  \gamma^{\mu}\delta\tilde{G}_{M}(\varepsilon
,Q^{2})-\frac{P^{\mu}}{m}\delta\tilde{F}_{2}(\varepsilon,Q^{2})+\frac{P^{\mu}}{m^{2}%
}\Dslash{K}~\tilde{F}_{3}(\varepsilon,Q^{2})\right]  N(p),
\end{align}
with
\begin{equation}
\delta\tilde{G}_{M}(\varepsilon,Q^{2})=\tilde{G}_{M}(\varepsilon,Q^{2})-G_{M}%
(Q^{2}),~\delta\tilde{F}_{2}(\varepsilon,Q^{2})=\tilde{F}_{2}(\varepsilon,Q^{2})-F_{2}(Q^{2}).
\end{equation}
The amplitudes $\delta\tilde{G}_{M}$, $~\delta\tilde{F}_{2}$ and $\tilde
{F}_{3}$ obtain different contributions from  all diagrams associated with
the QED radiative corrections to elastic $ep$-scattering.  In the present paper we only consider 
the calculation of the leading in $1/Q^{2}$ corrections  arising from the TPE contribution, see Fig.\ref{tpe}.

Some useful constraints on the behavior of these amplitudes  can be established from the consideration of their  analytical properties. 
For that purpose, following   \cite{Borisyuk:2008es}  we introduce the  functions:
\begin{equation}
G_{1}(\nu,Q^{2})=\delta\tilde{G}_{E}(\varepsilon,Q^{2})+\frac{\nu}{m^{2}%
}\tilde{F}_{3}(\varepsilon,Q^{2}), \label{def:G1}%
\end{equation}
\begin{equation}
G_{2}(\nu,Q^{2})=\delta\tilde{G}_{M}(\varepsilon,Q^{2})+\frac{\nu}{m^{2}%
}\tilde{F}_{3}(\varepsilon,Q^{2}), \label{def:G2}%
\end{equation}%
\begin{equation}
G_{3}(\nu,Q^{2})=\tilde{F}_{3}(\varepsilon,Q^{2}), \label{def:G3}%
\end{equation}
where we use the shorthand notation  $\varepsilon\equiv\varepsilon(\nu,Q^{2})$ and defined
$\delta\tilde{G}_{E}\equiv \delta\tilde{G}_{M}-(1+\tau)\delta\tilde{F}_{2}$.

An analysis of $t$-channel helicity amplitudes for the $ep\rightarrow ep$ process shows that in the Regge limit
$s\rightarrow\infty$, $Q^{2}/s\rightarrow0$, which  is equivalent to
$\nu\rightarrow\infty$, $Q^{2}/\nu\rightarrow0$,  the functions $G_{i}$  vanish 
\begin{equation}
\lim_{\nu\rightarrow\infty}G_{i}(\nu,Q^{2})=0. \label{null}%
\end{equation}
This higher energy behavior allows one to write down  unsubtracted
dispersion relations for the amplitudes $G_{i}$ as \cite{Borisyuk:2008es}:
\begin{equation}
G_{i}(\nu,Q^{2})=\int_{\nu_{th}}^{\infty}d\nu^{\prime}\frac{\operatorname{Im}
G_{i}(\nu',Q^{2})}{\nu^{\prime}-\nu}-\int_{-\infty}^{-\nu_{th}}d\nu^{\prime}
\frac{\operatorname{Im}G_{i}(\nu',Q^{2})}{\nu^{\prime}-\nu}.
\end{equation}
The Regge limit can easily be translated into a boundary condition for the
practically important variable $\varepsilon$, and corresponds to the limit
$\varepsilon\rightarrow1$ at fixed $Q^{2}$. From Eqs.(\ref{def:G1}
-\ref{def:G3}) one then obtains:
\begin{equation}
\lim_{\varepsilon\rightarrow1}~\left[  \delta\tilde{G}_{E,M}(\varepsilon,Q^{2})+\frac{\nu}{m^{2}}\tilde{F}_{3}(\varepsilon,Q^{2})\right]  =0,
\label{GnuF0}
\end{equation}
\begin{equation}
\lim_{\varepsilon\rightarrow1}~\tilde{F}_{3}(\varepsilon,Q^{2})=0.
\label{F30}
\end{equation}
Eqs.(\ref{GnuF0},\ref{F30}) imply  that  for certain observables the TPE corrections must vanish in the forward limit. 

The analytical expressions for various observables are well known in the
literature, see e.g. \cite{Guichon:2003qm}. For the convenience of the reader we provide some
of them here. The unpolarized cross section reads
\begin{equation}
\frac{d\sigma}{d\Omega_{\text{Lab}}}=\frac{d\sigma_{\text{NS}}}{d\Omega
_{\text{Lab}}}\frac{\tau}{\varepsilon(1+\tau)}\sigma_{R}(\varepsilon,Q),
\end{equation}
with the structureless part%
\begin{equation}
\frac{d\sigma_{\text{NS}}}{d\Omega_{\text{Lab}}}=\frac{4\alpha^{2}}{Q^{4}}%
\cos^{2}\frac{\theta}{2}\frac{E^{\prime3}}{E}. \label{def:sNS}%
\end{equation}
The variables $E$ and $E^{\prime}$ in Eq.(\ref{def:sNS}) denote the incoming
and outgoing electron Lab energies.
The {\it elastic} contribution to the  reduced cross $\sigma_{R}$ section is given by:
\begin{align}
\sigma^{el}_{R}(\varepsilon,Q)  &  =G_{M}^{2}+\frac{\varepsilon}{\tau}G_{E}^{2}
+2G_{M}\operatorname{Re}\left[  \delta\tilde{G}%
_{M}+\varepsilon\frac{\nu}{m^{2}}\tilde{F}_{3}\right]  +2\frac{\varepsilon
}{\tau}G_{E}\operatorname{Re}\left[  \delta\tilde{G}_{E}+\frac{\nu}{m^{2}%
}\tilde{F}_{3}\right]. \label{sigmR}%
\end{align}

The elastic contribution to the polarization observables measured in the recoil polarization experiments reads
\begin{align}
\sigma^{el}_{R}~P_{t}  & =-\sqrt{\frac{2\varepsilon(1-\varepsilon)}{\tau}}\left\{
G_{E}G_{M}+G_{E}\operatorname{Re}  \delta\tilde{G}_{M} 
+G_{M}\operatorname{Re}\left(
\delta\tilde{G}_{E}+\frac{\nu}{m^{2}}\tilde{F}_{3}\right)\right\}  ,
 \label{Pt}
\end{align}
\begin{equation}
\sigma^{el}_{R}~P_{l}=\sqrt{1-\varepsilon^{2}}\left\{  
G_{M}^{2}+2G_{M}%
\operatorname{Re}\left( \delta\tilde{G}_{M}+\frac{\varepsilon}{1+\varepsilon}\frac{\nu}{m^{2}}\tilde{F}_{3}
\right) \right\} ,
\label{Pl}%
\end{equation}
where $P_{l,t}$ correspond to the recoil proton  polarization along or
perpendicular to its momentum, respectively. 
Besides the elastic contributions, shown in Eqs.(\ref{sigmR}-\ref{Pl}),  the observables  also 
include the contribution from the inelastic processes
corresponding with the emission of soft photons  
which provide  the cancellation of 
IR-divergent terms in the elastic next-to-leading QED amplitudes. 

Our task is to compute the TPE amplitudes in the limit of large $Q^{2}%
\gg\Lambda^{2}$ and, if possible, for arbitrary values of $\varepsilon$.
However such task includes the analysis of different kinematical regions which
can be associated with different underlying QCD dynamics. Therefore we will 
split the $\varepsilon$-interval into three regions which can be described as follows.

The forward limit $\varepsilon\sim1$ which can be associated with the Regge limit, in which
$s\rightarrow\infty$ and $Q^{2}$ is fixed. In this case as one can conclude from
Eqs.(\ref{GnuF0},\ref{F30}) and Eq.(\ref{sigmR}) the TPE corrections to
the reduced cross section are vanishing.

The second situation is associated with the backward scattering where $s\sim
Q^{2}\gg|u|\sim\Lambda^{2}$ and corresponds to the region of small
$\varepsilon\ll1$. From the unpolarized data we expect that in this case the
effect of TPE corrections are largest. However,  the development of a 
theoretical approach in a systematic way for this region is a difficult task. 

 The third region is described by the kinematics where all Mandelstam variables
are large and of the same order: 
\bea
s\sim|u|~\sim Q^{2}\gg\Lambda^{2}.
\label{stu}
\eea
This region can be associated with wide-angle scattering.  In this situation one can try to compute the TPE
amplitudes by performing an expansion with respect to the large scale $Q^{2}$ with a
fixed ratio $Q^{2}/s$. Below we are going to realize this idea using the QCD
factorization approach. In this case  the values of $\varepsilon$ are
restricted to some interval $\varepsilon_{\min}<\varepsilon<\varepsilon_{\max
}$ where we define the boundaries $\varepsilon_{\min}$ and  $\varepsilon_{\max}$ 
from the phenomenological criteria that the minimal value of $|u|$ is given  by $|u_{\min}|=2.5$\ GeV$^{2}$,
 which guarantees some suppression of subleading power corrections, see Table~\ref{tabeps}.
\begin{table}[h]
\label{tabeps}
\begin{center}
\begin{tabular}
[c]{|c|c|c|c|c|}\hline
$Q^{2},$GeV$^{2}$ & $3$ & $4$ & $5$ & $6$\\\hline
$\varepsilon_{\min}$ & $0.60$ & $0.52$ & $0.45$ & $0.42$\\\hline
\end{tabular}
\end{center}
\end{table}
The upper boundary can be formally  defined by the relevance of the Regge dynamics.
For simplicity we will not introduce $\varepsilon_{\max}$ assuming an
extrapolation to the value $\varepsilon=1$.

\begin{figure}[h]%
\centering
\includegraphics[
height=0.8501in,
width=4.4973in
]%
{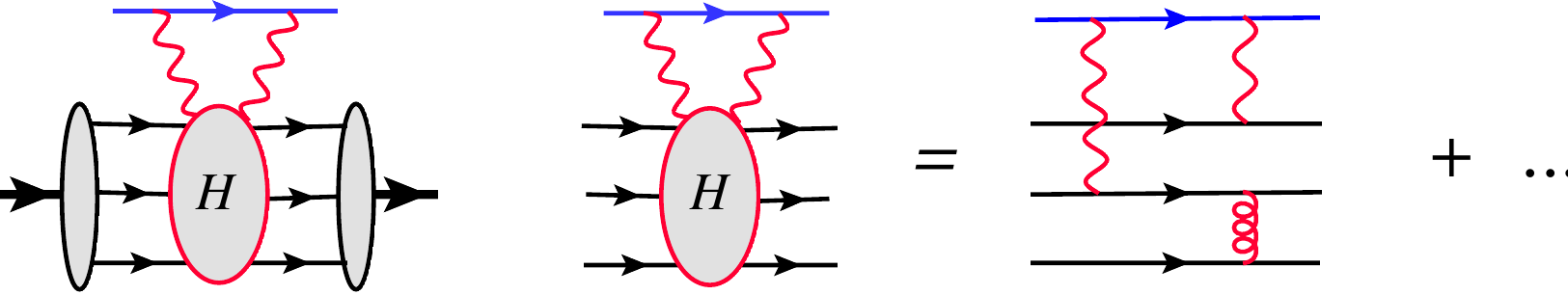}%
\caption{Reduced diagram describing the hard spectator scattering contribution
and the leading order diagram with one gluon exchange.}%
\label{hss-diagrams}%
\end{figure}

The idea to apply the QCD factorization approach for the wide-angle region has  already
been used in \cite{Borisyuk:2008db, Kivel:2009eg} where the hard spectator scattering contribution
was computed at leading order. Because both photons in this case are hard one
needs only one-gluon exchange  as shown in Fig.\ref{hss-diagrams}. The
nonperturbative input is described by the nucleon distribution amplitudes
(DAs), see the details in Refs.\cite{Borisyuk:2008db, Kivel:2009eg} . 
This calculation allows us to estimate the asymptotic behavior of TPE
amplitudes at large $Q^{2}$%
\begin{equation}
\delta G_{M}\sim\frac{\nu}{m^{2}}\delta F_{3}\sim\frac{\Lambda^{4}}{Q^{4}%
},~\ \delta F_{2}\sim\frac{\Lambda^{6}}{Q^{6}}.
\end{equation}
The behavior of the amplitudes $\delta G_{M}$ and $\delta F_{3}$ is similar to the
Dirac FF $F_{1}\sim\Lambda^{4}/Q^{4}$ and can also be described within the
collinear factorization approach. The helicity flip amplitudes $\delta F_{2}$
is suppressed by a power $Q^{-2}$ similar to the corresponding Pauli FF $F_{2}$
and  can not be described by collinear factorization due to end-point
divergencies.  Therefore qualitatively, upon neglecting the logarithmic structure, the
situation is quite similar to the nucleon FFs. \ 

On the other hand the analysis of the \ soft spectator scattering for the
nucleon FFs shows that these terms are not suppressed by inverse powers of $Q$ 
\cite{Dun1980, Fadin1981} and therefore can also provide sizable contributions especially
in the region of intermediate $Q$ where $Q\Lambda\sim m^{2}$. In Refs.\cite{Kivel:2010ns, Kivel:2012mf} we
investigated the soft spectator contribution and suggested the generalization
of the factorization  which includes both hard and soft spectator terms.
 Taking into account that the TPE dynamics is quite similar to the FF case
one may expect that the same  situation is relevant for this case too.  Then
the soft spectator scattering must also be included into the consideration when calculating the TPE amplitudes. 

Such an attempt has been developed in \cite{Chen:2004tw, Afanasev:2005mp}  where the diagrams  in
Fig.\ref{hss-diagrams}  have been evaluated within the framework of the so-called
handbag approach \cite{Radyushkin:1998rt}.
\begin{figure}[h]%
\centering
\includegraphics[
height=0.5604in,
width=2.6567in
]%
{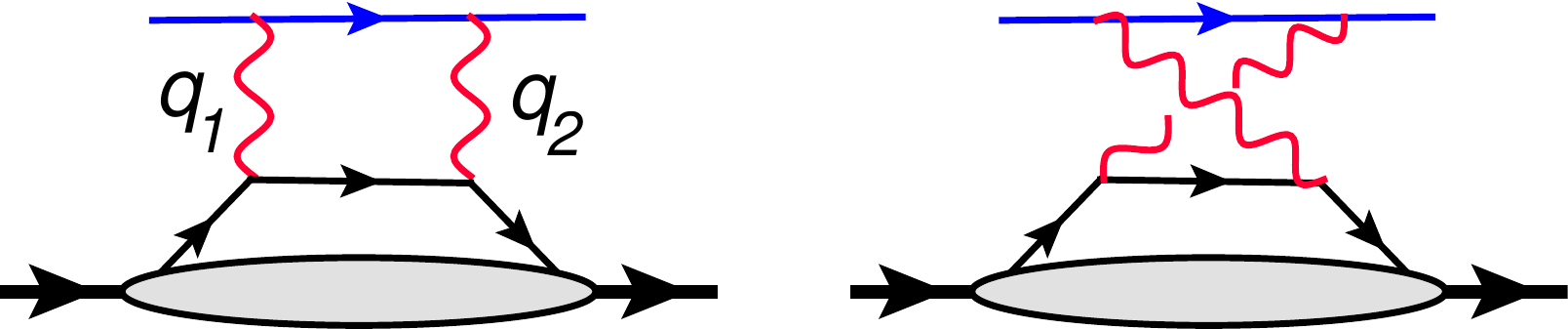}%
\caption{Diagrams describing the soft spectator scattering.  Both photons
interact with the same quark. }%
\label{sss-diagrams}%
\end{figure}
In this model the overlap of the hadronic states are described by the
generalized parton distribution (GPD) introduced as matrix element of the
light-cone twist-2  matrix element. The GPD which arises here at large $Q$
is considered as a natural generalization of the matrix element which appears
in the description of the deeply virtual Compton scattering (DVCS) at small
momentum transfer $Q\sim\Lambda$.  However in DVCS kinematics, GPDs describe
the soft dynamics of the system constructed from collinear partons with a
small invariant mass (small $Q\sim\Lambda$ ).  But at large momentum
transfer ( $Q\gg\Lambda$ ) one is faced with a different situation which may
be better associated with the dynamics of jets. In that case  the
factorization of the soft dynamics is described 
within the framwork of 
soft collinear effective theory (SCET) \cite{Bauer:2000ew,Bauer2000,Bauer:2001ct,Bauer2001,BenCh, BenFeld03}. 
In Ref. \cite{Kivel:2010ns}
we carried out a detailed consideration of the nucleon FFs within the SCET formalism. 
In the present paper we will  show that for the TPE amplitude the situation is quite similar, 
allowing for a QCD factorization along the same lines. 

\section{Soft spectator interaction for TPE in SCET}
\label{sec-scet}

The main feature of the SCET factorization is the presence of two large
scales: hard $\sim Q^{2}$ and hard-collinear $\sim\Lambda Q$.  As a first
step, one integrates out the hard modes and passes from the QCD to the
effective theory SCET-I. If the value of the hard-collinear scale is not
large, for instance, $\Lambda Q\sim m^{2}$ then one can not go on further
with the factorization. We shall define such a region of momentum transfers, $Q^{2}\sim
2.5-16~$GeV$^{2}$, as intermediate momentum transfer region. 
At present, this  region covers all
existing and planned experiments  of  nucleon FFs.

 Our task is to construct a SCET-I factorization for the TPE amplitudes in this
region. Such a factorization scheme includes the description of the soft
spectator contribution within SCET-I and also, if it is necessary, the hard
spectator contributions. These terms can be considered as part of the so-called SCET-II
factorization. However, as their factorization scheme is different, e.g. they do not have the Sudakov suppression, they can be considered separately.

Technically the SCET factorization is constructed by matching of QCD diagrams
onto appropriate operators in the SCET-I effective theory. The matrix elements of
the SCET-I operators describe the low scale processes which can not be
computed in perturbative QCD.  The simple power counting arguments allow one to expect
that the leading order SCET-I contribution is given by the diagrams in
Fig.\ref{hss-diagrams}. Obviously, these contributions are leading order with
respect to the strong coupling $\alpha_{s}$ but subleading with respect to
the electromagnetic constant $\alpha$. Suppose that the incoming and outgoing quarks
are hard-collinear. Then in the hard region where all momenta of the particles
in the box diagrams are hard $q^{\mu}_{1,2}\sim Q$  this process can be described by
perturbative QCD and we obtain the \ leading order contribution into
coefficient function in front of certain SCET-I operators. However, this region is
only a part of the full result. For instance, one can  also  expect that the
contribution from the region where only one photon is hard but the second is
soft, for instance, $q^{\mu}_{1}\sim \Lambda$, $q^{\mu}_{2}\sim Q$ can also be
relevant. Therefore  the SCET factorization must properly  take into account
all possible dominant regions associated with factorization of the hard modes.

 In the following we use two light-cone vectors $n=(1,0,0,-1)$ and $\bar
{n}=(1,0,0,1)$ and choose the Breit frame where the initial and final proton
 are collinear to the $z$-axis. Then the light-cone expansion of the external
momenta read%
\begin{equation}
q=Q\frac{n}{2}-Q\frac{\bar{n}}{2},~\ p\simeq Q\frac{\bar{n}}{2}%
,~\ \ \ p^{\prime}\simeq Q\frac{n}{2}, \label{mom1}%
\end{equation}%
\begin{equation}
k\simeq\frac{\bar{z}}{z}Q\frac{\bar{n}}{2}+\frac{1}{z}Q\frac{n}{2}+k_{\bot
},~\ ~k^{\prime}\simeq\frac{1}{z}Q\frac{\bar{n}}{2}+\frac{\bar{z}}{z}Q\frac
{n}{2}+k_{\bot}, \label{mom2}%
\end{equation}
where we introduced for convenience the dimensionless parameter 
$z={-t}/{s}$, with $\bar z\equiv 1-z$.  The
Breit system is convenient for the description of the large $Q^{2}$ behavior of the
nucleon FFs and it is natural to use it for the consideration of the TPE
contributions too.

Let us briefly describe the SCET notations used in our paper. We will use the
SCET formulation in coordinate space developed in \cite{BenCh}. \ For the
SCET fields we use the following notations. The fields  $\xi_{n},$ $A_{\mu}^{(n)}$ and
$\xi_{\bar{n}},~A_{\mu}^{(\bar{n})}~$ denote hard-collinear quark and gluon
fields associated with momentum $p^{\prime}$ and $p$, respectively, see
Eq.(\ref{mom1}).  As usually,  the hard-collinear quark fields satisfy  
\bea
\Dslash{n}\xi_{n}=0,\,\,  \Dslash{\bar n}\xi_{\bar n}=0.
\label{scetxi}
\ee
The fields $q$ and $A_{\mu}^{(s)}$  denote the soft quarks and
gluons which also enter in the SCET Lagrangian. These fields describe particles 
with soft momenta $k_{\mu}\sim \Lambda$.\footnote{In Ref.\cite{BenCh}  these 
modes  are introduced as ultra-soft.  
In this work we use the different terminology  suggested in  Refs.\cite{Hill:2002vw,Beneke:2003pa} . }  
  We also use the convenient
notation for the gauge invariant combinations often appearing in SCET such as
\begin{equation}
\chi_{n}(\lambda\bar{n})\equiv~W_{n}(\lambda \bar n)\xi_{n}(\lambda\bar{n}),
\label{qjet}
\end{equation}
where the hard-collinear gluon Wilson line (WL) reads~:
\begin{equation}
W_{n}(z)=\text{P}\exp\left\{  ig\int_{-\infty}^{0}ds~\bar{n}\cdot A^{(n)}%
(z+s\bar{n})\right\}  .
\end{equation}

In the QED sector we also split the fields according to the same SCET
prescription. Let us notice that the pure QED radiative corrections (electron vertex and
self-energy) can be computed exactly and for such calculations such
representations are not required. However in case of the TPE corrections, the hard and
soft photons correspond with different dynamics, making the SCET representation useful.

In the wide-angle kinematics we have four directions with a large energy
flow. It is therefore useful to introduce two more auxiliary light-cone vectors
associated with the lepton momenta: $k$ and $k^{\prime}$%
\begin{equation}
\bar{v}^{\mu}=\frac{2k^{\mu}}{Q},~\ v^{\mu}=\frac{2k^{\prime\mu}}{Q},~(\bar
{v}\cdot v)=2.
\end{equation}
Then, an arbitrary vector can be decomposed as
\begin{equation}
V^{\mu}=(V\cdot\bar{v})\frac{v^{\mu}}{2}+\left(  V\cdot v\right)  \frac
{\bar{v}^{\mu}}{2}+V_{T}, 
\end{equation}
where we denoted the transverse components with respect
to the $v,\bar{v}$ basis by $V_{T}$ with subscript $T$, in order to distinguish them from the transverse
components in the $n,\bar{n}$ basis which we denote as $V_{\bot}$.  Using the vectors
$\bar{v},v$ \ we introduce the hard-collinear lepton fields $\zeta_{\bar{v}}$
and$~\zeta_{v}$, satisfying~:  
\bea
\Dslash{v}\zeta_{v}=0, \,\, \Dslash{\bar v}\zeta_{\bar v}=0,
\label{scetzeta}
\eea
and the soft $\ $photon field by $B_{\mu}^{(s)}$.  This is a
minimal set of the auxiliary fields in QED which we need for our
considerations. We do not introduce the collinear photon fields because they
are not required for a description of the TPE amplitudes. We always use the Feynman
gauge for the gluon and photon fields. 

We start our consideration from the one-photon exchange. In the following, we will only consider 
the soft-spectator scattering contribution.
The soft-spectator scattering for nucleon FFs has been studied in \cite{Kivel:2010ns, Kivel:2012mf}.
We briefly repeat the results relevant for our calculations. To the
leading order, the SCET expression for the amplitude $A_{ep}$ can be
written as%
\begin{equation}
A_{ep}^{\gamma}\simeq\frac{4\pi\alpha}{Q^{2}}~\bar{u}(k^{\prime})\gamma^{\mu
}u(k)~C_{+}(Q,\mu)~\left\langle p^{\prime}\right\vert O_{+}^{\mu
}(0)~\left\vert p\right\rangle _{\text{{\footnotesize SCET}}} \label{AgSCET}%
\end{equation}
where the leading power SCET-I operator reads%
\begin{equation}
O_{+}^{\mu}(0)=\sum_{q}e_{q}~~\left\{  \bar{\chi}%
_{n}(0)\gamma_{\bot}^{\mu}~\chi_{\bar{n}}(0)+\bar{\chi}_{\bar{n}}%
(0)\gamma_{\bot}^{\mu}\chi_{n}(0)\right\}  , \label{Opl}%
\end{equation}
and $C_{+}^{\mu\nu}$ denotes the hard coefficient
function. Let us note that the operator in Eq.(\ref{Opl}) has an additional
term compared to the operator considered in \cite{Kivel:2010ns}. This  term  given by $\bar{\chi
}_{\bar{n}}\gamma_{\bot}^{\mu}\chi_{n}$ in Eq.(\ref{Opl}) can be associated
with the scattering of the hard photon on the antiquark. Such contribution describes 
 the  hadronization of the hard-collinear antiquark into proton. 
 Because the proton is dominated by three quark state
 such  term   is suppressed by the
 powers of the hard-collinear scale in the transition from SCET-I to SCET-II.
If the hard-collinear scale is not large and we restrict our consideration
only to the SCET-I theory, we can not neglect  such operators. The relative sign between the
two terms in Eq.(\ref{Opl}) is dictated by $C$-parity. The subscript SCET
in Eq.(\ref{AgSCET}) means that the matrix element has to be computed with
the SCET Lagrangian $\mathcal{L}_{\text{{\tiny SCET}}}=\mathcal{L}%
_{\text{{\tiny SCET}}}^{(n)}+\mathcal{L}_{\text{{\tiny SCET}}}^{(\bar{n})}+\mathcal{L}_{\text{soft}}$
which represent by the sum of Lagrangians describing each collinear sector.
The explicit expressions for $\mathcal{L}_{\text{{\tiny SCET}}}^{(n)}$ with
the fields defined in position space can be found in Refs. \cite{BenCh, BenFeld03}.  The SCET
matrix element in Eq.(\ref{AgSCET})  describes the interaction of the jets of
the hard-collinear particles represented by$~\chi_{n,\bar{n}}$ with the soft
background described by the soft quark and gluons in the SCET Lagrangians.
This dynamics can also be understood as a soft-overlap of the initial and
final hadronic states. The parametrization of this matrix element reads
\begin{equation}
\left\langle p^{\prime}\right\vert O_{+}^{\mu}(0)~\left\vert p\right\rangle
_{\text{{\footnotesize SCET}}}=\bar{N}(p^{\prime})\nbn%
\gamma_{\bot}^{\mu}N(p)~f_{1}(Q,\mu), \label{def:f1}%
\end{equation}
where $f_{1}$ is the SCET FF. By construction it depends on the hard-collinear
scale, referred to by the argument $Q$, whereas the second
argument $\mu$ denotes the renormalization scale. Performing the matching it
is convenient to put $\mu=Q$ and then to evolve it down to the values of the
hard-collinear scale $\mu\sim\sqrt{\Lambda Q}$.  The evolution is described
by the renormalization of the SCET operator, see e.g. \cite{Manohar:2003vb}. In what
follow we suggest to use simple notation
\begin{equation}
~f_{1}(Q,\mu=Q)\equiv~f_{1}(Q).
\end{equation}
Taking into account that the tree level coefficient function reads
\begin{equation}
C_{+}(Q,\mu=Q)=1+\mathcal{O}(\alpha_{s}),
\end{equation}
one obtains 
\begin{equation}
A_{ep}^{\gamma}=\frac{4\pi\alpha}{Q^{2}}~\bar{u}(k^{\prime})\gamma^{\mu}u(k)~~\bar
{N}(p^{\prime})\nbn\gamma_{\bot}^{\mu}N(p)~f_{1}(Q).
\end{equation}
Comparing this with the $A_{\gamma}$ in Eq.(\ref{def:Ag}) one finds (see  details in
Appendix \ref{tree})%
\begin{equation}
G_{M}(Q^{2})=f_{1}(Q),~~F_{2}(Q^{2})=\frac{4m^{2}}{Q^{2}}~f_{1}(Q). \label{GMf1}%
\end{equation}
The Pauli FF $F_{2}$ is suppressed by the factor $m^{2}/Q^{2}$ due to one unit of helicity flip.
Moreover, this expression is not complete because it also includes the
contribution from the subleading operator, see e.g.\cite{Kivel:2010ns},  which we do not
consider now for simplicity.\footnote{In
\cite{Kivel:2010ns}  the kinematical power corrections to the SCET\ FF $f_{1}$ were missed.
The complete leading power contribution is given by the sum $F_{2}=4m^{2}f_{1}
/Q^{2}+m^{2}C_{B}\ast f_{2}/Q^{2}$ where $C_{B}$ is the subleading coefficient functions, 
in the notation of Ref.~\cite{Kivel:2010ns}.}

The generalization of this approach to next-to-leading  order in QED can be
done along the same lines. It is convenient to include the soft-photon
field into the SCET Lagrangian similar to the soft-gluon field and perform the
expansion at the last step.  Then for the elastic amplitude we can write%
\begin{equation}
A_{ep}\simeq\left\langle p^{\prime},k^{\prime}\right\vert T\{\tilde{C}%
_{+}^{\mu\nu}\ast O_{+}^{\mu}O_{e}^{\nu}+\tilde{C}_{-}^{\mu\nu}\ast O_{-}%
^{\mu}O_{e}^{\nu}\}~\left\vert k,p\right\rangle _{\text{{\footnotesize SCET}}%
}, \label{Aep}%
\end{equation}
where the hard coefficient functions $C_{\pm}^{\mu\nu}$describe the  hard
subprocesses, the asterisk denotes the convolution integrals in position space
and the SCET operators $O_{\pm}^{\mu}$ describe the dynamics of
hard-collinear and soft particles and read:
\begin{equation}
O_{\pm}^{\mu}\equiv O_{\pm}^{\mu}(\lambda_{1},\lambda_{2})=\sum
_{\text{flavors}}\left\{  \bar{\chi}_{n}(\lambda_{1}\bar{n})\gamma_{\bot}%
^{\mu}\chi_{\bar{n}}(\lambda_{2}n)\pm\bar{\chi}_{\bar{n}}(\lambda_{2}%
n)\gamma_{\bot}^{\mu}\chi_{n}(\lambda_{1}\bar{n})\right\}  , \label{def:Opm}%
\end{equation}%
\begin{equation}
O_{e}^{\nu}\equiv O_{e}^{\nu}(\eta_{1},\eta_{2})=\bar{\chi}_{v}(\eta_{1}%
\bar{v})\gamma^{\nu}\chi_{\bar{v}}(\eta_{2}v),
\end{equation}
where we introduced
\begin{equation}
~\bar{\chi}_{v}=\bar{\zeta}_{v}W_{v},~\chi_{\bar{v}}=W_{\bar{v}}^{\dag}%
\zeta_{\bar{v}},
\end{equation}
with the hard-collinear photon WLs
\bea
W_{\bar{v}}(z)=\text{P}\exp\left\{  -ie\int_{-\infty}^{0}dt~v\cdot B_{hc}%
^{(\bar{v})}(z+ vt)\right\} ,
\\
W_{v}(z)=\text{P}\exp\left\{  -ie\int_{-\infty}%
^{0}dt~\bar{v}\cdot B_{hc}^{(v)}(z+\bar{v}t)\right\}  .
\eea
Taking into account that these WLs are not required for the TPE calculation we
can skip them assuming%
\begin{equation}
O_{\zeta}^{\nu}(\eta_{1},\eta_{2})\simeq\bar{\zeta}_{v}(\eta_{1}\bar{v}%
)\gamma^{\nu}\zeta_{\bar{v}}(\eta_{2}v).
\end{equation}
We also assume a similar simplification for the  operators $O_{\pm}^{\mu}$.
 This implies that we reduce our considerations only to TPE diagrams in
Fig.\ref{hss-diagrams}.\footnote{Let us note that each individual box diagram
has the regions associated with collinear photons. However, in the sum these
contributions cancel due to gauge invariance.} The  operators $O_{\pm}^{\mu}$
 represent the operators which arise at leading power and leading order in
$\alpha_{s}$.  The other operators are subleading and  provide only subleading  contributions
according to SCET power counting  or are subleading in $\alpha_{s}$.  A more detailed
consideration of this observation is provided in Appendix~\ref{power}.

The $C$-even operator $O_{-}^{\mu}$ can appear only due to the hard two-photon
exchange. The corresponding hard coefficient function $\tilde{C}%
_{-}^{\mu\nu}$ can be associated with the \ hard region in the TPE diagrams in
Fig.\ref{hss-diagrams}, i.e. $\tilde{C}_{-}^{\mu\nu}\sim\mathcal{O}(\alpha
^{2})$. Hence computing the  matrix element $\langle O^{\mu}_{-} O_{e}^{\nu}\rangle$ we can neglect 
the soft photons in the SCET Lagrangian because they provide only higher order $\mathcal{O}(\alpha^{3})$ contributions.   
Therefore the leptonic matrix element can be factorized as~:
\begin{equation}
\left\langle p^{\prime},k^{\prime}\right\vert T\{\tilde{C}_{-}^{\mu\nu}\ast
O_{-}^{\mu}O_{e}^{\nu}\}~\left\vert k,p\right\rangle
_{\text{{\footnotesize SCET}}}
=~\left\langle k^{\prime}\right\vert O_{e}^{\nu
}~\left\vert k\right\rangle_{\text{{\footnotesize SCET}}} \ast\tilde{C}_{-}^{\mu\nu}\ast\left\langle
p^{\prime}\right\vert O_{-}^{\mu}~\left\vert p\right\rangle
_{\text{{\footnotesize SCET}}}%
\label{EqOO}
\end{equation}%
The hadronic matrix element  gives the new
SCET FF which we define as follows~:%
\begin{equation}
\left\langle p^{\prime}\right\vert O_{-}^{\mu}(\lambda_{1},\lambda
_{2})\left\vert p\right\rangle _{\text{{\footnotesize SCET}}}=e^{i(p^{\prime
}\bar{n})\lambda_{1}-i(pn)\lambda_{2}}~\bar{N}(p^{\prime})\nbn
{4}\gamma_{\bot}^{\mu}N(p)~\mathcal{F}_{1}(Q,\mu). \label{def:F1}%
\end{equation}
The dependence on $Q$ and $\mu$ has to be understood in the same way as in
case FF $f_{1}$ defined in Eq.(\ref{def:f1}). 
Substituting this into Eq.(\ref{EqOO}) yields 
\bea
&&\left\langle p^{\prime},k^{\prime}\right\vert T\{\tilde{C}_{-}^{\mu\nu}\ast
O_{-}^{\mu}O_{e}^{\nu}\}~\left\vert k,p\right\rangle
_{\text{{\footnotesize SCET}}}=
\nonumber \\ 
&&\phantom{empty space }\frac{e^{2}}{Q^{2}}\bar{u}_{v}\gamma^{\nu}u_{\bar v}~\bar
{N}(p^{\prime})\nbn\gamma_{\bot}^{\mu}N(p)
 C_{-}^{\mu\nu}(z,Q^{2},\mu_{F},\mu)\mathcal{F}_{1}
(Q,\mu),~
\label{Fterm}
\eea
where $C_{-}^{\mu\nu}$ is the Fourier transformation of the coefficient
function $\tilde{C}_{-}^{\mu\nu}$ defined in position space, $u_{v,\bar{v}}$ 
denote large components of the electron spinors  $\Dslash{v}u_{v}=0,\  \Dslash{\bar v}u_{\bar v}=0$. Notice that such
approximations is exact in the limit of  massless electrons, $m_{e}=0$.   The coefficient function $C_{-}^{\mu\nu}$  depends 
on the two different  factorization scales $\mu_{F}$ and $\mu$.  The factorization scale  $\mu_{F}$  describes the factorization of the 
QED loop and is closely associated with the virtualities of the photons.  The factorization scale $\mu$  describes the factorization for QCD degrees of freedom.
We will set $\mu=Q$ and assume $\mathcal{F}_{1}(Q,\mu=Q)\equiv \mathcal{F}_{1}(Q)$. 
 The contribution given in Eq.(\ref{Fterm})  corresponds to  the first graph on $rhs$ of 
 Fig.\ref{tpe}. 

Consider now the first term in Eq.(\ref{Aep}).  This contribution must be
computed  taking into account the presence of the soft photons in the SCET
Lagrangians. The situation in the leptonic sector can be simplified if we
factorize the soft photons using the fields redefinition, see e.g.\cite{Bauer:2002nz}
\begin{equation}
\bar{\zeta}_{v}(\eta_{1}\bar{v})=\bar{\zeta}_{v}^{(0)}(\eta_{1}\bar{v}
)Y_{v}^{\dag}(0),~\ \zeta_{\bar{v}}(\eta_{2}v)=S_{\bar{v}}(0)\zeta_{\bar{v}
}^{(0)}(\eta_{2}v), 
\label{spdec}
\end{equation}
where the soft photon WLs reads
\bea
Y_{v}^{\dag}(0)=\text{P}\exp\left\{  -ie\int_{0}^{\infty}dt~v\cdot
B^{(s)}(tv)\right\} \\
 S_{\bar{v}}(0)=\text{P}\exp\left\{  -ie\int%
_{-\infty}^{0}dt~\bar{v}\cdot B^{(s)}(t\bar{v})\right\}  .
\eea
The new electron fields $\zeta_{v,\bar{v}}^{(0)}$ do not interact with the
soft photons anymore and therefore the leptonic  matrix element is factorized
\begin{align}
\left\langle p^{\prime},k^{\prime}\right\vert T\{\tilde{C}_{+}^{\mu\nu}\ast
O_{+}^{\mu}O_{e}^{\nu}\}~\left\vert k,p\right\rangle
_{\text{{\footnotesize SCET}}}  &  \simeq\tilde{C}_{+}^{\mu\nu}\ast
\left\langle p^{\prime}\right\vert T\{O_{+}^{\mu}(\lambda_{1},\lambda
_{2})Y_{v}^{\dag}(0)S_{\bar{v}}(0)\}~\left\vert p\right\rangle
_{\text{{\footnotesize SCET}}}~\nonumber\\
&  ~\ \ \ \ \  \times\left\langle k^{\prime}\right\vert
\{\bar{\zeta}_{v}^{(0)}(\eta_{1}\bar{v})\gamma^{\nu}\zeta_{\bar{v}}^{(0)}%
(\eta_{2}v)\}~\left\vert k\right\rangle _{\text{{\footnotesize SCET}}},
\label{firstme}
\end{align}
and can be easily computed\footnote{Here we do not
describe the formal details related to transition from SCET-I to SCET-II in
the leptonic sector.}
\begin{equation}
\left\langle k^{\prime}\right\vert \{\bar{\zeta}_{v}^{(0)}(\eta_{1}\bar
{v})\gamma^{\nu}\zeta_{\bar{v}}^{(0)}(\eta_{2}v)\}~\left\vert k\right\rangle
_{\text{{\footnotesize SCET}}}=e^{i\eta_{1}(\bar{v}\cdot k^{\prime})-i\eta
_{2}(v\cdot k)}~\bar{u}_{v}\gamma^{\nu}u_{\bar{v}}. \label{lept-me}%
\end{equation}
 The  hadronic matrix element  can be considered as a generalization of the leading
order  matrix element  defined in Eq.(\ref{def:f1}) in the presence of the
soft photons created by lepton source. Therefore we can write
\bea
&&\left\langle p^{\prime}\right\vert T\{O_{+}^{\mu}(\lambda_{1},\lambda
_{2})Y_{v}^{\dag}(0)S_{\bar{v}}(0)\}~\left\vert p\right\rangle
_{\text{{\footnotesize SCET}}}    =e^{i(p^{\prime}\bar n)\lambda_{1}
-i(pn)\lambda_{2}}~\bar{N}(p^{\prime})\nbn\gamma_{\bot}^{\sigma}N(p)
\nonumber\\
&&\phantom{empty} \times\left\{ 
g_{\bot}^{\mu\sigma} \left[
f_{1}(Q)+\frac{\alpha}{\pi}
g_{1}(z,Q)
\right]
+v_{\bot}^{\sigma}v_{\bot}^{\mu} \frac{\alpha}{\pi} \frac{4(z-2)}{z^{2}} g_{3}(z,Q)
\right\} ,
\label{g1:def}%
\eea
where the SCET amplitudes $g_{1,3}$ represent the next-to-leading QED correction arising
from the interaction of the soft photons with the hard-collinear and soft
spectator quarks. The coefficient $4 (z-2) / z^{2}$ is introduced for convenience. 
These functions also depend on the  factorization scales $\mu_{F}$ and $\mu$ which we 
do not show in Eq.(\ref{g1:def}) for simplicity. 
 Substituting (\ref{lept-me}) and (\ref{g1:def}) into
(\ref{firstme})  we obtain%
\begin{align}
&\left\langle p^{\prime},k^{\prime}\right\vert T\{\tilde{C}_{+}^{\mu\nu}\ast
O_{+}^{\mu}O_{e}^{\nu}\}~\left\vert k,p\right\rangle_{\text{{\footnotesize SCET}}} 
   =C_{+}^{\mu\nu}\bar{u}_{v}\gamma^{\nu}u_{\bar{v}}~\bar{N}(p^{\prime})\nbn\gamma_{\bot}^{\sigma}N(p)
\nonumber\\ &  ~\ \ \ \ \ \ \ \ \ \ \ \ \ \ \ \ \ \ \times
\left\{ 
g_{\bot}^{\mu\sigma} \left[
f_{1}(Q)+\frac{\alpha}{\pi}
g_{1}(z,Q)
\right]
+v_{\bot}^{\sigma}v_{\bot}^{\mu} \frac{\alpha}{\pi} \frac{4(z-2)}{z^{2}} g_{3}(z,Q)
\right\}  . \label{1st-me}%
\end{align}
The contributions with $g_{1,3}$  in Eq.(\ref{1st-me})  corresponds to  the second graph on $rhs$ 
of Fig.\ref{tpe}. 
The  coefficient function $C_{+}^{\mu\nu}~$ is fixed by the one-photon approximation. Using Eqs.(\ref{AgSCET}) one obtains
\begin{equation}
C_{+}^{\mu\nu}=g^{\mu\nu}\frac{4\pi\alpha}{Q^{2}}C_{+}(Q,\mu)=g^{\mu\nu
}\frac{4\pi\alpha}{Q^{2}}+\mathcal{O}(\alpha_{s})+\mathcal{O}(\alpha^{2}).
\label{CplLO}
\end{equation}
The terms of order $\mathcal{O}(\alpha^{2})$ includes pure  QED corrections like electron and photon self energies, photon-electron vertex corrections 
and also  hadron vertex and self-energy  which we will not recompute in this work. 
Using Eq.(\ref{CplLO}),  the TPE contribution to the amplitude $A_{ep}^{\gamma\gamma}$  can be written as
\begin{align}
A^{2 \gamma}_{ep}  &  =\frac{4\pi\alpha}{Q^{2}}\bar{u}_{v}\gamma^{\nu}u_{\bar{v}}~\bar
{N}(p^{\prime})\nbn\gamma_{\bot}^{\sigma}N(p)
\left\{ 
g_{\bot}^{\nu\sigma} \left[
f_{1}(Q)+\frac{\alpha}{\pi}
g_{1}(z,Q)
\right]
+\frac1s  K_{\bot}^{\sigma}P^{\nu} \frac{\alpha}{\pi} g_{3}(z,Q)
\right\} 
\nonumber\\
&  ~\ \ \ \ \ \ \ \ \ \ \ \ \ \ \ \ \ \ \ \ \ \ \ \ \ +%
\bar{u}_{v}\gamma^{\nu}u_{\bar v}~\bar{N}(p^{\prime})\nbn\gamma_{\bot}^{\mu}N(p)~C_{-}^{\mu\nu}(z,Q^{2})~\mathcal{F}_{1}(Q), \label{AepSCET}%
\end{align}
where we used that
\be{vv}
v_{\bot}^{\sigma}v_{\bot}^{\mu}\, \frac{\alpha}{\pi} \frac{4(z-2)}{z^{2}}\bar{u}_{v}\gamma^{\mu}u_{\bar{v}} \bar{N}(p^{\prime})\nbn\gamma_{\bot}^{\sigma}N(p)
=
\frac1s K_{\bot}^{\sigma}P^{\mu}\, \bar{u}_{v}\gamma^{\mu}u_{\bar{v}} \bar{N}(p^{\prime})\nbn\gamma_{\bot}^{\sigma}N(p).
\ee

The expression in (\ref{AepSCET}) is
our final expression for the soft spectator scattering contribution describing
 elastic $ep$-scattering amplitudes with the TPE corrections as in
Fig.\ref{hss-diagrams}.  In order to use this formula in phenomenological
 applications, we have to compute the hard coefficient function $C_{-}^{\mu\nu}$ 
 and estimate the SCET FFs which enter in Eq.(\ref{AepSCET}). 

\section{Calculation of the hard coefficient function}

\begin{figure}[h]
\centering
\includegraphics[
height=1.0212in,
width=2.4873in
]%
{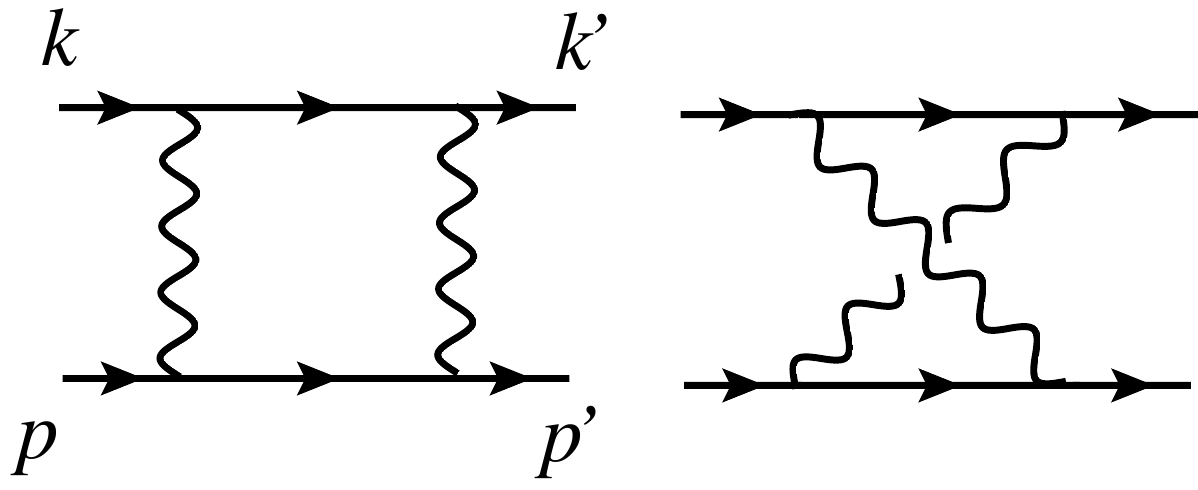}%
\caption{Two-photon exchange diagrams with external momenta required for
calculation of the hard coefficient function $C_{-}^{\mu\nu}$}%
\label{box-quarks}%
\end{figure}

The calculation of the hard coefficient function $C_{-}^{\mu\nu}$ can be done by
matching of the box diagram in Fig.\ref{box-quarks}.
In order to perform the matching we consider quark instead of
proton states.\footnote{More explicitly, we must consider quark plus antiquark
states. However antiquark contributions can be easily restored from the quark
calculation. We also consider one flavor with electric charge $e_{q}$.}. The
expressions for the diagrams with massless quarks are well known, see e.g.
\cite{VanNieuwenhuizen:1971yn, Chen:2004tw, Afanasev:2005mp} and read%
\begin{equation}
A_{eq}^{2 \gamma}=\frac{4\pi\alpha}{Q^{2}}\bar{u}_{v}\gamma_{\mu}u_{\bar  v}~
\bar{q}_{n}
\left\{  
\gamma^{\mu}\frac{\alpha}{\pi}e_{q}^{2}~\tilde{f}_{1}~+\frac{1}{s}P^{\nu}\Dslash{K}~\frac{\alpha}{\pi}e_{q}^{2}~\tilde{f}_{3}
\right\}  q_{\bar n}
+\bar v_{n}\{\dots\}v_{\bar n}, 
 \label{Aeqtpe}%
\end{equation}
where  and $q_{n,\bar{n}}$ and $v_{n,\bar{n}}$ denote the SCET spinors
for the quark and antiquark, respectively. The amplitudes $\tilde{f}%
_{i}$ read (recalling $z=-t/s$)%
\begin{equation}
\operatorname{Re}\tilde{f}_{1}=\ln\bar{z}\ln\frac{s}{\lambda^{2}}+\frac{1}%
{2}\ln^{2}\bar{z}-\frac{1}{4}\frac{z\ln^{2}z}{1-z}-\frac{z}{4} \ln^{2}  \frac{\bar
{z}}{z}-\frac{1}{2}\ln\bar{z}-\frac{z\pi^{2}}{4}+\frac{\pi^{2}}{2},
\label{f1t}%
\end{equation}%
\begin{equation}
\operatorname{Re}\tilde{f}_{3}=\frac{2-z}{2\bar{z}^{2}}\ln^{2}z-\frac{2-z}%
{2}\ln^{2}\frac{\bar{z}}{z}+\frac{1}{\bar{z}}\ln z+\ln\frac{\bar{z}}{z}%
-\frac{2-z}{2}\pi^{2}, \label{f3t}%
\end{equation}
where $\lambda^{2}$ denotes the artificial photon mass which plays the role of
IR scale. Let us note that $\lambda^{2}$ is an IR QCD
regulator and is not related with the IR QED regularization. We also do
not consider the imaginary parts because all discussed observables are sensitive
only to the real part of the TPE amplitudes.

\begin{figure}[h]%
\centering
\includegraphics[
height=0.7812in,
width=4.968in
]%
{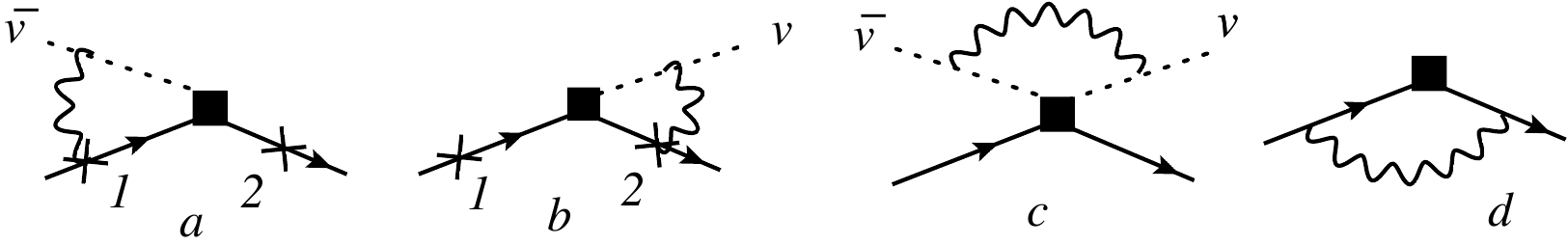}%
\caption{The  one-loop diagrams $a,b$ contribute to the
SCET perturbative function $g_{1}^{q}$ associated with the TPE contribution. The dashed line
with label $v$ and $\bar{v}$ denote the soft WLs, the crossed quark lines show
the different attachments of the soft gluon.
The diagrams $c$ and $d$ are related to the vertex corrections and  therefore were
neglected. }
\label{g1-renorm}
\end{figure}

From the discussion in the previous section, it follows that
the corresponding amplitude can be written as
\begin{eqnarray}
A_{eq}^{2 \gamma} &=& \frac{4\pi\alpha}{Q^{2}}\bar{u}_{v}\gamma^{\nu}u_{\bar{v}%
}~\left(  ~\bar{q}_{n}\gamma_{\bot}^{\sigma}q_{\bar{n}}+\bar{v}_{n}\gamma_{\bot
}^{\sigma}v_{\bar{n}}\right)  \nonumber \\
&\times&\left\{  
\frac{\alpha}{\pi}g_{1}^{q}(z,Q)
+ \frac 1s P^{\nu}K_{\perp}^{\sigma} g_{3}^{q}(z,Q)
+C_{-}^{\sigma\nu}(z,Q^{2})~\mathcal{F}_{1}^{q}(Q)
\right\}  ,
\label{Aeq-tpe}%
\end{eqnarray}
where the FFs $\mathcal{F}_{1}^{q}$ and  $g_{1,3}^{q}$ denote the SCET amplitudes computed
with quarks. FF $\mathcal{F}_{1}^{q}$ is easily
computed from the tree level diagrams and reads%
\begin{equation}
\mathcal{F}_{1}^{q}=e_{q}^{2}+\mathcal{O}(\alpha_{s}).
\end{equation}
The calculation of $g_{1}^{q}$ involves one-loop diagrams shown in
Fig.\ref{g1-renorm}.%
 These diagrams must be computed in SCET-I which describes the coupling
of the soft photon with the hard-collinear quarks. The one-loop diagrams in
Fig.\ref{g1-renorm} have an UV-divergence which is regularized using dimensional
regularization $D=4-2\varepsilon$. We present the details of this calculation
in Appendix~\ref{calculation}. After UV renormalization the result reads%
\begin{equation}
g_{1}^{q}(z,Q)=\ln\bar{z}\ln\frac{\mu_{F}^{2}}{\lambda^{2}}~\mathcal{F}_{1}^{q},\,  g_{3}^{q}(z,Q)=0.
\label{YSO}
\end{equation}
Formally, such contribution arises form the  mixing between the operators
$Y_{v}^{\dagger}S_{v} O_{+}^{\mu}$ and $O_{-}^{\mu}$ due to soft photon\ interactions. 

It is convenient to define the two independent coefficient functions $C_{M,3}$ as 
\begin{align}
C_{-}^{\mu\nu}(z,Q^{2})  &  =
g^{\mu\nu}_{\perp}\frac{\alpha}{\pi}C_{M}(z,Q^{2},\mu_{F})+\frac{1}{s}P^{\nu}K^{\mu}_{\perp}\frac{\alpha}{\pi} C_{3}(z).
\end{align}
Using the explicit results for $\mathcal{F}_{1}^{q}$ and $g_{1,3}^{q}$ and
comparing two expressions for $A_{eq}^{2 \gamma}$ given in Eqs.(\ref{Aeqtpe}%
) and (\ref{Aeq-tpe}) we obtain
\begin{align}
g^{\mu\nu}_{\perp}\frac{\alpha}{\pi}C_{M}(z,Q^{2},\mu_{F})+\frac{1}{s}P^{\nu}K^{\mu}_{\perp}\frac{\alpha}{\pi} C_{3}(z,Q^{2})
  =g^{\mu\nu}_{\perp}\frac{\alpha}{\pi}~\left(  \tilde{f}_{1}-\ln\bar{z}\ln\frac {\mu_{F}^{2}}{\lambda^{2}}\right)  ~+\frac{1}{s}P^{\nu}K^{\mu}_{\perp}~\frac{\alpha
}{\pi}~\tilde{f}_{3},
\end{align}
which yields%
\begin{equation}
C_{M}(z,Q^{2},\mu_{F})=\ln\bar{z}\ln\frac{s}{\mu_{F}^{2}}+\frac{1}{2}\ln^{2}%
\bar{z}-\frac{1}{4}\frac{z\ln^{2}z}{1-z}-\frac{z\ln^{2}\bar{z}/z}{4}-\frac
{1}{2}\ln\bar{z}-\frac{z\pi^{2}}{4}+\frac{\pi^{2}}{2}, \label{CM}%
\end{equation}%
\begin{equation}
C_{3}(z,Q^{2})=\frac{2-z}{2\bar{z}^{2}}\ln^{2}z-\frac{2-z}{2}\ln^{2}\frac{\bar{z}%
}{z}+\frac{1}{\bar{z}}\ln z+\ln\frac{\bar{z}}{z}-\frac{2-z}{2}\pi^{2}.
\label{C3}%
\end{equation}
Notice that the soft scale $\lambda^{2}$ cancel in the final expression for
$C_{M}$ as it should be. In Eq.(\ref{CM}) we write $\ln[s/\mu_{F}^{2}]$ with the total energy $s$ for convenience. 

Therefore our final result for the soft spectator contribution  to the TPE
amplitude can be written as%
\begin{align}
A^{2 \gamma}_{ep}  &  =\frac{4\pi\alpha}{Q^{2}}\bar{u}_{v}\gamma^{\mu}u_{\bar{v}}~\bar
{N}(p^{\prime})\nbn\gamma_{\bot}^{\mu}N(p)\left\{
~f_{1}(Q)+\frac{\alpha}{\pi}g_{1}(z,Q,\mu_{F})+\frac{\alpha}{\pi}C_{M}(z,Q^{2},\mu_{F})\mathcal{F}_{1}(Q)\right\} \nonumber\\
&  ~\ \ \ \ \ \ \ \ \ \ \ \ \ \ \ \ \ \ \ \ \ \ \ \ \ +
\bar{u}_ {v}\Dslash{P} u_{\bar{v}}~\bar{N}(p^{\prime})\nbn\Dslash{K}_{\bot}N(p)~\frac
{1}{s}\frac{\alpha}{\pi}
\left\{
g_{3}(z,Q)+C_{3}(z,Q^{2})\mathcal{F}_{1}(Q) 
\right \}. \label{Aep:fin}%
\end{align}

Let us now discuss the complete factorization formula describing the  TPE contribution.   We suggest  that 
 the TPE  amplitudes in  Eq.(\ref{def:Agg}) in the region of wide-angle scattering  (\ref{stu})  can be described by the following tentative factorization formula
\bea
\delta \tilde G^{2 \gamma}_{M}(\varepsilon,Q^{2})&=&\delta \tilde G^{(s)}_{M}(\varepsilon,Q^{2})+\delta \tilde G^{(h)}_{M}(\varepsilon,Q^{2}),
\label{def:dGM}\\
\tilde F_{3}(\varepsilon,Q^{2})&=&\tilde F^{(s)}_{3}(\varepsilon,Q^{2})+\tilde F^{(h)}_{3}(\varepsilon,Q^{2}),
\label{def:F3}\\
\delta\tilde  F^{2 \gamma}_{2}(\varepsilon,Q^{2})&=&\delta\tilde  F^{(s)}_{2}(\varepsilon,Q^{2}) +\delta\tilde F^{(h)}_{2}(\varepsilon,Q^{2}),
\label{def:dF2}
\eea
where the indices $(s)$ and $(h)$ denote the contributions related to  the soft and hard spectator scattering, respectively. 
 
 The hard spectator  contributions describe the scattering where  the all spectator quarks are involved  into the hard subprocess. 
 Schematically these contributions are shown by the third diagram on $rhs$ in Fig.\ref{tpe}.
 Schematically one can write%
 \bea
 && F_{3}^{(h)}    =\boldsymbol{\Psi}(x_{i})\ast {H}_{3}(z,Q^{2};\  x_{i},y_{i})\ast\boldsymbol{\Psi}(y_{i}),
  \label{H3}\\
&&\delta\tilde G_{M}^{(h)}   =\boldsymbol{\Psi}(x_{i})\ast {H}_{M} (z,Q^{2};\  x_{i},y_{i}) \ast\boldsymbol{\Psi}(y_{i}),
~\label{HM} 
\eea
where symbols  ${H}_{3,M}$ denote the hard coefficient functions and
the asterisk is used as notation for the  convolution integrals over the collinear
fractions $x_{i}, y_{i}$.  
 The leading order  coefficient functions $H_{3,M}$  were 
 computed  in Refs.~\cite{Borisyuk:2008db, Kivel:2009eg}.  
 The hard spectator contributions  are dominated by one hard gluon exchange, see Fig.\ref{hss-diagrams}, and 
are of order   $\alpha_{s}$. 
 The  nucleon distribution amplitude  $\boldsymbol{\Psi}(x_{i})$  describes
 the  nonperturbative  overlap of the collinear quarks with the nucleon state.  
  The hard spectator contribution to the helicity flip amplitude $F_{2}^{(h)}$ can be written in a  
  similar way but it is power suppressed. Furthermore, its 
the convolution integrals are ill defined due to the end-point  singularities.  Therefore we will not consider 
the helicity flip amplitude  in the present publication.   

The soft spectator contributions describe the scattering where the spectator quarks are soft.  
These TPE contributions have been discussed above.  
Comparing  Eq.~(\ref{def:Agg}) with  Eq.~(\ref{Aep:fin}) (the matching is quite
similar to one described in Appendix A) we obtain
\begin{equation}
\delta \tilde G^{(s)}_{M}(\varepsilon,Q^{2})=\frac{\alpha}{\pi}\left\{  ~g_{1}(z,Q,\mu_F%
)+C_{M}(z,Q^{2}, \mu_F)\mathcal{F}_{1}(Q)\right\}  ,~ \label{dGM}%
\end{equation}%
\begin{equation}
\frac{\nu}{m^{2}}\tilde F_{3}^{(s)}=\frac{\alpha}{\pi}\frac{\nu}{s}
\left\{
g_{3}(z,Q)+
C_{3}(z,Q^{2})\mathcal{F}_{1}(Q)
\right\}, \label{F3}
\end{equation}%
where we assume $z\equiv z(\varepsilon)$ and $\nu$ defined in Eq.~(\ref{nu}).

The soft contribution to the helicity flip  amplitude is also suppressed by the power of 
the large scale $1/Q^{2}$ and can be written as
 (for simplicity we do not write the arguments on the $rhs$)
\bea
\delta \tilde F^{(s)}_{2}(\varepsilon,Q^{2})=\left[\delta \tilde F^{(s)}_{2}\right]_{\text{subl}}
+\frac{\alpha}{\pi}\frac{4m^{2}}{Q^{2}}
\left[
~g_{1}+C_{M}\mathcal{F}_{1}+
\frac{\nu}{s}\left\{ C_{3}\mathcal{F}_{1}+g_{3}\right\}
\right].
\label{dF2s}
\eea
  In  this expression we computed  only the kinematical
corrections similar to the Pauli FF $F_{2}$ in Eq.(\ref{GMf1}).  The complete answer includes also
contribution  with the matrix elements of the subleading SCET operators $\left[\delta \tilde F^{(s)}_{2}\right]_{\text{subl}}$  which we do not consider  for simplicity. 

As we can see from Eqs.(\ref{dGM},\ref{F3}) the two leading power amplitudes are described by the three nonperturbative functions: the SCET FF $\mathcal{F}_{1}(Q)$  and 
the SCET amplitudes $g_{1,3}(z,Q)$.  If   both 
photons are hard then the soft spectator scattering is described by the pure  QCD  sector. 
In this case  the hard-collinear  dynamics involves  only the one hard scale $Q$
and is described  by the SCET FF   $\mathcal{F}_{1}(Q)$. Therefore the $\varepsilon$-dependence  is described by the hard coefficient functions   $C_{3,M}$ and can 
be computed in  perturbative QCD.   
The situation is different when  the one photon is soft. Such soft photon interacts with  the hard-collinear and  soft  constituents 
and therefore the soft-overlap  contribution is more complicated and is described by the SCET amplitudes $g_{1,3}(z,Q^{2})$.  
We recall that we assume that the hard-collinear scale  is not large and  we can not use it for  perturbative calculations. 
 In this case the $\varepsilon$($z$)-dependence originates from the soft dynamics  and can not be computed  from pQCD.

Another difficulty is related to the separation of the  amplitudes for the soft and hard spectator contributions.  
The  simple sums  as  in Eqs.(\ref{def:dGM}-\ref{def:dF2}) 
are motivated by the  structure of the SCET-I  operators. However  the  overlap of the soft and collinear sectors in SCET-II  makes such separation ambiguous.  
As a rule this leads to  the end-point singularities in the collinear integrals defining the hard spectator terms.  In Ref.\cite{Kivel:2012mf} it was discussed that such situation 
even occurs for the  FF $F_{1}$ where one could expect that the hard spectator contribution  is well defined.  
 In general,  the separation between the soft and hard spectators contributions can be well 
formulated  only within a certain regularization scheme  which allows to treat  the soft and collinear sectors separately and  consistently.  
From this point of view our results  are
still not  complete  because we  did not provide  such a separation scheme.  
This question will be discussed  in the next two sections.

Finally let us note that the QED dynamics considered here is closely connected
with the underlying QCD dynamics. In particular  the energy and virtuality  of the soft
photons can not be smaller than the typical QCD soft scale $\Lambda$, because
otherwise the photon interacts with the total proton charge and can not resolve
the constituent charges inside the target. Therefore our  IR regulator
$\lambda$ (introduced for convenience as photon mass) must cancel in the
hard coefficient function which describes the scattering of the hard photons.
The standard QED IR-divergence  is included in the SCET amplitude $g_{1}$ which is
defined via the matrix element in Eq.(\ref{g1:def}). 

Let us also note that each separate diagram in Fig.\ref{box-quarks} also receives 
contributions associated with the collinear regions (one of the photons is
collinear to  one of the external momenta). Moreover these collinear regions
overlap with the soft one. However in the sum of the diagrams, the collinear
contributions cancel and this cancellation is required by gauge invariance
because the electric charges of the quarks and leptons are different.  A 
detailed  discussion of this point can be found in Appendix~\ref{cancellation}.

In conclusion, we suggested a factorization formula for the TPE soft spectator
scattering contribution. We performed the matching and computed the coefficient functions in leading
order in approximation $\alpha_{s}$.  In order to apply this result
in a phenomenological analysis, we next need to define  the SCET amplitudes  which arise in
our approach.

\section{Estimate of the soft spectator contribution with  $\mathcal{F}_{1}$  from wide-angle Compton scattering data}
\label{sec-wacs}

\subsection{QCD factorization for the wide-angle Compton scattering process}

The SCET FF  $\mathcal{F}_{1}(Q)$  defined   in   Eq.(\ref{def:F1})  describes the long distance interactions in 
 the  soft spectator  contribution with  the hard photons.
  In order to estimate this quantity for different values
of $ Q^{2}$  one  can use the universality of its definition. In other words if
one can find another process which is described by the same SCET FF then one can use the corresponding observables in order 
to estimate  $\mathcal{F}_{1}(Q)$.  
In  our  case  such analysis can be carried out for  the  wide-angle Compton scattering (WACS).  The cross section of this
reaction and  a polarization asymmetry in the region of large Mandelstam
variables (large scattering angles $\theta_{cm}\sim90^{o}$ in the cms frame )
 have been measured at JLab \cite{Danagoulian:2007gs}.  

\begin{figure}[h]%
\centering
\includegraphics[
height=0.7572in,
width=1.4545in
]%
{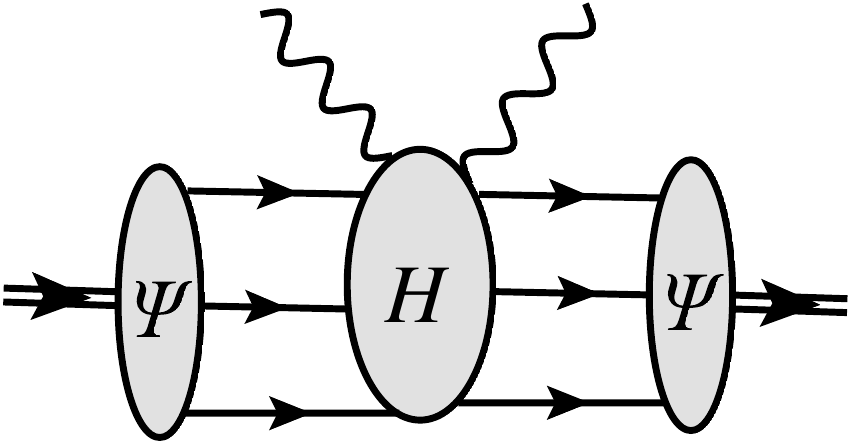}%
\caption{Reduced diagram describing the hard spectator scattering contribution
in the WACS.}%
\label{hss-wacs}%
\end{figure}

\begin{figure}[h]%
\centering
\includegraphics[
height=0.9323in,
width=2.4989in
]%
{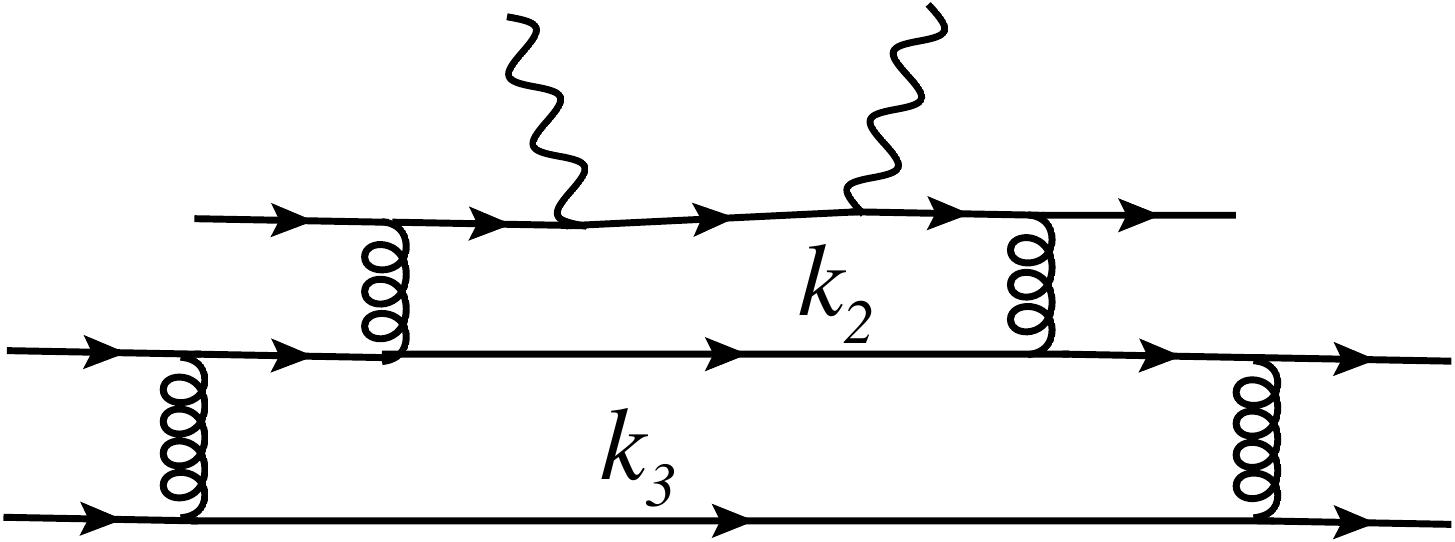}%
\caption{Example of the diagram which has the leading power contribution from
the region with the soft spectators $k_{2\mu}\sim k_{3\mu}\sim\Lambda$}%
\label{wacs-2loop}%
\end{figure}

The theoretical  description of this reaction has been
presented long time ago \cite{Brodsky:1973kr, Matveev:1973ra}  and it is based on the
dominance of the hard spectator scattering which leads to the collinear
factorization for the dominant amplitudes.  In this case the situation is
very similar to the situation with the nucleon electromagnetic FFs. At large
energy and momentum transfer $s\sim-t\sim-u\gg\Lambda^{2}$  the leading power
contribution to the WACS amplitude $T_{i}$ can be described as a collinear convolution of the hard
coefficient function $H_{i}(s,t)$ with the nucleon DAs $\boldsymbol{\Psi}$ 
\be{Thss}
T^{(h)}_{i}(s,t)=\boldsymbol{\Psi}(y_{i})*H_{i}(s,t;~ x_{i},y_{i})*\boldsymbol{\Psi}(x_{i})
\ee
where the asterisk denotes the  convolution integrals with respect to the collinear fractions $x_{i}, y_{i}$.  The reduced diagram describing the
hard spectator scattering is shown in Fig.\ref{hss-wacs}.%
Besides the hard spectator scattering, one can easily find that the soft spectator scattering
can also contribute to the same power and that this situation is quite similar
to the one discussed for the FF $F_{1}$ in \cite{Dun1980, Kivel:2010ns}. For instance, one
can investigate the soft region for the diagram in Fig.\ref{wacs-2loop} in the
same way as it was done for the analogous diagram in case of the FF $F_{1}$ in
\cite{Dun1980, Kivel:2010ns}.%
Such analysis allows to see
that the contribution from the domain where  momenta
$k_{2\mu}\sim k_{3\mu}\sim\Lambda$ are soft is suppressed by the same power of
$Q^{2}$ as from the domain where these momenta are hard: $k_{2\mu}\sim k_{3\mu}\sim Q$.
 This observation provides the arguments   that the complete factorization  is described by the soft   
and hard spectator contributions where the last one is  described by  the collinear factorization suggested in \cite{Brodsky:1973kr}.
   Moreover, a phenomenological analysis of
the experimental data \cite{Hamilton:2004fq, Danagoulian:2007gs} allows 
one to conclude that the dominant
contribution in the cross section is provided  by the soft spectator
scattering mechanism.  The best description of the existing WACS data has been
achieved using  the so-called handbag or GPD approach 
\cite{Radyushkin:1998rt,Diehl:1998kh,Huang:2001ej, Miller:2004rc}  which
 can be considered as a model for the soft spectator scattering. However, the
systematic analysis within this approach is rather problematic because  the
parametrization of the  soft overlap contribution by the GPD matrix element
 is not consistent with the IR-structure of the QCD diagrams, see for
instance \cite{Huang:2001ej}.

In the current section  we consider the description of the WACS  in the SCET
framework. 
\begin{figure}[h]%
\centering
\includegraphics[
width=5.0in
]%
{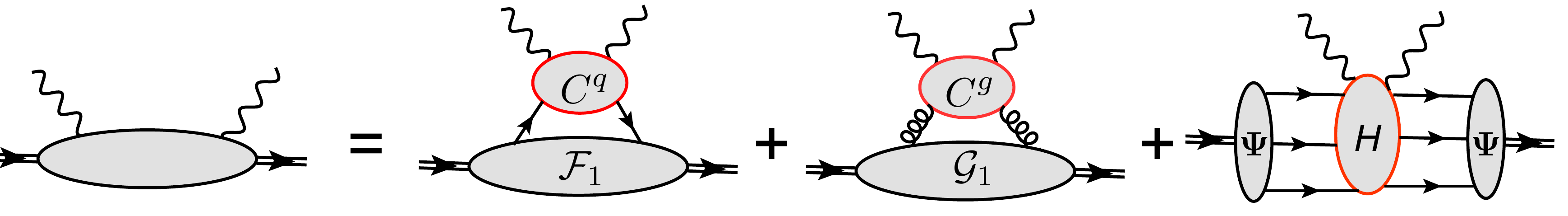}%
\caption{Schematic representation of the WACS factorization.  The
soft-overlap contributions are described by the SCET FFs  $\mathcal{F}_{1}$ and $\mathcal{G}_{1}$. }%
\label{wacs-scet}%
\end{figure}
Such consideration is quite similar to the  analysis of the nucleon FFs carried out in
 \cite{Kivel:2010ns}.  
We expect that in the region $s\sim-t\sim-u\gg\Lambda^{2}$   the WACS amplitudes, 
denoted by $T_{i}$, can be described
within SCET approach by  the following tentative factorization formula
\be{LOfact}
T_{i}(s,t)=T^{(s)}_{i}(s,t)+T^{(h)}_{i}(s,t), 
\ee
where the hard spectator contribution $T^{(h)}$  is given by Eq.(\ref{Thss}).  The soft spectator term  reads~:
\be{Tisoft}
T^{(s)}_{i}(s,t)=C^{q}_{i}(s,t,\mu) \mathcal{F}_{1}(Q,\mu)+C^{g}_{i}(s,t,\mu) \mathcal{G}_{1}(Q,\mu),
\ee
where the hard coefficient functions  $C_{i}^{q,g}$  describe a hard subprocess, the functions  $ \mathcal{F}_{1}$ and $\mathcal{G}_{1}$ 
describe the hard-collinear and soft interactions and can be defined  in the framework of  
the SCET-I approach. 
We provide their explicit definitions below.   
In Fig.\ref{wacs-scet} we illustrate  the WACS factorization in terms of  reduced diagrams.

In order to proceed  further let us introduce the following  definitions.    
The amplitude of the process $\gamma(k)+p(p)\rightarrow\gamma(k^{\prime
})+p(p^{\prime})$ can be written as \cite{Babusci:1998ww}%
\begin{equation}
\left\langle p^{\prime},k^{\prime}~out\right\vert \left.  in~k,p\right\rangle
=i\left(  2\pi\right)  ^{4}~\delta(p+k+p^{\prime}+k^{\prime})~\varepsilon
^{\ast\mu}\left(  k^{\prime}\right)  \varepsilon^{\nu}\left(  k\right)
e^{2}\, T^{\mu\nu},
\end{equation}
with 
\begin{align}
 T^{\mu\nu}   =i\int d^{4}x~e^{-i(kx)}%
\left\langle p^{\prime}\right\vert T\{J_{em}^{\mu}(0)J_{em}^{\nu
}(x)\}\left\vert p\right\rangle\ ,
\label{TJJ}
\end{align}
where $J_{em}^{\mu}$ represents the quark electromagnetic current
\begin{equation}
J_{em}^{\mu}(x)=\sum_{q}e_{q}~\bar{q}(x)\gamma^{\mu}q(x).
\end{equation}
The hadronic matrix element $T^{\mu\nu }$ is parametrized in terms of six independent scalar amplitudes $T_{i}$ as 
\begin{align}
T^{\mu\nu} = & \bar{N}(p^{\prime})\left\{ -\frac{P^{\prime\mu}P^{\prime\nu}}{P^{\prime2}%
}\left(  T_{1}+\Dslash{K}~T_{2}\right)  -\frac{N^{\mu}N^{\nu}}{N^{2}}\left(
T_{3}+\Dslash{K}~T_{4}\right)\right. \nonumber\\
&\left.  +\frac{P^{\prime\mu}N^{\nu}-P^{\prime\nu}N^{\mu}}{P^{\prime2}K^{2}}%
i\gamma_{5}~T_{5}+\frac{P^{\prime\mu}N^{\nu}+P^{\prime\nu}N^{\mu}}{P^{\prime
2}K^{2}}~i\gamma_{5}\Dslash{K}~T_{6}\right\} N(p). \label{Ai}%
\end{align}
In the last formula we used
\[
P=\frac{1}{2}(p+p^{\prime}),~P^{\prime}=P-K\frac{(P.K)}{K^{2}},~K=\frac{1}%
{2}(k+k^{\prime}),
\]%
\[
N^{\mu}=\varepsilon\lbrack\mu\alpha\beta\gamma]P^{\alpha}\frac{1}%
{2}(p-p^{\prime})^{\beta}K^{\gamma},~\ \varepsilon_{0123}=+1,
\]%
\begin{equation}
s=(p+k)^{2},~t=(p^{\prime}-p)^{2}=-Q^{2},~z=\frac{-t}{s}\simeq\frac{1}%
{2}(1-\cos\theta_{cm}).~\
\end{equation}
One can see from Eq.(\ref{Ai}) that three  amplitudes $T_{2,4,6}$ describe the Compton scattering without nucleon helicity flip, while the three others describe the amplitudes with  helicity flip. 

We again consider a Breit-like frame and therefore the light-cone expansions of the
momenta can be easily obtained from  Eq. (\ref{mom2})  substituting  the lepton momenta $k$ and $k'$ by photon momenta $q$ and $q'$, respectively. 
Therefore the formulation of the appropriate SCET  degrees of freedom in the collinear sectors associated with $n$ and $\bar n$ 
light-cone directions is the same as discussed
in Section~\ref{sec-scet}.

In the following  we   only discuss the factorization for the soft spectator contribution. 
The SCET-I  factorization  implies that the $T$-product of the  electromagnetic currents in Eq.(\ref{TJJ}) can be represented as the convolution of the coefficient function with
a certain SCET-I operator constructed from the SCET-I fields  $\xi_{n},\ A^{(n)}_{\mu},\ \xi_{\bar n},\ A^{(\bar n)}_{\mu}$.  This schematically can be written as 
\begin{align}
T\{J_{em}^{\mu}(0)J_{em}^{\nu}(x)\}  =   \sum_{i}  \tilde C^{\mu\nu}_{i}\ast O_{i}[\xi_{n},A^{(n)}, \xi_{\bar n}, A^{(\bar n)}],
\label{TJJscet}
\end{align}
where the asterisk  denotes the convolution integrals in the coordinate space. 
We assume that  the dominant contribution in the wide-angle kinematics  is provided by appropriate  operators with the minimal
dimension according to SCET-I counting rules, see  Refs.\cite{Bauer:2000ew,Bauer2000,Bauer:2001ct,Bauer2001,BenCh, BenFeld03}.  
Furthermore in our  analysis of the soft spectator scattering contribution we consider as leading power terms  only the contributions 
associated with the SCET-I counting, i.e.  we consider the powers of the large scale $Q$ arising from factorization of the hard modes. 
We do not consider the full power counting analysis which includes also the SCET-II counting. 
We consider the region where the hard-collinear scale is still small, and therefore we can not perform an expansion  with respect to this  scale.     

In our case we find that the set of the  suitable operators which is consistent with the symmetries  and has the minimal SCET-I dimension consists   
 of the quark  operator
$O_{-}^{\alpha}(\lambda_{1},\lambda_{2})$ defined in Eq.(\ref{def:Opm}) and  of the pure gluon operator $O_{g}^{\alpha\beta}(\lambda_{1},\lambda_{2})$ 
which includes  the transverse gluon  fields   $\ A^{(n)}_{\mu\perp},\ A^{(\bar n)}_{\mu\perp}$ and can be written as 
\be{Og}
O_{g}^{\alpha\beta}(\lambda_{1},\lambda_{2})= \text{Tr} \{ [W^{\dagger}_{n}(\lambda_{1}\bar n)D_{\alpha\perp}W_{n}(\lambda_{1}\bar n)]
[W^{\dagger}_{\bar n}(\lambda_{2} n)D_{\beta\perp}W_{\bar n}(\lambda_{2} n)] \},
\ee 
where the covariant derivative $iD_{\mu}=i\partial_{\mu}+gA^{(n,\bar n)}_{\mu}$ acts on the fields only inside the square brackets $[...]$.  
Below we restrict our discussion only to the leading order in $\alpha_{s}$. Then we can skip the 
discussion of the gluon operator 
because the corresponding hard  coefficient function arises only at the next-to-leading order. Therefore at leading order,  we only have a  
contribution only due to the quark operator   $O_{-}^{\alpha}(\lambda_{1},\lambda_{2})$ that yields
\begin{align}
T\{ J_{em}^{\mu}(0)J_{em}^{\nu}(x) \}  \simeq   \tilde C_{h}^{\mu\nu\alpha}\ast O_{-}^{\alpha}(\lambda_{1},\lambda_{2}).
\end{align}
Substituting this into the matrix element in Eq.(\ref{TJJ}) and using  Eq.(\ref{def:F1})  we obtain   
 \begin{align}
T^{\mu\nu} \simeq  C_{h}^{\mu\nu\alpha}(s,t)~\bar{N}(p^{\prime})\nbn\mathcal{\gamma}_{\bot}^{\alpha}N(p)\mathcal{F}_{1}(Q),
\label{TCF}
\end{align}
where $ C_{h}^{\mu\nu\alpha}(s,t)$ corresponds to the coefficient function in momentum space. From Eq.(\ref{TCF})  we can conclude  that  
 all  six amplitudes $T_{i}$  introduced in Eq.(\ref{TJJ}) are described in terms of one SCET FF $\mathcal{F}_{1}$. However helicity flip structures 
 in this case  can be obtained only as subleading contributions.  We also observe that the 
 dependence on the total energy $s$  
 is completely described   by  the hard coefficient function.    
 
\begin{figure}[ptb]%
\centering
\includegraphics[
height=0.3811in,
width=3.0394in
]%
{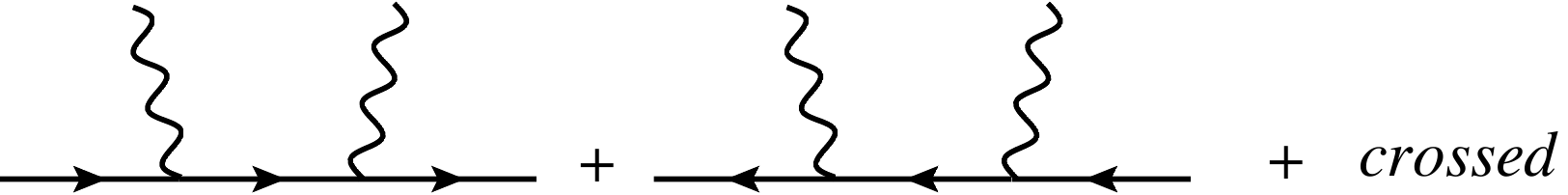}%
\caption{Leading order diagrams to the hard coefficient function describing the
soft spectator contribution  in WACS. }%
\label{wacs-born}%
\end{figure}

 In order to compute  $ C_{h}^{\mu\nu\alpha}$ one has to perform the matching for the tree level diagrams shown in 
 Fig.\ref{wacs-born}.
We skip the details of this relatively simple calculation 
 and provide the  results for the WACS amplitudes defined in Eq.(\ref{Ai}).
 For  the invariant amplitudes  $T^{(s)}_{i}$  we obtain ($Q=\sqrt{-t}$, $m$ is nucleon mass)
\begin{equation}
T^{(s)}_{2}=-T^{(s)}_{4}=-\frac{2s+t}{s(s+t)}
~\mathcal{F}_{1}(Q),~~~T^{(s)}_{6}=-\frac{t}{s(s+t)}\mathcal{F}_{1}(Q),\ \label{A246}%
\end{equation}%
\begin{align}
T^{(s)}_{1}    =\frac{m}{-t}\frac{1}{2}\frac{(2s+t)^{2}}{s(s+t)}~\mathcal{F}
_{1}(Q)+\bar T^{(s)}_{1}, ~
T^{(s)}_{3}  =\frac{m}{t}\frac{1}{2}\frac{(2s+t)^{2}}{s(s+t)}~\mathcal{F}_{1}(Q)+\bar T^{(s)}_{3},~ 
 T^{(s)}_{5}=\bar T^{(s)}_{5}. \label{A135}, 
\end{align}%
where the terms $\bar T^{(s)}_{i}$ appearing  on $rhs$ of Eqs.(\ref{A135}) 
denote contributions  associated with the subleading SCET operators  which have not been considered in our analysis. 
From these equations one can conclude  that  helicity flip amplitudes $T^{(s)}_{1,3,5}$ are suppressed by a power of
$1/Q$ (to see this explicitly one has to pass to dimensionless amplitudes in Eq.(\ref{Ai})).


The separation of the hard and soft contributions as described  in Eq.(\ref{LOfact}) is  not  simple  because it  usually implies 
an additional regularization. Such regularization  is required  in order to separate the  soft and collinear sectors and to describe correctly 
the so-called large rapidity logarithms, see for instance  Refs.\cite{Collins:1999dz,Manohar:2006nz,Becher:2011pf,Chiu:2012ir} and references therein.   
Such problem is clearly seen when one computes the hard spectator 
contributions, resulting in collinear convolution integrals which are divergent in the end-point region. 
This is a clear indication that one has to consider the overlap of the soft and hard spectator contributions. 

 In Ref.\cite{Kivel:2012mf}   the soft and hard spectator contributions 
 describing the FF $F_{1}$ has been investigated  and  it was demonstrated that in this case the problem 
 of the soft-collinear overlap also exists.  
We suppose that  collinear factorization describing the hard spectator contribution in the TPE  and WACS  
 is also violated  due to specific  end-point divergencies.    
 Therefore  we  suppose that the  most optimal way  to proceed further is to explore  the so-called  physical subtraction scheme suggested in 
  \cite{Beneke:2000ry, Beneke:2000wa}  and used in the description of  different   $B$-decay processes.  
    
  The  idea of this approach is  to exclude  the FF  $\mathcal{F}_{1}$  from the expression for the physical amplitudes and write the relations between the physical
  amplitudes directly. In our case, using Eqs.(\ref{Tisoft}) (neglecting the gluon contribution at leading order) and  Eqs.(\ref{Thss}) we can write
  \be{F1T2}
  \mathcal{F}_{1}(Q)= \frac{T_{2}(s,t)}{C_{2}(s,t)}-\boldsymbol{\Psi}(y_{i})*\frac{H_{2}(s,t;~ x_{i},y_{i})}{{C_{2}(s,t)}}*\boldsymbol{\Psi}(x_{i}).
  \label{F1T2C2}
  \ee  
  Such formal expression implies that we use some regularization in order to  define the  divergent quantities in both sides of  (\ref{F1T2}).  
  Substituting expression  (\ref{F1T2})  into   Eq.~(\ref{Tisoft}) for $T_{4,6}$ we obtain
 \bea
 &&T_{i}(s,t)=C_{i}(s,t)\frac{T_{2}(s,t)}{C_{2}(s,t)}
 \nonumber \\ &&
\phantom{empty} +\boldsymbol{\Psi}(y_{i})*\left\{H_{i}(s,t;~ x_{i},y_{i})-\frac{H_{2}(s,t;~ x_{i},y_{i})}{{C_{2}(s,t)}}\right\}*\boldsymbol{\Psi}(x_{i}), \ i=4,6.
 \label{TiT2}
 \eea
 The  end-point singularities  arising in the collinear  convolution 
 integrals  $\boldsymbol{\Psi}*H_{i}*\boldsymbol{\Psi}$ must cancel in the expression in 
 Eq.(\ref{TiT2}) because the other terms with the physical amplitudes  do not have  any  problems.  
    Hence  we can study the relations between the physical amplitudes $T_{i}$  systematically order by order in $\alpha_{s}$. 
 Taking into account that the coefficient functions $H_{i}$ are of order $\alpha^{2}_{s}$ and therefore  assuming that the corresponding corrections in Eq.~(\ref{TiT2})   are  
 small we obtain the leading order relations
 \bea
 T_{4}(s,t)&=&C_{4}(s,t)\frac{T_{2}(s,t)}{C_{2}(s,t)}=-T_{2}(s,t)+\mathcal{O}(\alpha_{s}), 
\label{T4T2LO} \\  
 T_{6}(s,t)&=&C_{6}(s,t)\frac{T_{2}(s,t)}{C_{2}(s,t)}=\frac{t}{2s+t} T_{2}(s,t)+\mathcal{O}(\alpha_{s}).
 \label{T6T2LO} 
 \eea
  where the coefficient functions $C_{2,4,6}$  have been computed above to the leading order accuracy.  
  Let us also note that the  next-to-leading  corrections to these relations arise from the corrections  to the coefficient functions $C_{2,4,6}$. 
  
 The nice feature  is that the SCET FF  $\mathcal{F}_{1}$ does not depend on $s$. 
 Therefore this allows one  to substitute  Eq.(\ref{F1T2}) into other expressions 
 involving the FF  $\mathcal{F}_{1}$, allowing to make predictions 
  for  different values of $s$.  This is  important for our case because the cross sections  for the elastic $ep$-scattering have been measured  for different values of $s$.  
  
  
  From Eqs.(\ref{T4T2LO},\ref{T6T2LO}) one can easily  find that to leading order accuracy the ratios $T_{i}/C_{i}$  satisfy
   \be{TC}
   \frac{T_{2}(s,t)}{C_{2}(s,t)}\simeq \frac{T_{4}(s,t)}{C_{4}(s,t)}\simeq \frac{T_{6}(s,t)}{C_{6}(s,t)}.
   \ee
  Therefore it is convenient  to introduce the following notation
  \be{def:R2}
  \mathcal{R}(s,t)=\frac{T_{2}(s,t)}{C_{2}(s,t)}=-\frac{s(s+t)}{2s+t}T_{2}(s,t)+\mathcal{O}(\alpha_{s}).
  \ee
  Contrary to the SCET FF  $\mathcal{F}_{1}$  the ratio $\mathcal{R}$  is  a well defined 
  quantity  free from the rapidity  regularization dependence.\footnote{ We assume that the coefficient function $C_{2}$ in Eq.(\ref{def:R2}) is computed at  factorization scale $\mu^{2}=-t$. } 
  This ratio formally depends on the 
two variables $s$ and $t$. However, in the region where the QCD factorization is applicable and 
where power and higher order in $\alpha_{s}$ corrections are small  the dependence on $s$ must be small, i.e. we must observe that 
\be{dR2}
\frac{d \mathcal{R}(s,t)}{ds}\simeq 0,
\ee
which provides a good criterium of the applicability of our approximations.  The ratio  $\mathcal{R}$ for the different values of  $s$ can be extracted from  the cross section  for
WACS measured at JLab   \cite{Danagoulian:2007gs}.  
Then this ratio can be used for the calculation of the TPE amplitude  defined  in  Eqs.(\ref{dGM}, \ref{F3}).

\subsection{Phenomenological analysis of the WACS observables  and extraction of the ratio $\mathcal{R}$}

The unpolarized cross section describing Compton scattering  reads \cite{Babusci:1998ww}
\begin{equation}
\ \frac{d\sigma}{dt}=\frac{\pi\alpha^{2}}{(s-m^{2})^{2}}W_{00}, \label{dsig}%
\end{equation}
with%
\begin{align}
W_{00}  &  =\frac{1}{2}(s-m^{2})(m^{2}-u)(|T_{2}|^{2}+|T_{4}|^{2}%
)+(m^{4}-su)|T_{6}|^{2}\nonumber\\
&  +m(s-u)\operatorname{Re}\left[  T_{1}T_{2}^{\ast}+T_{3}T_{4}^{\ast}\right]
+\frac{1}{2}(4m^{2}-t)(|T_{1}|^{2}+|T_{3}|^{2})-t|T_{5}|^{2}. \label{W00}%
\end{align}
Substituting the expressions for $T_{i}$ into Eq.~(\ref{W00}) and neglecting the
subleading power corrections we obtain
\begin{equation}
\frac{d\sigma}{dt}\simeq\frac{2\pi\alpha^{2}}{(s-m^{2})^{2}}\left(  \frac
{1}{1-z}+1-z\right)  |\mathcal{R}|^{2}=\frac{d\sigma^{\text{KN}}}%
{dt}|\mathcal{R}|^{2}, \label{wacs-c-s}%
\end{equation}
where $d\sigma^{\text{KN}}$ is the Klein-Nishina cross section corresponding with a 
point-like massless particle. This formula is very similar to the one obtained in
the  handbag approach \cite{Radyushkin:1998rt} except for the definition of the form factor.

 There are two more observables which describe the correlations of the recoil
polarization with the polarization of the photons
\begin{equation}
K_{LL}=\frac{\sigma_{\Vert}^{R}-\sigma_{\Vert}^{L}}{\sigma_{\Vert}^{R}%
+\sigma_{\Vert}^{L}},~\ K_{LS}=\frac{\sigma_{\bot}^{R}-\sigma_{\perp}^{L}%
}{\sigma_{\bot}^{R}+\sigma_{\perp}^{L}},
\end{equation}
where $\Vert (\perp)$ refers to a longitudinally (transversely) polarized nucleon and $R(L)$  denotes a right (left) handed photon.
For the first one we obtain (again neglecting  the power corrections)%
\begin{equation}
K_{LL}\simeq K_{LL}^{\text{KN}}=\frac{2~z\left(  2-z\right)  }{\left(
2-z\right)  ^{2}+z^{2}}. \label{KLL}%
\end{equation}
In this case the SCET FF cancels out and we simply reproduce the Klein-Nishina result for
the point-like massless particle.  The second asymmetry $K_{LS}$ also depends 
on the nucleon helicity flip amplitudes $T_{1,3,5}$.  In order to estimate $K_{LS}$  
we use for these amplitudes the  expressions obtained in Eqs. (\ref{A135}) 
where we neglect the contributions associated with the subleading SCET operators $\bar T^{(s)}_{i}=0$. 
Then the leading order expression reads
\begin{equation}
K_{LS}^{\text{LO}}\simeq-\frac{m}{\sqrt{s}}~\sqrt{\frac{z}{1-z}}%
~\frac{\left(  2-z\right)  \left(  3z-2\right)  }{\left(  2-z\right)
^{2}+z^{2}}, 
\label{KLS}%
\end{equation}
where $m$ denotes the nucleon mass. These asymmetries have been measured in JLab \cite{Hamilton:2004fq} at $s=6.9$%
~GeV$^{2}$ and $t=-4.0$~GeV$^{2}$
\begin{equation}
K_{LL}^{\text{exp}}=0.678\pm0.083,~\ K_{LS}^{\text{exp}}=0.114\pm0.078.
\end{equation}
Using our approximate formulas (\ref{KLL}) and (\ref{KLS}) \ we obtain for
these kinematics%
\begin{equation}
K_{LL}^{\text{KN}}=0.70,~\ K_{LS}^{\text{LO}}=0.066.
\end{equation}
The expression for the longitudinal polarization is in good agreement while
the transverse polarization is about a factor two smaller but still within the
error bars. \ Notice however, that in these kinematics the value of the
$u\simeq-1.14$~GeV$^{2}$ is not large compared to the nucleon mass, so that 
potentially one can expect relatively large power corrections. Therefore the
good agreement with experiment is probably due to a 
compensation of such corrections in higher orders.
 Nevertheless this analysis of the recoil polarizations allows us to make the
following qualitative conclusion: it seems that the symmetry  relations (\ref{T4T2LO}) and (\ref{T6T2LO})
provides a reasonable approximation for WACS. Therefore it allows us to use the
data for the unpolarized cross section and formula (\ref{wacs-c-s}) in order
to extract information about the function $\mathcal{R}$.

In order to extract the values $|\mathcal{R}|$ from the data we take the
data for the WACS from \cite{Danagoulian:2007gs} for  which  $|t|\geq2.5$~GeV$^{2}$ and
$|u|\geq2.5$~GeV$^{2}$.   Then  for these points we extract the quantity
\begin{equation}
|\mathcal{R}^{\text{exp}}|=\sqrt{\frac{d\sigma^{\text{exp}}(s,t)}%
{d\sigma^{\text{KN}}(s,t)}}.
\end{equation}
In Fig.\ref{fitF1} we show the results for the
$|\mathcal{R}^{\text{exp}}|$ extracted from the data at three different
values of energy: $s=6.8,8.9$ and $10.9$~GeV$^{2}$ and $2.5< -t <6.5$~GeV$^{2}$. 
 As one can see from this plot, the extracted values 
 $|\mathcal{R}^{\text{exp}}|$  do not show  any  dependence 
on the values of $s$ as required  by factorization.  
Therefore we fit the  $Q^{2}$ dependence of the data  by the simple empirical function
\begin{equation}
\mathcal{R}(Q)=\frac{c}{\ln^{2}{Q^{2}}/{\Lambda^{2}}}, \label{F1fit}%
\end{equation}
which gives $c=0.10\pm0.01$ and $\Lambda=1.08\pm0.03$~GeV. 

 In order to
see the sensitivity of our approach on the possible power corrections we also
perform the same extractions keeping the exact kinematics in the formula for
the cross section. In this case we see that the obtained values for $|\mathcal{R}^{\text{exp}}|$ 
are  larger,  especially in the region of smaller  $-t$ and
the extracted values also show higher sensitivity on $s$.
The solid line in this panel shows the same fit (\ref{F1fit}) which in this
case yields $c=0.13\pm0.01$ and $\Lambda=1.09\pm0.02$~GeV. From this
 analysis we can conclude that the data at smaller values of  $|u|$ or $|t|$
 potentially may obtain sizable corrections from the power suppressed
contributions. However because our calculation is incomplete this estimate can only 
be considered as a  qualitative approximation. In  our numerical
calculation of TPE amplitudes we will use the results obtained without subleading
power corrections. 
\begin{figure}[h]
\centering
\includegraphics[height=2.0in]{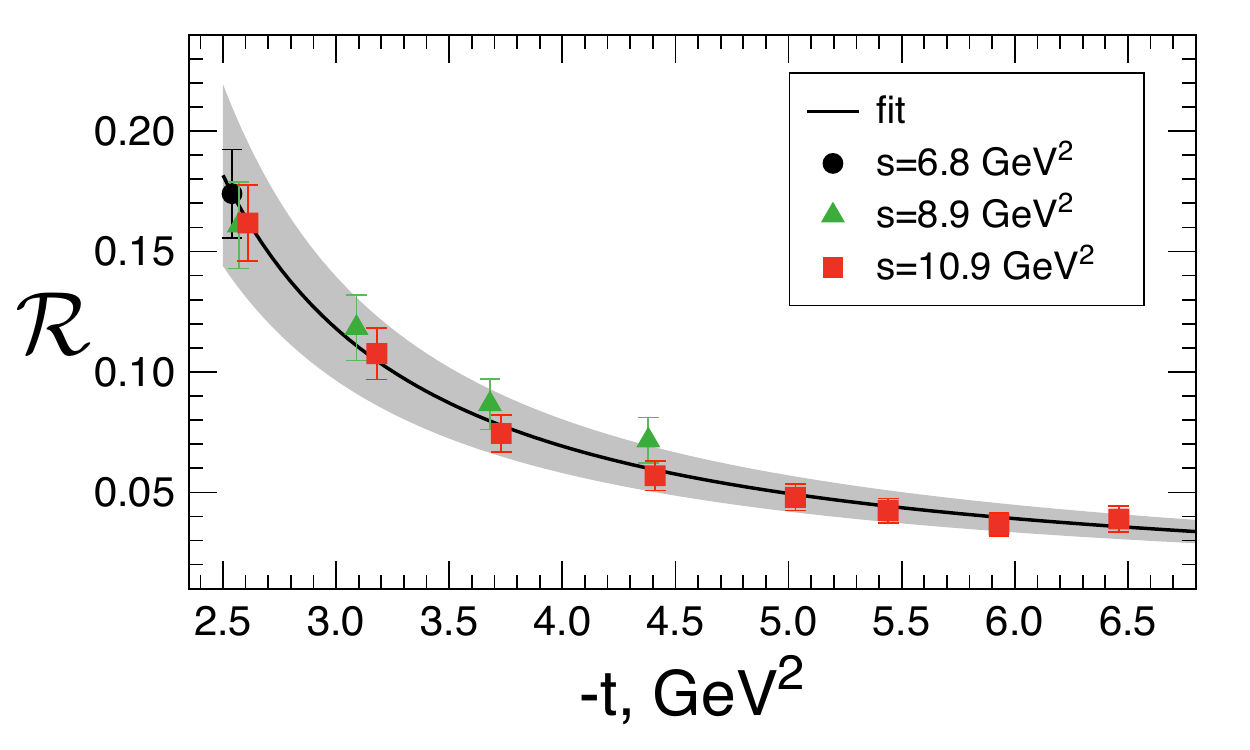}
\caption{The ratio $\mathcal{R}$ extracted from the WACS data \cite{Danagoulian:2007gs} and the fit
(\ref{F1fit}) without  kinematic power corrections . The gray band shows the  $1\sigma$ error bands. }
\label{fitF1LO}
\end{figure}
\begin{figure}[h]
\centering
\includegraphics[height=2.0in]{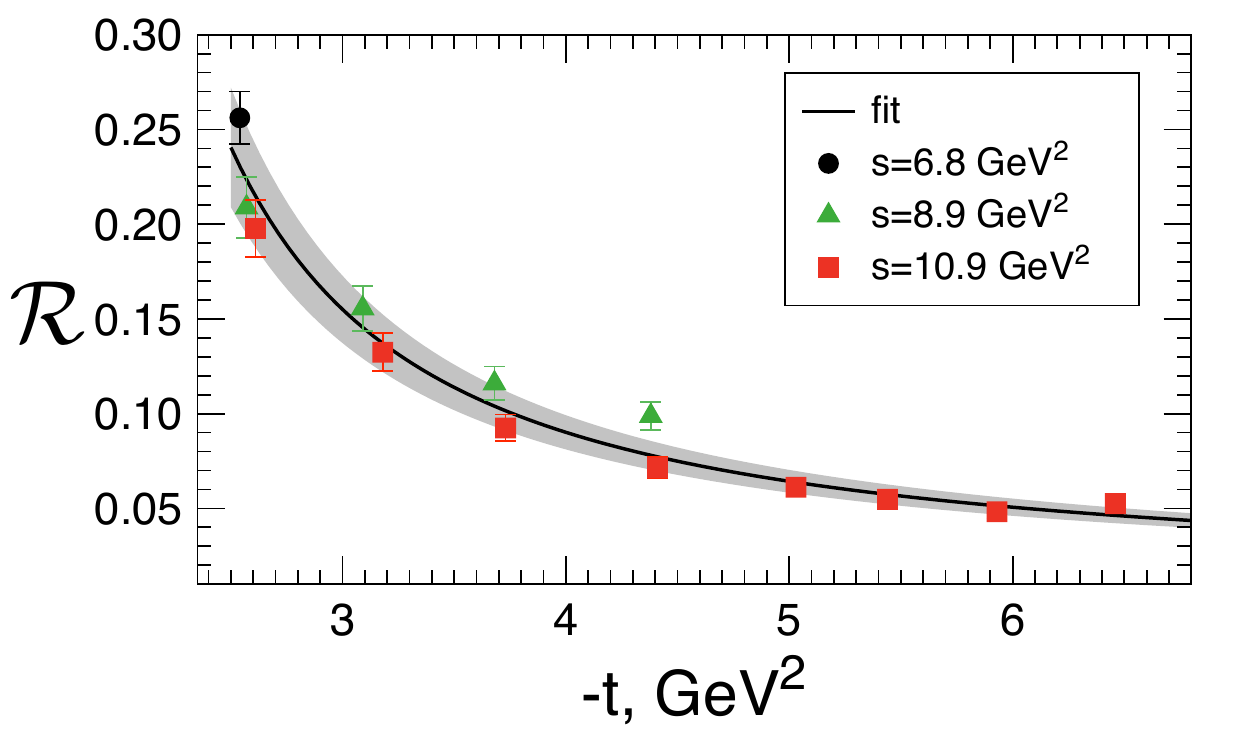}
\caption{ The same as in Fig.\ref{fitF1LO} but including the  kinematic power corrections.}
\label{fitF1}%
\end{figure}

\section{Estimate of the amplitudes  $g_{1,3}$  in a hadronic model}
\label{sec-g1}
 
 Unfortunately  the extraction of the amplitudes  $g_{1,3}$ from another 
 measurable process is at present not feasible. 
 Therefore one has to  build a model in order  to estimate these quantities. 
 
 As we discussed  above, the splitting  of the physical amplitudes into soft and hard spectator contributions  implies a certain rapidity regularization 
 which   allows one to consider the collinear and soft sectors independently.    In the  present paper  we  do not consider such scenario for $g_{1,3}$ 
 and perform the  numerical estimates of the hard spectator contributions defined in Eqs.(\ref{HM},\ref{H3}) using the nucleon distribution amplitude which does not  provide
 the end-point singularities in  the collinear convolution integrals.   
This allows us to consider the SCET matrix element which defines  the amplitudes   $g_{1,3}$  in (\ref{g1:def}) as a well defined quantity.

The SCET amplitude $g_{1}$ depends on the factorization scale $\mu_{F}$.  This factorization scale arises  from the factorization of  the QED box graphs
where $\mu_{F}$  separates the hard photon configuration from  the regions with soft  virtualities. 
The evolution equation with respect to this scale  can  be obtained from the QED renormalization of the
SCET operator $Y_{v}^{\dagger}S_{v} O_{+}^{\mu}$, see Eq.(\ref{YSO}). 
Equivalently it can be obtained from the requirement  that the physical amplitude does not
depend on $\mu_{F}$. Using Eq.(\ref{Aep:fin}) yields%
\begin{equation}
\mu_{F}\frac{d}{d\mu_{F}}A_{ep}^{2 \gamma}=0\Leftrightarrow~\mu_{F}\frac{d}{d\mu_{F}%
}g_{1}(z,Q,\mu_{F})+\mu_{F}\frac{d}{d\mu_{F}}C_{M}\left(  z,Q^{2},\mu_{F}\right)
\mathcal{F}_{1}=0.
\end{equation}
From the expression for $C_{M}$ in Eq.(\ref{CM}) we obtain%
\begin{equation}
\mu_{F}\frac{d}{d\mu_{F}}g_{1}(z, Q,\mu_{F})=2\ln\bar{z}~\mathcal{F}_{1}(Q).
\end{equation}
The solution of this evolution equation reads
\begin{equation}
g_{1}(z,Q,\mu_{F})=g_{1}(z,Q,\mu_{0})+\ln\bar{z}~\ln\frac{\mu_{F}^{2}}{\mu_{0}^{2}}\mathcal{F}_{1}(Q), \label{g1mu}%
\end{equation}
where $g_{1}(z,Q,\mu_{0})$  denotes some initial condition.  The logarithmic
contribution in Eq.(\ref{g1mu})  can be absorbed into the coefficient function
$C_{M}$.  In order to estimate the contribution associated with the amplitude  $g_{1}$ we have to
fix $\mu_{F}=\mu_{0}$ and  evaluate  $g_{1}(z,Q,\mu_{0})$.

Let us firstly consider an appropriate choice of the scale $\mu_{0}$.  As we have
mentioned before, in factorizing the hard modes we pass to the  SCET-I which
describes the hard-collinear dynamics. Therefore one can expect that it is
natural  to assume that $\mu_{0}\sim\sqrt{\Lambda Q}$.  However the
contributions associated with the hard-collinear photons cancel  in the sum of
the TPE diagrams. This allows us to assume that  
the factorization scale $\mu_{F}~$separates the hard and
soft electromagnetic configurations in the box diagrams. Using this
interpretation we can fix the scale $\mu_{0}$ to be of order $\Lambda$. With
 such choice $g_{1}(\mu_{0}\simeq\Lambda)$ describes the  TPE configurations
where the soft photon carries momentum $q_{i\mu}\lesssim\Lambda$.
Notice that in this case the virtuality of the involved active quark remains
hard-collinear as it should be.  Such choice for $\mu_{F}$ explains its
decoupling from the QCD factorization scale $\mu\sim\sqrt{\Lambda Q}$ which
enters in the definition of the QCD SCET FFs.  Let us also mention that
choosing  $\mu_{0}\simeq\Lambda$ one obtains  the large logarithm in the
coefficient function: $C_{M}(\mu_{F}=\mu_{0})\sim\mathcal{F}_{1} \ln\bar{z}~\ln\frac{s}{\mu_{0}^{2}}$. 
However the combination $\frac{\alpha}{\pi}\ln\frac{s}{\mu_{0}^{2}}$
 is still quite small for the practically relevant
values of $s$ therefore we do not need to sum such contributions.

Taking into account that the amplitude  $g_{1}$  describes the highly
asymmetrical configuration where one of the photons carries the momentum
$q_{i\mu}\lesssim\Lambda$  one can conclude that it must also include the
IR-sensitive QED contribution. Such term is very important for the correct
treatment of the QED IR-divergencies.  In what follows we suppose that the
soft photon virtualities which are  relevant for the QED IR-divergent part
are much smaller than the  QCD  soft virtualities of order $\Lambda$: 
$q_{i}^{2}\ll\Lambda^{2}$.  Such ultrasoft photons can interact only with the
point-like proton and  their dynamics can be described in the framework of
 some effective theory with the nucleon degrees of freedom.  These arguments
suggest that we can try to estimate $g_{1}(\mu_{0}\sim\Lambda)$ using a model
with a nucleon and excited resonance states.  We may expect that such model
can provide a reasonable estimate if the dominant contribution to $g_{1,3}(\mu_{0}\sim\Lambda)$ 
arises from the region of ultrasoft photon. From this
point of view such  calculation is quite similar  to hadronic model calculations performed before, 
with the one important difference: the virtuality of the soft photon is
constrained by $\mu_{0}^{2}\sim\Lambda^{2}$ and the hard dynamics is 
factorized into the hard coefficient function $C_{+}$. 

\begin{figure}[h]%
\centering
\includegraphics[
height=0.8178in,
width=5.6554in
]%
{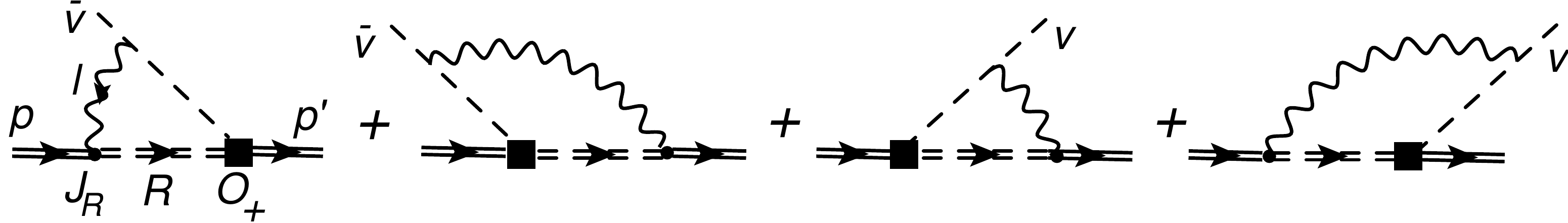}%
\caption{The diagrams in the low-energy effective theory describing the expression
for the amplitude proportional to $g_{1}$. The simple dashed line denotes WLs associated with vectors
$v$ and $\bar{v}$, the black square denotes the vertex with $O_{\mu}^{+}$.}%
\label{resonans}%
\end{figure}

In order to perform our calculation we proceed as follows. First, we insert the
intermediate nucleon states  approximating the matrix element
\bea
&&\left\langle p^{\prime}\right\vert T\{O_{+}^{\mu},\left(  Y_{v}^{\dag}%
S_{\bar{v}}-1\right)  \}\left\vert p\right\rangle _{\text{{\footnotesize SCET}%
}}\simeq
\nonumber \\ && \phantom{empty}
\int\frac{d^{D}r}{(2\pi)^{D}}\sum_{R,s}\left\langle p^{\prime
}\right\vert O_{+}^{\mu}\left\vert r,s\right\rangle
_{\text{{\footnotesize SCET}}}\frac{i}{r^{2}-m_{R}^{2}}\left\langle
r,s\right\vert \left(  Y_{v}^{\dag}S_{\bar{v}}-1\right)  \left\vert
p\right\rangle \label{apprx1}
\nonumber \\ 
&& \phantom{empty}  +\int\frac{d^{D}r}{(2\pi)^{D}}\sum_{R,s}\left\langle p^{\prime}\right\vert
\left(  Y_{v}^{\dag}S_{\bar{v}}-1\right)  \left\vert r,s\right\rangle
\frac{i}{r^{2}-m_{R}^{2}}\left\langle
r,s\right\vert O_{+}^{\mu}\left\vert p\right\rangle_{\text{{\footnotesize SCET}}} ,
\eea
where $R$ denotes the nucleon resonance state and $s$ describes its 
polarization. In Eq.(\ref{apprx1}) we wrote the combination $Y_{v}^{\dag
}S_{\bar{v}}-1$ in order to stress that we consider only  the
next-to-leading QED contributions. Let us again mention  that we consider only
the  TPE interactions of the soft photons with hadrons neglecting  the
electron and hadron vertex contributions.  The matrix elements with the
soft photon WLs can be computed in the effective theory with the nucleon degrees of freedom. 
\begin{equation}
\left\langle r,s\right\vert \left(  Y_{v}^{\dag}S_{\bar{v}}-1\right)
\left\vert p\right\rangle \simeq\int dx~\left\langle r,s\right\vert T\left\{
\left(  Y_{v}^{\dag}S_{\bar{v}}-1\right)  ,J_{R}^{\alpha}(x)ieB_{\alpha}%
^{(s)}(x)\right\}  \left\vert p\right\rangle , \label{YS-1}%
\end{equation}
where $J_{R}^{\alpha}(x)$ denotes the electromagnetic current of the hadron $R$.
Contracting the photon fields we obtain%
\bea
&&T\left\{  \left(  Y_{v}^{\dag}S_{\bar{v}}-1\right)  ,ieB_{\alpha}%
^{(s)}(x)\right\}
\simeq 
\nonumber \\ && \phantom{empty}
T\left\{  -ie\int_{-\infty}^{0}dt~\bar{v}\cdot B^{(s)}(t\bar{v}%
)-ie\int_{0}^{\infty}dt~v\cdot B^{(s)}(tv),~ieB_{\alpha}^{(s)}(x)\right\}
\\ &&  \phantom{empty}
=e^{2}\int\frac{d^{D}l}{\left(  2\pi\right)  ^{D}}~e^{-i\left(  lx\right)
}\frac{(-i)}{\left[  l^{2}+i\varepsilon\right]  }~\left\{  \frac{i\bar
{v}^{\alpha}}{\left[  -(\bar{v}l)+i\varepsilon\right]  }+\frac{iv^{\alpha}%
}{\left[  (vl)+i\varepsilon\right]  }\right\}  .
\label{Bcont}
\eea
Substitution (\ref{Bcont}) into (\ref{YS-1}) gives
\bea
&& \left\langle r,s\right\vert \left(  Y_{v}^{\dag}S_{\bar{v}}-1\right)
\left\vert p\right\rangle  \simeq 
\nonumber \\ && \phantom{empty}
e^{2}\left\langle r,s\right\vert
J_{R}^{\alpha}(0)\left\vert p\right\rangle \int\frac{d^{D}l}{\left[
l^{2}+i\varepsilon\right]  }~\delta^{D}(r-l+p)
\left\{
\frac{\bar{v}^{\alpha}}{\left[  -(\bar{v}l)+i\varepsilon\right]  }%
+\frac{v^{\alpha}}{\left[  (vl)+i\varepsilon\right]  }\right\}  .
\eea
An analogous expression can also be obtained for the second matrix element in
Eq.(\ref{apprx1}).  Substituting these expressions into Eq.(\ref{apprx1}) and
integrating over $d^{D}r$ yields
\bea
\left\langle p^{\prime}\right\vert T\{O_{+}^{\mu},\left(  Y_{v}^{\dag}%
S_{\bar{v}}-1\right)  \}\left\vert p\right\rangle _{\text{{\footnotesize SCET}%
}}   \simeq\frac{e^{2}\mu_{\text{uv}}^{2\varepsilon}}{\left(  2\pi\right)
^{D}}\int\frac{d^{D}l~}{\left[  l^{2}+i\varepsilon\right]  } 
&& \left\{ \frac{\bar{v}^{\alpha}}{\left[  -(\bar{v}l)+i\varepsilon\right]  }%
+\frac{v^{\alpha}}{\left[  (vl)+i\varepsilon\right]  }\right\} 
\nonumber\\ 
&&\times \sum_{R}M_{R}^{\mu\alpha}(l),
\eea
where
\begin{align}
M_{R}^{\mu\alpha}(l)  &  =\sum_{s}\left\langle p^{\prime}\right\vert
O_{+}^{\mu}\left\vert p+l,s\right\rangle _{\text{{\footnotesize SCET}}}%
\frac{i}{\left(  p+l\right)  ^{2}-m_{R}^{2}}\left\langle p+l,s\right\vert
J_{R}^{\alpha}(0)\left\vert p\right\rangle\nonumber \\
&  +\sum_{s}\left\langle p^{\prime}\right\vert J_{R}^{\alpha}(0)\left\vert
p^{\prime}-l,s\right\rangle \frac{i}{\left(  p^{\prime}-l\right)  ^{2}%
-m_{R}^{2}}\left\langle p^{\prime}-l,s\right\vert O_{+}^{\mu}\left\vert
p\right\rangle _{\text{{\footnotesize SCET}}}.
\label{MR}
\end{align}
This representation  can be easily
understood as a calculation of the diagrams in Fig.\ref{resonans}.  
 We use dimensional regularization with $D=4-2\varepsilon$  in order to regularize  
 the UV-divergencies, see further discussion.  

In order to describe correctly the region associated with the small momentum $l$ (soft photon)
we  need to perform in  Eq.(\ref{MR}) some approximations.  Namely, we neglect the small momentum $l$
in the numerators and denominators in  Eq.(\ref{MR}).
\begin{equation}
\left\langle p^{\prime}\right\vert O_{+}^{\mu}\left\vert p+l,s\right\rangle
\simeq\left\langle p^{\prime}\right\vert O_{+}^{\mu}\left\vert
p,s\right\rangle _{\text{{\footnotesize SCET}}},\ \ \left\langle
p+l,s\right\vert J_{R}^{\alpha}(0)\left\vert p\right\rangle \simeq\left\langle
p,s\right\vert J_{R}^{\alpha}(0)\left\vert p\right\rangle .
\end{equation}
\begin{equation}
\frac{i}{\left(  p+l\right)  ^{2}-m_{R}^{2}}\simeq\frac{i}{p_{+}l_{-}%
+m^{2}-m_{R}^{2}},~~\frac{i}{\left(  p^{\prime}-l\right)  ^{2}-m_{R}^{2}%
}\simeq\frac{i}{-p_{-}^{\prime}l_{+}+m^{2}-m_{R}^{2}}. \label{denom}%
\end{equation}
In these formulas we do not skip the masses because
\begin{equation}
p_{+}l_{-}\lesssim\Lambda Q,~\ m^{2}-m_{R}^{2}\simeq\left(  m-m_{R}\right)
\left(  m+m_{R}\right)  \sim\Lambda\left(  m+m_{R}\right)  ,
\end{equation}
and therefore the difference $m^{2}-m_{R}^{2}$  is potentially comparable with the
hard-collinear scale in the region of moderate values of $Q$ where $\Lambda
Q\sim m^{2}$. Moreover this difference of the masses provides also a natural
regulator  for the QED IR-divergence. In this case only the elastic contribution has an IR-divergence.   
Therefore we can write%
\bea
\left\langle p^{\prime}\right\vert O_{+}^{\mu}\left(  Y_{v}^{\dag}S_{\bar{v}%
}-1\right)  \left\vert p\right\rangle _{\text{{\footnotesize SCET}}}  
=\frac{e^{2}\mu_{\text{uv}}^{2\varepsilon}}{\left(  2\pi\right)  ^{D}}%
&&\int\frac{d^{D}l}{\left[  l^{2}-\lambda^{2}+i\varepsilon\right]  }\left\{
\frac{\bar{v}^{\alpha}}{\left[  -(\bar{v}l)+i\varepsilon\right]  }%
+\frac{v^{\alpha}}{\left[  (vl)+i\varepsilon\right]  }\right\} 
\nonumber\\ 
&&\times
\left(  M_{p}^{\mu\alpha}(l)+M_{\Delta}^{\mu\alpha}(l)+~...\right)  ,
\label{intMp}
\eea
where we introduced the IR-regularization by the photon mass $\lambda$. In what
follow we restrict our considerations  computing only the elastic contribution
$M_{p}^{\mu\alpha}$. This contribution is required for the correct treatment
of the QED divergencies in the physical cross section and \ usually it
provides the largest numerical effect. Using (\ref{def:f1}) and 
\begin{equation}
 \left\langle p\right\vert J^{\alpha}(0)\left\vert p\right\rangle
\simeq\bar{N}(p)\gamma^{\alpha}N(p),
\end{equation}
we obtain
\bea
M_{p}^{\mu\alpha}&\simeq&
\bar{N}(p^{\prime})\left[  \left\{ \nbn\gamma_{\perp}^{\mu}~f_{1}(Q)\right\}  \frac{i\left(  \Dslash{p}+m\right)
}{p_{+}l_{-}+i\varepsilon}\gamma^{\alpha}+\gamma^{\alpha}\frac{i\left(
\Dslash{p}^{\prime}+m\right)  }{-p_{-}^{\prime}l_{+}+i\varepsilon}\left\{
\nbn\gamma_{\perp}^{\mu}~f_{1}(Q)\right\}  \right] N(p), 
\nonumber \\
&\simeq& i~\left[  \frac{\bar{n}^{\alpha}}{l_{-}+i\varepsilon}+\frac{n^{\alpha}%
}{-l_{+}+i\varepsilon}\right]  \bar{N}(p^{\prime})~\nbn
\gamma_{\perp}^{\mu}N(p)~f_{1}(Q).
\eea
Substituting this expression in Eq.(\ref{intMp})  we obtain
\begin{equation}
\left\langle p^{\prime}\right\vert O_{+}^{\mu}\left(  Y_{v}^{\dag}S_{\bar{v}%
}-1\right)  \left\vert p\right\rangle _{\text{{\footnotesize SCET}}}\simeq
e^{2}~\bar{N}(p^{\prime})\nbn \gamma_{\perp}^{\mu}N(p)~f_{1}%
(Q)~J_{el},
\end{equation}
with the scalar integral%
\begin{align}
J_{el}  &  =\frac{ie^{2}\mu_{\text{uv}}^{2\varepsilon}}{\left(  2\pi\right)
^{D}}\int\frac{d^{D}l}{\left[  l^{2}-\lambda^{2}+i\varepsilon\right]
}\left\{  \frac{\bar{v}^{\alpha}}{\left[  -(\bar{v}l)+i\varepsilon\right]
}+\frac{v^{\alpha}}{\left[  (vl)+i\varepsilon\right]  }\right\} 
\left\{  \frac{\bar{n}^{\alpha}}{\left(  l\bar{n}\right)  +i\varepsilon}%
+\frac{n^{\alpha}}{-\left(  ln\right)  +i\varepsilon}\right\}  .
\end{align}
This integral is the same as the one derived for quarks in Appendix~\ref{calculation}. This is not
surprising because we consider the soft photon exchange between the highly
energetic particles (lepton and proton). The effective action
describing such a process can be written in the form of the soft
photon WLs. The only difference is in the electric charges:  the soft photon interacts  with the total proton charge.  
Therefore for the real part we  obtain
\begin{equation}
\operatorname{Re}J_{el}=-\frac{1}{4\pi^{2}}\ln\left\vert \frac{\left(  \bar
{v}\bar{n}\right)  }{\left(  \bar{v}n\right)  }\right\vert \left\{  \frac
{1}{\varepsilon_{\text{uv}}}+\ln\frac{\mu_{\text{uv}}^{2}}{\lambda^{2}%
}\right\}  ,
\end{equation}
where \ $\mu_{\text{uv}}$ is the UV-renormalization scale in the $\overline{MS}%
$-scheme. The result can be rewritten as
\begin{equation}
\ln\left\vert \frac{\left(  \bar{v}\bar{n}\right)  }{\left(  \bar{v}n\right)
}\right\vert \approx\ln\left\vert \frac{2(kp)}{2(kp^{\prime})}\right\vert
=\ln\left\vert \frac{s-m^{2}}{-u+m^{2}}\right\vert \equiv\ln\left\vert
\frac{\tilde{s}}{\tilde{u}}\right\vert , \label{Log}%
\end{equation}
where we introduced $\tilde{s}=s-m_{N}^{2}$, $\tilde{u}=-u+m^{2}$. After
UV-renormalization we find%
\begin{equation}
\left\langle p^{\prime}\right\vert O_{+}^{\mu}\left(  Y_{v}^{\dag}S_{\bar{v}%
}-1\right)  \left\vert p\right\rangle _{\text{{\footnotesize SCET}}}%
=\frac{\alpha}{\pi}~\bar{N}(p^{\prime})\nbn \gamma_{\perp}^{\mu
}N(p)~f_{1}(Q)\ln\frac{\lambda^{2}}{\mu_{\text{uv}}^{2}}\ln\left\vert
\frac{\tilde{s}}{\tilde{u}}\right\vert . \label{appr2}%
\end{equation}
We will make a few comments on this result. The UV-divergence arises as soon as  we neglect  the
small terms $l^{2}$ in the denominators of the nucleon in Eq.(\ref{denom}).
Such approximation can be understood as a transition to the
effective theory which describes the interaction of the energetic
particles\footnote{Remind that in our frame the leptons and protons move with
the energies of order $s\sim Q$.} with the soft photons. The appropriate
 effective action can be naturally described by the
soft photon WLs directed along the proton momenta. Formally the UV-divergence is
related to the multiplicative renormalization of WLs.  The
UV-renormalization induces the dependence on the scale $\mu_{\text{uv}}$ which
we can be understood as highest possible virtuality of the soft photon
consistent with our approximations. It is naturally to expect that
we can put $\mu_{\text{uv}}=\mu_{0}\sim\Lambda$.  In the final expression
(\ref{appr2}) we keep the exact expression $\ln\left\vert \frac{\tilde{s}%
}{\tilde{u}}\right\vert $ because  this expression provides the exact
cancellation of the IR-divergent contribution.  This point is discussed in
detail in the next section. 

Our calculation has many common features with the soft approximation made 
in  Ref.\cite{Tsai:1961zz}.  In that work the TPE diagram with the nucleon
intermediate state has also been computed neglecting  the small photon
momentum $l$ in the numerator of the integrand (including the FF of the hard
vertex) and in the hard photon propagator but not in the denominators of the
electron and proton lines. As a result, the integrand is reduced to a three-point
function. The resulting integral is UV-finite but the answer unavoidably includes
the region where the virtuality $l^{2}$ of the photon (and  other
particles)  can be large $l^{2}\sim Q^{2}$. But such description of this
hard region is incorrect because the dynamics of the hard particles must be described
using QCD.  As a result 
the result of Refs.~\cite{Tsai:1961zz, Afanasev:2005mp} includes  large logarithms 
$\ln \lambda^{2} / s$ 
\begin{equation}
\bar\delta^{2 \gamma}[\text{Tsai}]\sim\ln\frac{\lambda^{2}}{s}\ln\left\vert
\frac{\tilde{s}}{\tilde{u}}\right\vert G^{2}_{M}+~...~.
\end{equation}
because in this case the hard scale $s$  provides the required ``UV cut-off''.
Therefore we expect that such description of the TPE contribution in the
region of large energies is  incorrect in spite of the fact that  it correctly treats the QED IR
behavior.  

In our calculation we should not have any region 
corresponding with hard virtualities because we already factorized
the hard dynamics. Therefore it is important to perform the expansion in
small momenta $l$ as in (\ref{denom}) in order to avoid the double counting of
the hard region. Then we do not obtain any large logarithms. The uncertainty in
the choice of the renormalization scale $\mu_{\text{uv}}\sim\mu_{0}$ has a clear
physical meaning and is associated with the applicability of the our soft
approximation and also reflects  an ambiguity in the calculation of the
initial condition $g_{1}(z,\mu_{0})$.

Hence  assuming the dominance of the elastic contribution and comparing Eq.(\ref{appr2}) with (\ref{g1:def}) we obtain
\begin{equation}
g_{1}(z,Q,\mu_{0})\simeq\frac{\alpha}{\pi}\ f_{1}(Q)\ln\frac{\mu_{0}^{2}%
}{\lambda^{2}}\ln\bar{z}\simeq
\frac{\alpha}{\pi}\ G_{M}(Q^{2})\ln\frac{\lambda^{2}}{\mu_{0}^{2}}\ln\left\vert \frac{\tilde{s}}{\tilde{u}}\right\vert, ~~~   g_{3}(z,Q)\simeq 0. 
\label{g1fin}
\end{equation}
where we used Eq.(\ref{GMf1}) to re-express the SCET FF $f_{1}$ in term of the physical
FF $G_{M}$. 

We obtained that the second  SCET amplitude $g_{3}$ in this approximation is very small. We  suppose   that this is a feature of our  
model in which the soft photon interacts with the hadron as  with a structureless point-like particle.  We expect that contributions with higher spins    
 in Eq.(\ref{intMp})  also  do not provide any significant contributions  to  $g_{3}$.   Probably, this amplitude
is sensitive to the region with the relatively higher virtualities  for which   the soft photon can still resolve  the soft and hard-collinear structures, 
i.e.  where the soft photon  is sensitive to the rapidity gap arising  due to the soft spectator scattering.

\section{Phenomenological analysis}
\label{sec-phnm}

In order to make a numerical analysis we  eliminate 
the SCET FF $\mathcal{F}_{1}$  using the  expression in Eq.(\ref{F1T2}).
Consider for simplicity the amplitude  $\tilde F_{3}$.    
Using   the soft spectator contributions given in Eq.(\ref{F3}) we obtain
 \bea
\frac{\nu}{m^{2}} \tilde F_{3}(\varepsilon,Q^{2})=g_{3}(z,Q)+\frac{\alpha}{\pi
}~\frac{\nu}{s}C_{3}(s,t)\mathcal{F}_{1}+\boldsymbol{\Psi}(x_{i})\ast{H}_{3}(z,Q^{2},\ x_{i},y_{i})\ast\boldsymbol{\Psi}(y_{i}),
\label{tF3tpe}
\eea
where for the hard spectator scattering contribution we used expressions given in Eqs.(\ref{H3}).   Substituting  Eq.(\ref{F1T2})  for $\mathcal{F}_{1}$  
into Eqs.(\ref{tF3tpe}) (we assume the different value of $s=s'$  for WACS)  we obtain
\bea
\frac{\nu}{m^{2}} \tilde F_{3}(\varepsilon,Q^{2})   &=&      g_{3}(z,Q)
+ \frac{\alpha}{\pi}~\frac{\nu}{s}C_{3}(s,t) \frac{T_{2}(s',t)}{C_{2}(s',t)}
\nonumber \\
&& +{\boldsymbol\Psi}(y_{i}) \ast
\left\{H_{3}(z,Q^{2};~ x_{i},y_{i})-\frac{ H_{2}(s',t;~ x_{i},y_{i}) }{ C_{2}(s',t) }\right\} 
\ast\boldsymbol{\Psi}(x_{i}),
\\[2mm]
&\simeq&  \frac{\alpha}{\pi}~\frac{\nu}{s}C_{3}(s,t) \mathcal{R}(Q)
+{\boldsymbol\Psi}(y_{i})\ast  H_{3}(z,Q^{2};~ x_{i},y_{i}) \ast\boldsymbol{\Psi}(x_{i}).
\label{tF3tpeT2}
\eea
In order to obtain  the last equation we used  the estimate (\ref{g1fin}) in \ which \  $g_{3}$ \ is\  zero,  substituted  
$T_{2}(s',t)/C_{2}(s',t)=\mathcal{R}(s',t)\simeq \mathcal{R}(Q)$  and   we  also neglected  the contributions of order $\mathcal{O}(\alpha^{2}_{s})$  in the hard coefficient functions  $H_{2,3}$, keeping in mind  that  the leading order contributions  are   $H_{3,M}\sim \mathcal{O}(\alpha_{s})$   while  $H_{2}\sim \mathcal{O}(\alpha^{2}_{s})$ and    $C_{2}\sim \mathcal{O}(\alpha^{0}_{s})$. 
An analogous expression can be also obtained for the TPE amplitude $\delta \tilde G^{2 \gamma}_{M}$
and  reads
\bea
\delta \tilde G^{2 \gamma}_{M}(\varepsilon,Q^{2})\simeq \frac{\alpha}{\pi}\ G_{M}(Q^{2})\ln\frac{\lambda^{2}}{\mu_{0}^{2}}\ln\left\vert \frac{s-m^{2}}{u-m^{2}}\right\vert
+ \frac{\alpha}{\pi}C_{M}(z,Q^{2},\mu_0) \mathcal{R}(Q)
\nonumber \\  
+\ {\boldsymbol\Psi}(y_{i})\ast  H_{M}(z,Q^{2};~ x_{i},y_{i})\ast\boldsymbol{\Psi}(x_{i}).
\label{tGMtpeT2} 
\eea
where we  used Eq.(\ref{g1fin}) for $g_{1}$ and assumed that   $H_{M}\sim \mathcal{O}(\alpha_{s})$.  

The formulas presented in Eqs.(\ref{tF3tpeT2},\ref{tGMtpeT2}) will be used for the numerical estimates  presented below. 
We like to emphasize that  these expressions are of course model dependent.  
We expect that the largest ambiguity  arises from the amplitudes $g_{1,3}$  which were computed in the hadronic model  in Eqs. (\ref{g1fin}),  
and which describe the TPE  subprocess  with  one  soft photon.  

The other terms in  Eqs.(\ref{tF3tpeT2},\ref{tGMtpeT2})  describe the contributions  with 
two hard photons and can be fixed more accurately.  
For the  ratio $\mathcal{R}$ we  use the fit (\ref{F1fit}) with the parameters
$c=0.10\pm0.01$ and $\Lambda=1.08\pm0.03$~GeV.    In the numerical calculations we use for the scale $\mu_{0}= 1/\sqrt{-t}$ yielding $\mu_{0}=630-450~$MeV for $Q^{2}=2.5-5$~GeV$^{2}$.  We observed that
variations of the value of this scale by $\pm100$~MeV do not provide any   significant numerical effect. 
The explicit expressions for the coefficient  functions $H_{M,3}$   and distribution amplitudes $\boldsymbol{\Psi}$ can be 
 found in Ref. \cite{Borisyuk:2008db, Kivel:2009eg}.  
 The   DA $\boldsymbol{\Psi}$ can be 
expressed  in terms of one scalar  DA for which we use the model
\begin{equation}
\varphi_{N}(x_{1},x\,_{2},x_{3})=f_{N}~120x_{1}x_{2}x_{3}\left\{
1+r_{+}(1-3x_{3})+r_{-}(x_{2}-x_{1})\right\}  ,
\label{DA}
\end{equation}
where the low energy constants \ $f_{N},$ $r_{+}$ and $r_{-}$ have been
estimated at $\mu=1$GeV as \cite{Braun:2006hz}:
\begin{equation}
f_{N}=\left(  5.0\pm0.5\right)  \times10^{-3}\text{GeV}^{2},~r_{+}%
\simeq0.35,\ r_{-}~\simeq1.37.
\end{equation}
The scale of the running coupling $\alpha_{s}(\mu_{R}^{2})$ was set to be
$\mu_{R}^{2}\simeq0.6~Q^{2}$ and we use the two-loop running coupling with
$\alpha_{s}(\mu_{R}^{2}=1.5$~GeV$^{2})=0.360$.

In our numerical calculations we also use the  ratio
$G_{E}/G_{M}$ which is  fixed from the experimental 
data \cite{Jones00,Punjabi:2005wq,Gayou02,Puckett:2010ac,Puckett:2011xg} and using 
the following fit of the data for $Q^{2}$ in the interval
$0.5-8.5$~GeV$^{2}$:
\begin{equation}
R\equiv\frac{G_{E}}{G_{M}}=\mu_{p}^{-1}(1- a \ln^{2}[Q^{2}/\Lambda_{R}%
^{2}]), \label{def:R}%
\end{equation}
with $a = 0.070\pm0.007$, $\Lambda_{R}=0.54\pm0.03$~GeV. The other parameters
which we need are
\begin{equation}
m=0.938\text{ GeV},~\mu_{p}=2.7928,~\alpha=1/137.
\end{equation}

Finally let us inspect the behavior of the amplitudes (\ref{tF3tpeT2},\ref{tGMtpeT2})  in the limit $\varepsilon
\rightarrow1~(z \rightarrow 0)$ where TPE corrections must  be consistent with behavior 
Eq.(\ref{null}). Using the expressions for $C_{M,3}$ given in  Eqs.(\ref{CM},\ref{C3}), 
we obtain in this limit~:
\begin{equation}
\delta\tilde  G^{2 \gamma}_{M}(\varepsilon=1,Q^{2})=-\frac{\nu}{m^{2}}\tilde F_{3}(\varepsilon=1,Q^{2})=\frac{\alpha\pi}{2}\mathcal{F}_{1}(Q), 
\label{fwd}%
\end{equation}%
where we used that  hard spectator contributions   vanish separately   at $\varepsilon=1$ 
\begin{equation}
\lim_{z\rightarrow 0} {\boldsymbol\Psi}(y_{i})\ast  H_{M,3}(z,Q^{2};~ x_{i},y_{i})\ast\boldsymbol{\Psi}(x_{i})=0.
 \label{hard:fwd}
\end{equation}
This can be easily shown using the explicit expressions from Ref.\cite{Borisyuk:2008db,Kivel:2009eg}.
Using Eqs.(\ref{fwd}) one finds that  Eq.(\ref{null}) is fulfilled.

\subsection{Reduced cross section}
The formula  for the elastic reduced cross section presented in Eq.(\ref{sigmR}) can not be compared to the observable cross section because of
IR-divergent amplitudes.  For the real parts, the IR divergence arising from the elastic  next-to-leading (NLO) QED  diagrams is canceled 
 by IR-divergent contributions arising from the interference of the bremsstrahlung
 and Born diagrams.  Therefore  one must include  also the contribution of the inelastic diagrams.  
 After that  the reduced cross section can be presented as following
 \be{sgmRfull}
 \sigma^{\text{exp}}_{R}=\sigma^{1\gamma}_{R}
 \left(
 1+ \delta_{\text{el, virt}}+\delta_{\text{brems}} 
 \right), 
 \ee 
 where the Born cross section reads
 \be{sgm1g}
 \sigma^{1\gamma}_{R}=G_{M}^{2}+\frac{\varepsilon}{\tau}G_{E}^{2}.
 \ee
 The contribution of the virtual (real) radiative corrections to the elastic amplitudes in Eq.(\ref{sgmRfull}) is included in $\delta_{el, \text{virt}}$ ($\delta_{\text{brems}}$) respectively. 
 The analysis of the radiative corrections for the existing experimental data
has been carried out using the Mo and Tsai (MT) formulas  \cite{Tsai:1961zz}. 
Let us write  in this case  the reduced cross section as
\be{sgmRfullMT}
 \sigma^{\text{exp}}_{R}=\sigma^{1\gamma,\text{MT}}_{R}
 \left(
 1+ \delta^{\text{MT}}_{\text{virt}}+\delta^{\text{MT}}_{\text{brems}} 
 \right) 
 \ee 
In this case the extracted values of $\sigma^{1\gamma,\text{MT} }_{R}$ can be considered as experimental results. 
 If one uses a different approach to compute the radiative corrections  with a different elastic input then the 
 extracted value of the Born cross section is different.  This difference 
 must be considered as effect related to the TPE correction in our case. In what follows we use the same expressions for the  
 elastic contributions other than two-photon exchange, 
 and for the bremsstrahlung contribution  as in  MT-formulas, i.e.
 \begin{eqnarray}
 \delta_{el, \text{virt}} - \delta^{\text{MT}}_{el, \text{virt}} &=& 
 \delta_{2 \gamma} - \delta^{\text{MT}}_{2 \gamma}  \nonumber \\
 \delta_{\text{brems}}  &=& \delta^{\text{MT}}_{\text{brems}} 
 \end{eqnarray}
   Therefore from Eq.(\ref{sgmRfull}) and Eq.(\ref{sgmRfullMT}) we obtain
\be{sgmRfin}
 \sigma^{1\gamma,\text{MT} }_{R} =\sigma^{1\gamma}_{R}
  \left(
 1+\delta_{2\gamma}-\delta^{\text{MT}}_{2\gamma}
\right).
 \ee
 where the FFs $G_{M}$ and $G_{E}$ on the $rhs$ must be considered as unknown quantities.  
 Using Eq.(\ref{sigmR}) one can write the explicit expressions for the TPE contributions as 
 \be{dltel}
\sigma^{1\gamma}_{R}\  \delta_{2\gamma}=
2G_{M}\operatorname{Re}\left[  \delta\tilde{G}_{M}^{2 \gamma} +\varepsilon\frac{\nu}{m^{2}}\tilde{F}_{3}\right]  
+
2\frac{\varepsilon}{\tau}G_{E}\operatorname{Re}\left[  \delta\tilde{G}^{2 \gamma}_{E}+\frac{\nu}{m^{2}}\tilde{F}_{3}\right].
 \ee
It is convenient to present the analytical expression for the  MT term $\delta^{\text{MT}}_{2\gamma}$  in the similar form. 
Using the analytical  expression presented in  \cite{Afanasev:2005mp} we can write
\be{dltelMT}
\sigma^{1\gamma}_{R}\  \delta^{\text{MT}}_{2\gamma} =2G_{M}\operatorname{Re}
\left[
G_{M}\frac12 \delta^{\text{MT}}_{2\gamma}  
\right]
+2\frac{\varepsilon}{\tau}G_{E}\operatorname{Re}
\left[ 
G_{E}~\frac12 \delta^{\text{MT}}_{2\gamma}
\right].
\ee
The analytical expression for the factor $\delta^{\text{MT}}_{2\gamma}$ reads
\begin{align}
\operatorname{Re} \delta^{\text{MT}}_{2\gamma}  & =\frac{2\alpha}{\pi}\left\{  \ln\frac{\lambda^{2}}%
{\tilde{s}}\ln\left\vert \frac{\tilde{s}}{\tilde{u}}\right\vert +\frac{1}%
{2}\ln^{2}\left\vert \frac{\tilde{s}}{\tilde{u}}\right\vert -\text{Li}\left(
\frac{\tilde{s}}{s}\right) 
-\frac{1}{2}\ln^{2}\frac{\tilde{s}}{s}+\operatorname{Re}\text{Li}\left(  \frac{\tilde{u}}%
{u}\right)  +\frac{1}{2}\ln^{2}\frac{\tilde{u}}{u}\right\}  ,
 \label{dlt2g}
\end{align}
where $\lambda^{2}$ is the soft photon mass (IR regulator), $\tilde{s}%
=s-m^{2}$,$~\tilde{u}=u-m^{2}$ and Li$(z)$ is the Spence function (dilogarithm)
defined by%
\begin{equation}
\text{Li}(z)=-\int_{0}^{z}dt\frac{\ln(1-t)}{t}.
\end{equation}
The result presented  in Eq.(\ref{dlt2g}) is obtained by computing the box diagrams with nucleon intermediate state in the soft
limit  \cite{Tsai:1961zz}. However the  expression  in Eq.(\ref{dlt2g}) is incomplete  because the exact answer has one more term  \cite{Afanasev:2005mp}
\begin{equation}
\left[\delta^{\text{MT}}_{2\gamma}\right]_{\text{exact}}=\delta^{\text{MT}}_{2\gamma}+{\alpha\pi}.
\label{dltexct}
\end{equation}
However the factor $\alpha\pi$, which originates from the crossed box diagram\footnote{
More explicitly, the factor ${\alpha\pi}$  in Eq.(\ref{dltexct}) originates from analytic continuation of $\ln^{2}$ terms when evaluating crossed box diagram 
from the direct box diagram as shown in Ref.\cite{VanNieuwenhuizen:1971yn}}, 
is absent  in the formulas in Ref. \cite{Tsai:1961zz}
and therefore is excluded from the expression of $\delta^{\text{MT}}_{2\gamma}$.

Using Eqs.(\ref{dltel}), (\ref{dltelMT}) we can present  the $rhs$ in Eq.(\ref{sgmRfin}) as
\bea
&& \sigma^{1\gamma,\text{MT} }_{R}= \sigma^{1\gamma}_{R}+
2G_{M}\operatorname{Re}\left[  \delta\tilde{G}_{M}^{2 \gamma} +\varepsilon\frac{\nu}{m^{2}}\tilde{F}_{3}-G_{M} \frac12 \delta^{\text{MT}}_{2\gamma} \right]  
 \nonumber \\ &&  \phantom{empty space}
+  2\frac{\varepsilon}{\tau}G_{E}\operatorname{Re}\left[  \delta\tilde{G}^{2 \gamma}_{E}
+\frac{\nu}{m^{2}}\tilde{F}_{3}-G_{E} \frac12 \delta^{\text{MT}}_{2\gamma}\right]. 
\label{sgmRexpl}
\eea
The amplitudes $\delta\tilde{G}_{M}^{2 \gamma}$, $ \delta\tilde{G}^{2 \gamma}_{E}$ and $\delta^{\text{MT}}_{2\gamma}$ in Eq.(\ref{sgmRexpl})  
depend on the IR photon mass $\lambda^{2}$ which however cancels in the difference, we discuss these subject in detail in Appendix~\ref{compensation}.

The TPE contribution associated with the  $\delta\tilde{G}^{2 \gamma}_{E}$ in Eq.(\ref{sgmRexpl})
\begin{equation}
 \delta\tilde{G}^{2 \gamma}_{E}+\frac{\nu}{m^{2}}\tilde{F}_{3}=\delta\tilde{G}_{M}^{2 \gamma} 
-(1+\tau)\delta\tilde F_{2}+\frac{\nu}{m}\tilde F_{3},
\end{equation}
contains the unknown amplitude $\delta\tilde F_{2}$. This amplitude  depends on the matrix elements of  subleading SCET operators
which we did not consider in our analysis.  We will assume that  the corresponding difference 
$\delta\tilde{G}^{2 \gamma}_{E}+\frac{\nu}{m^{2}}\tilde{F}_{3}-G_{E}\frac12 \delta^{\text{MT}}_{2\gamma}$  in  Eq.(\ref{sgmRexpl}) behaves similar to the FF
 $G_{E}=G_{M}-(1+\tau)F_{2}=RG_{M}$ and  therefore \
\begin{equation}
\delta\tilde{G}^{2 \gamma}_{E}+\frac{\nu}{m^{2}}\tilde{F}_{3}-G_{E}\frac12 \delta^{\text{MT}}_{2\gamma}\sim\frac{\alpha}{\pi}\mathcal{O}(R).
\end{equation}
In this case the correction originating from such term in Eq.(\ref{sgmRexpl}) is relatively small
\begin{equation}
2\frac{\varepsilon}{\tau}G_{E}\operatorname{Re}\left[  \delta\tilde{G}^{2 \gamma}_{E}+\frac{\nu}{m^{2}}\tilde{F}_{3}-G_{E} \delta^{\text{MT}}_{2\gamma}\right]
\sim2\frac{\varepsilon}{\tau}G^{2}_{M}~\frac
{\alpha}{\pi}\times\mathcal{O}(R^{2}),\label{dGEsmall}%
\end{equation}
 because the ratio $R$ is small  and we  can neglect it. 
Therefore in our numerical analysis  we use for the reduced cross section the following expression
\be{sgmRexp}
 \sigma^{1\gamma,\text{MT} }_{R}= G_{M}^{2}+\frac{\varepsilon}{\tau}G_{E}^{2} +
2G_{M}\operatorname{Re}\left[  \delta\tilde{G}_{M}^{2 \gamma} +\varepsilon\frac{\nu}{m^{2}}\tilde{F}_{3}-G_{M}\frac12 \delta^{\text{MT}}_{2\gamma} \right].  
 \ee

\begin{figure}[h]
\centering
\includegraphics[height=1.5in]{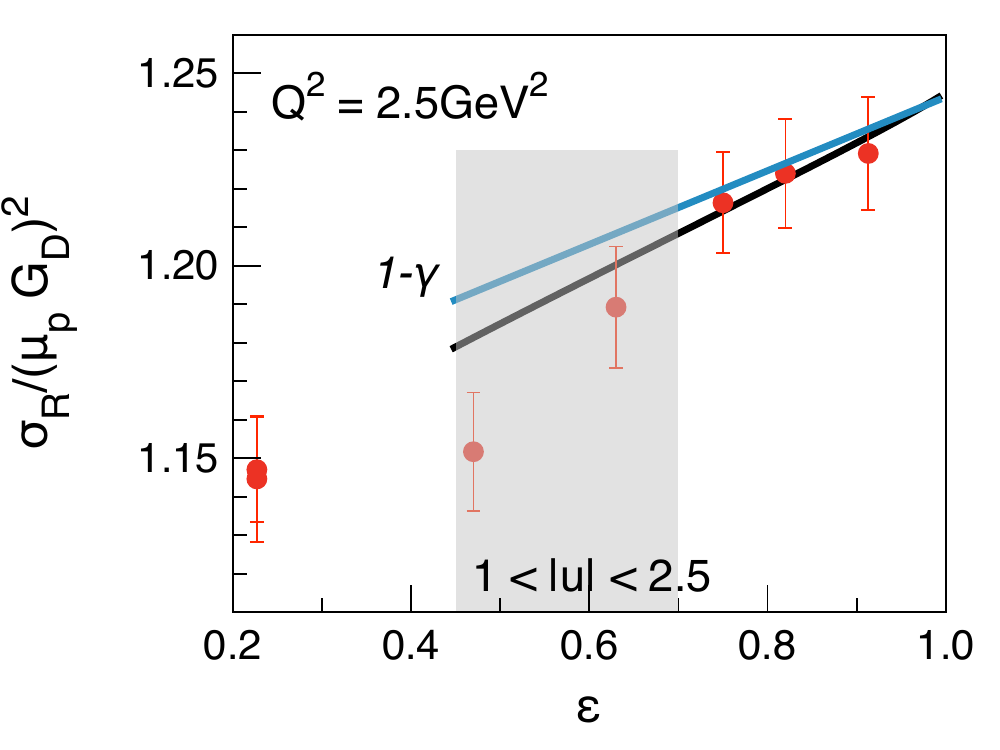}
\includegraphics[height=1.5in]{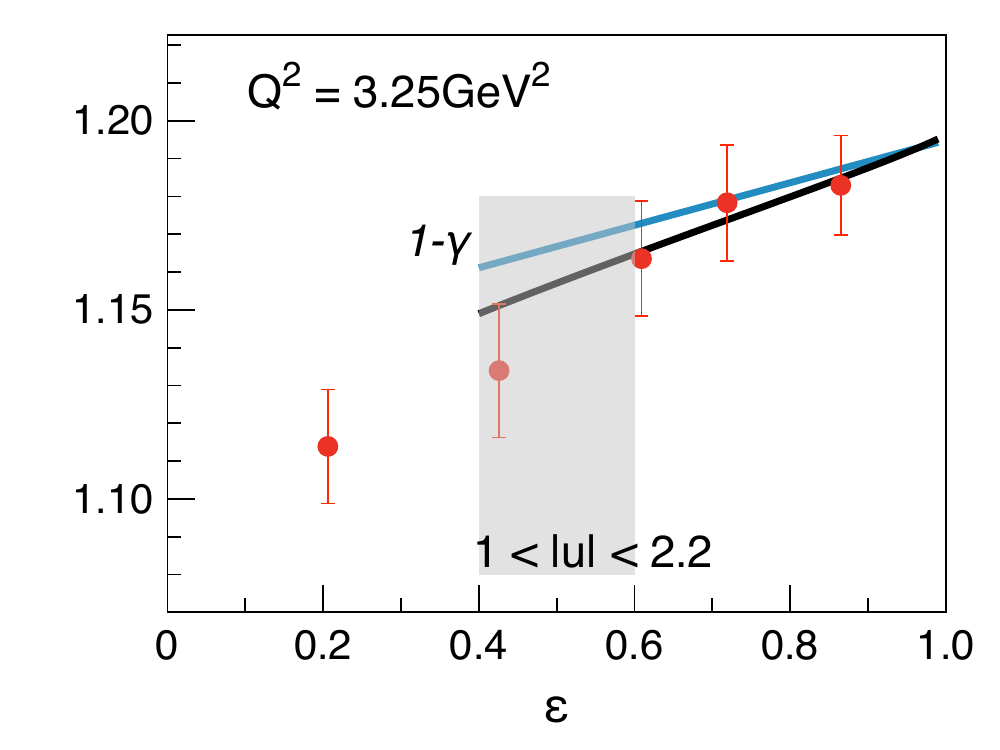}
\\
\includegraphics[height=1.5in]{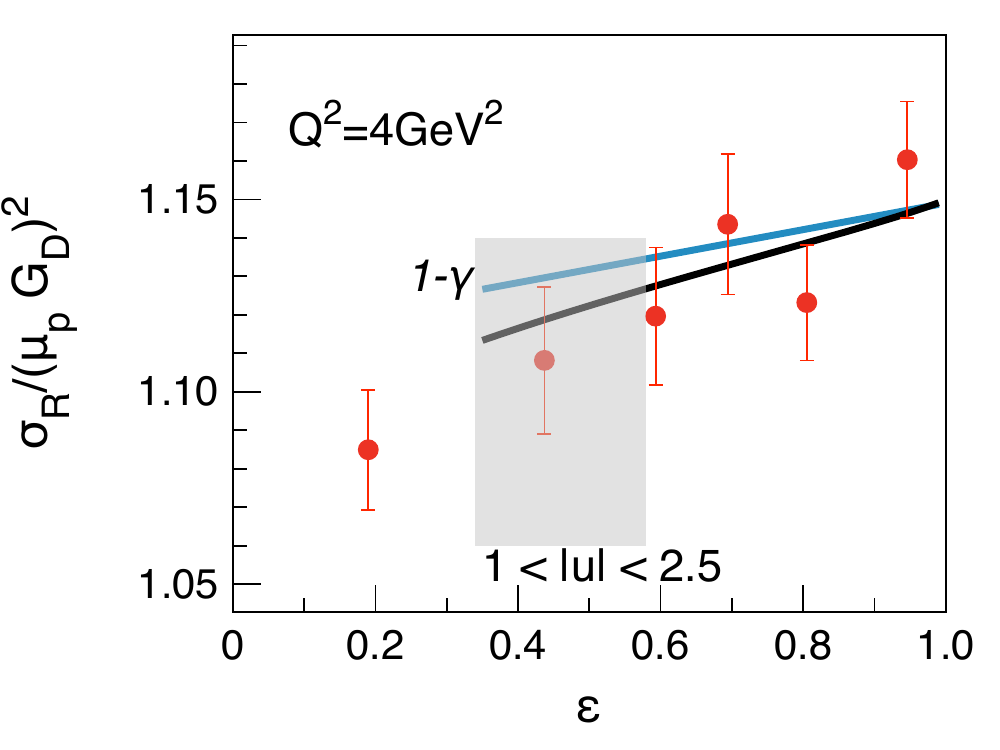}
\includegraphics[height=1.5in]{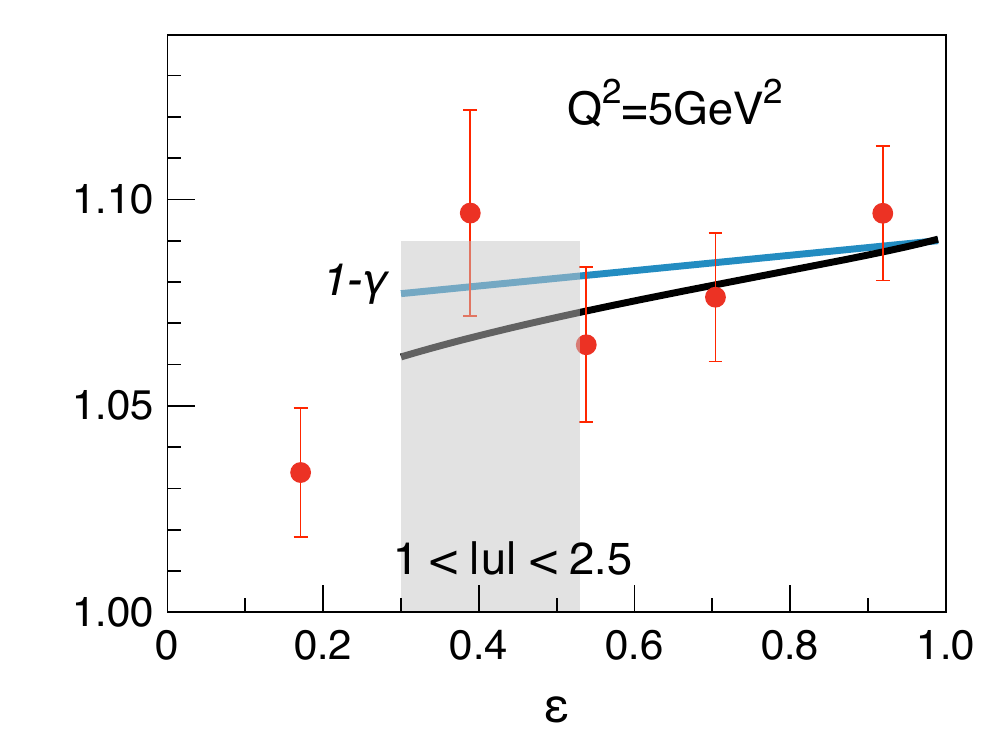}
\caption{ The fit of data for the $ep\rightarrow ep$ elastic scattering reduced cross section from \cite{Andivahis:1994rq} for different values of $Q^{2}$. 
For the fit we used only  the data points which are lying  on the right  of the shaded area. }%
\label{SLAC}%
\end{figure}

\begin{figure}[h]
\centering
\includegraphics[height=1.5in]{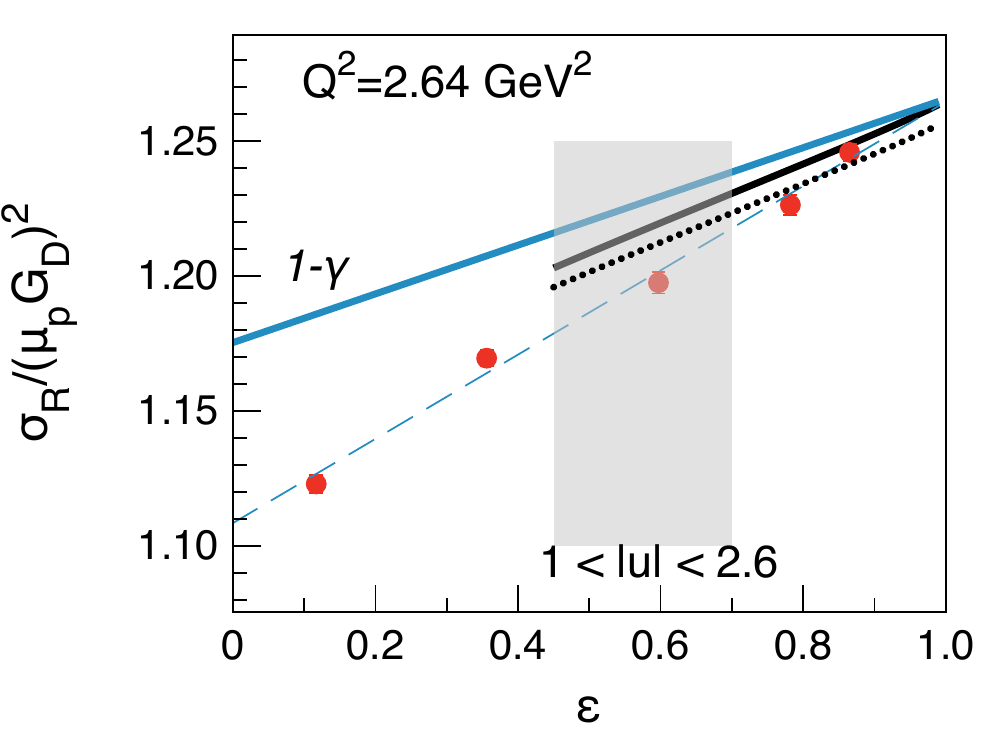}
\includegraphics[height=1.5in]{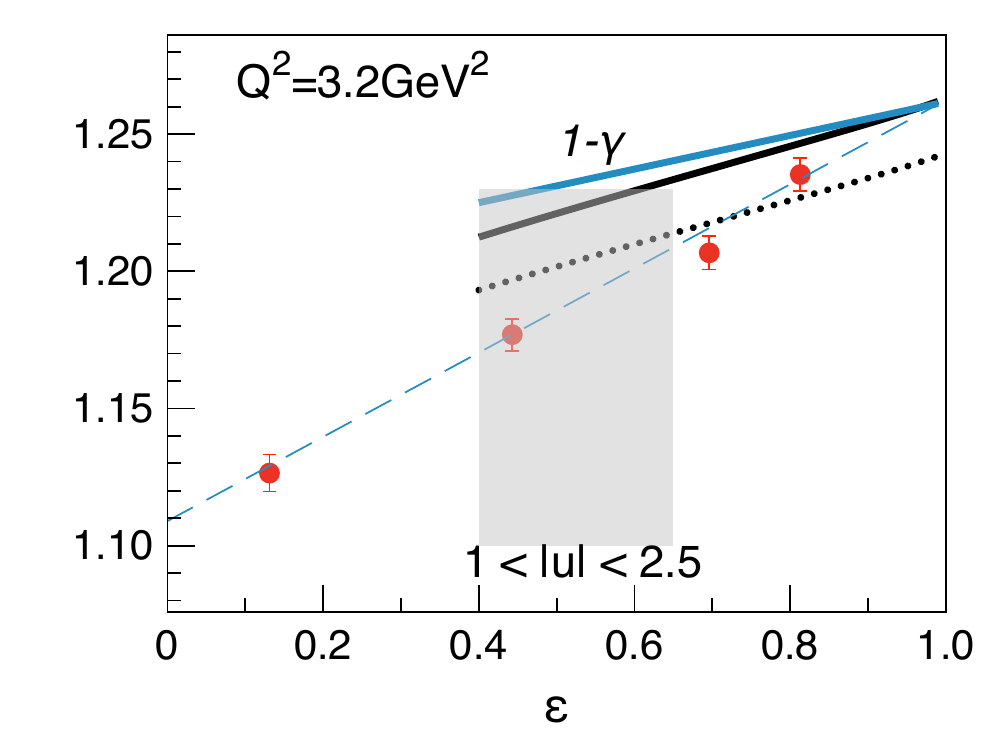}
\includegraphics[height=1.5in]{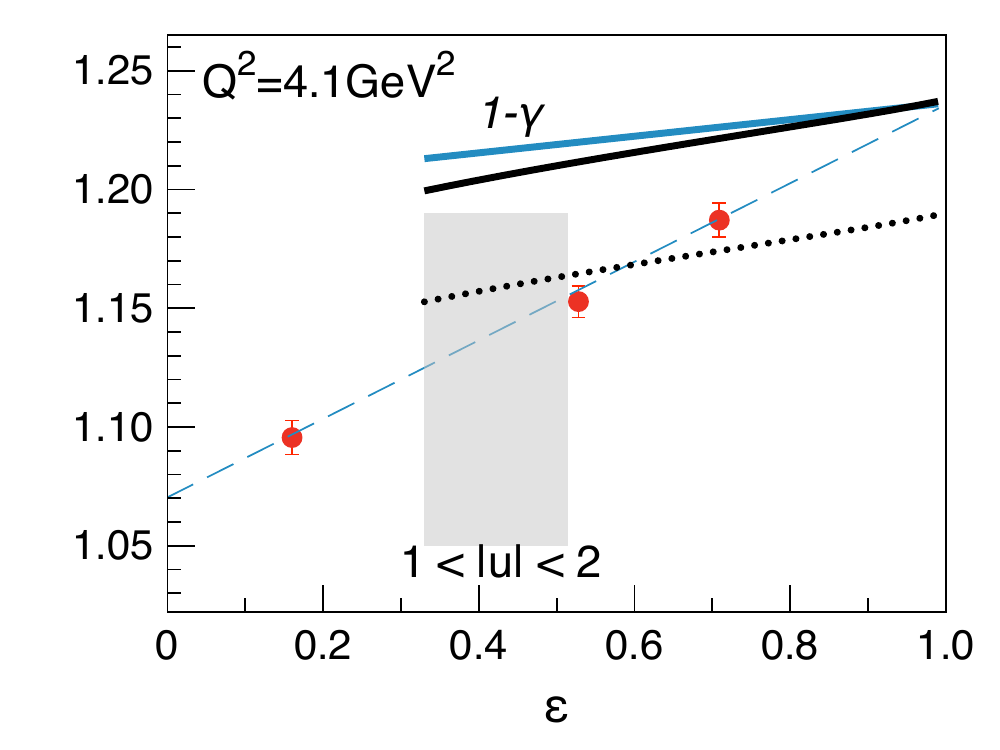}
\caption{ The fit of $G_{M}$  using the data  \cite{Qattan:2005zd}  with the computed  TPE corrections. 
The black solid line shows the  $\sigma_{R}$  with $G^{\text{[Gut]}}_{M}$ extracted from the linear fit in \cite{Guttmann:2010au}. 
The black dotted curve represents  the fit of the data points which are lying  on the right  of the shaded area. }%
\label{JLab}%
\end{figure}

The values $\sigma^{1\gamma,\text{MT} }_{R}$  must be considered as experimental data. 
Taking into account that the ratio $R=G_{E}/G_{M}$ is also measured, see Eq.(\ref{def:R}),   one  finds that only the FF $G_{M}$ is  an unknown quantity on the $rhs$  of the Eq.(\ref{sgmRexp}).  Therefore using Eq.(\ref{sgmRexp})  we can extract the value $G_{M}$ using the data from the kinematical region where our approximation is valid:
$s\sim -t\sim -u\gg \Lambda$.  We  use the data  from  Refs. \cite{Andivahis:1994rq, Qattan:2005zd}. 
In  Figs.\ref{SLAC},\ref{JLab} we show the fit of the experimental data for different
values of $Q$.  The shaded area shows a transition region where the Mandelstam variable  $u$ is
already quite small and one may expect  sizable contributions originating from the  higher order and, probably, power corrections. 
The corresponding  values of $|u|$ (in GeV$^2$) 
are shown at the bottom of the plot. For the fit we used only the points
which are lying on the right of the shaded area. First we observe that
the obtained TPE contribution  is relatively small.  It turns out that  the soft
spectator contribution by coincidence is numerically very close to the MT result for the considered range of $Q^{2}$.

 As we can
see from the plots the obtained behavior of $\sigma_{R}(\varepsilon)$ (solid
black line) is linear in the region of  $\varepsilon$  where the calculation can be applied and  describes
the difference in the slope between polarized and unpolarized data quite
reasonably for the SLAC data within the relatively large error bars. 
The comparison is  worse for the JLab data.
However for the JLab data we have to take into account that, for instance,  for
$Q^{2}=4.1$GeV$^{2}$ the second point corresponds already to relatively low values
of $u$. 

The main trend is that  in the region of small $\varepsilon$ the slope must
change and this means that the reduced cross section can not be described by
the linear function in $\varepsilon$ at large $Q$. Such nonlinearity is still relatively small
at $Q^{2}=2.64~$GeV$^{2}$ (the linear fit is shown by the blue dashed line in
Fig.\ref{JLab} ). But at  $Q^{2}=4.1~$GeV$^{2}$ this effect might  already
be  sizable.  Such a scenario implies a large corrections when
$- u\rightarrow1$~GeV$^{2}$, i.e. in the backward region. 
Theoretically it
may be interesting  to 
go beyond the formalism presented in this work and
compute the TPE corrections using QCD factorization in
the region with small $|u|\lesssim m^{2}$ in order to  have  description for
the whole region of $\varepsilon$.  

The extracted values of $G_{M}$ are shown in Table~\ref{GMtable} and for
comparison we also provide the results of different phenomenological fits
made in Refs. \cite{Guttmann:2010au,Brash:2001qq,Venkat:2010by}. 
\begin{table}[h]
\caption{Results  for magnetic form factor  obtained in this work   from the fit of SLAC data \cite{Andivahis:1994rq} (black solid line in Fig.\ref{SLAC}) 
and JLab data \cite{Qattan:2005zd} (dotted curve in Fig.\ref{JLab}).  The extracted values denoted as $G_{M}^{[\text{fact}]}$.
For comparison we show  the phenomenological  extractions  presented in Refs. \cite{Guttmann:2010au} ($G_{M}^{[\text{Gut}]}$), \cite{Brash:2001qq} ($G_{M}^{[\text{Br}]}$)  
and  \cite{Venkat:2010by}  ($G_{M}^{[\text{Ven}]}$)}.
\label{GMtable}
\begin{center}%
\begin{tabular}
[c]{|c|c|c|c|c|}\hline
{$Q^{2},$GeV$^{2}$} & {$G_{M}^{[\text{fact}]}$} & {$G_{M}^{[\text{Gut}]}$}
& {$G_{M}^{[\text{Br}]}$ } & {$G_{M}^{[\text{Ven}]}$ }\\\hline
\multicolumn{3}{|c|}{JLab} & \multicolumn{2}{|c|}{ }    \\\hline
{$2.64$} & {$0.1356$} & {$0.1360$} & {$0.1350$} & {$0.1352$}\\\hline
{$3.2$} & {$0.1001$} & {$0.1009$} & {$0.0985$} & {$0.0989$}\\\hline
{$4.1$ } & {$0.0654$ } & { $0.0667$} & { $0.0641$} & { $0.0647$}\\\hline
\multicolumn{3}{|c|}{SLAC}& \multicolumn{2}{|c|}{ } \\\hline
{$2.5$} & {$0.1464$} & {$-$} & {$0.1471$} & {$0.1472$}\\\hline
{$3.25$} & {$0.0958$} & {$-$} & {$0.0960$} & {$0.0963$}\\\hline
{$4.0$} & {$0.0670$} & {$-$} & {$0.0670$} & {$0.0675$}\\\hline
{$5.0$} & {$0.0447$} & {$-$} & {$0.0446$} & {$0.0452$}\\\hline
\end{tabular}
\end{center}
\end{table}

\begin{figure}[h]
\centering
\includegraphics[height=2.5in]{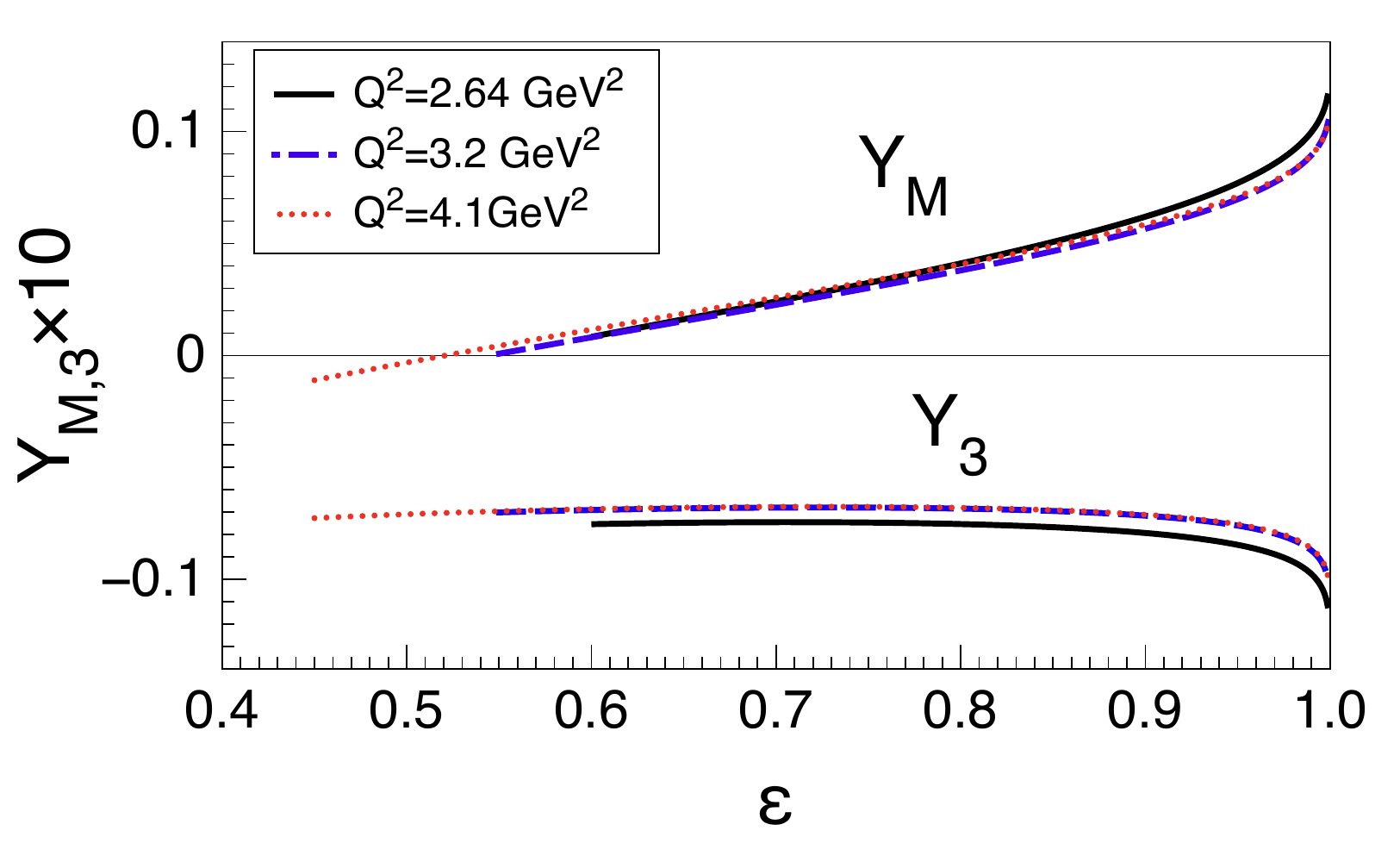}
\caption{ TPE amplitudes (\ref{YM3}) as a functions of $\varepsilon$ at fixed values of $Q^{2}$.   }%
\label{YM3theory}%
\end{figure}

In Fig.\ref{YM3theory} we show the TPE amplitudes which are defined as
 \bea
 Y_{M}=\left(\delta\tilde G^{2 \gamma}_{M}-G_{M}\frac12 \delta^{\text{MT}}_{2\gamma}\right)/G_{M}, \  Y_{3}=\delta\tilde F_{3}/G_{M}.
 \label{YM3}
 \eea
These ratios are plotted for three fixed values of $Q^{2}=2.64,3.2,4.1$~GeV$^{2}$ as  a function of $\varepsilon$. 
For the FF $G_{M}$ we take the values shown in Table~\ref{GMtable}. 
 As we can see these functions show a weak dependence  on  $Q^{2}$. 
At  $Q^{2}=2.64$~GeV$^{2}$ these functions have been estimated from the data using certain assumption about  their behavior at $\varepsilon=1$, 
see e.g.  \cite{Guttmann:2010au, Borisyuk:2010ep, Qattan:2011ke}.  Thus the present calculation shows that the assumption that the amplitudes $Y_{M,3}$ vanish
separately at  $\varepsilon\rightarrow 1$ is probably too strong.  The absolute values of the  $Y_{M,3}$ in Fig.\ref{YM3theory} are also much smaller than the  phenomenological 
ones shown in  \cite{Guttmann:2010au}.

\begin{figure}[h]
\centering
\includegraphics[height=2in]{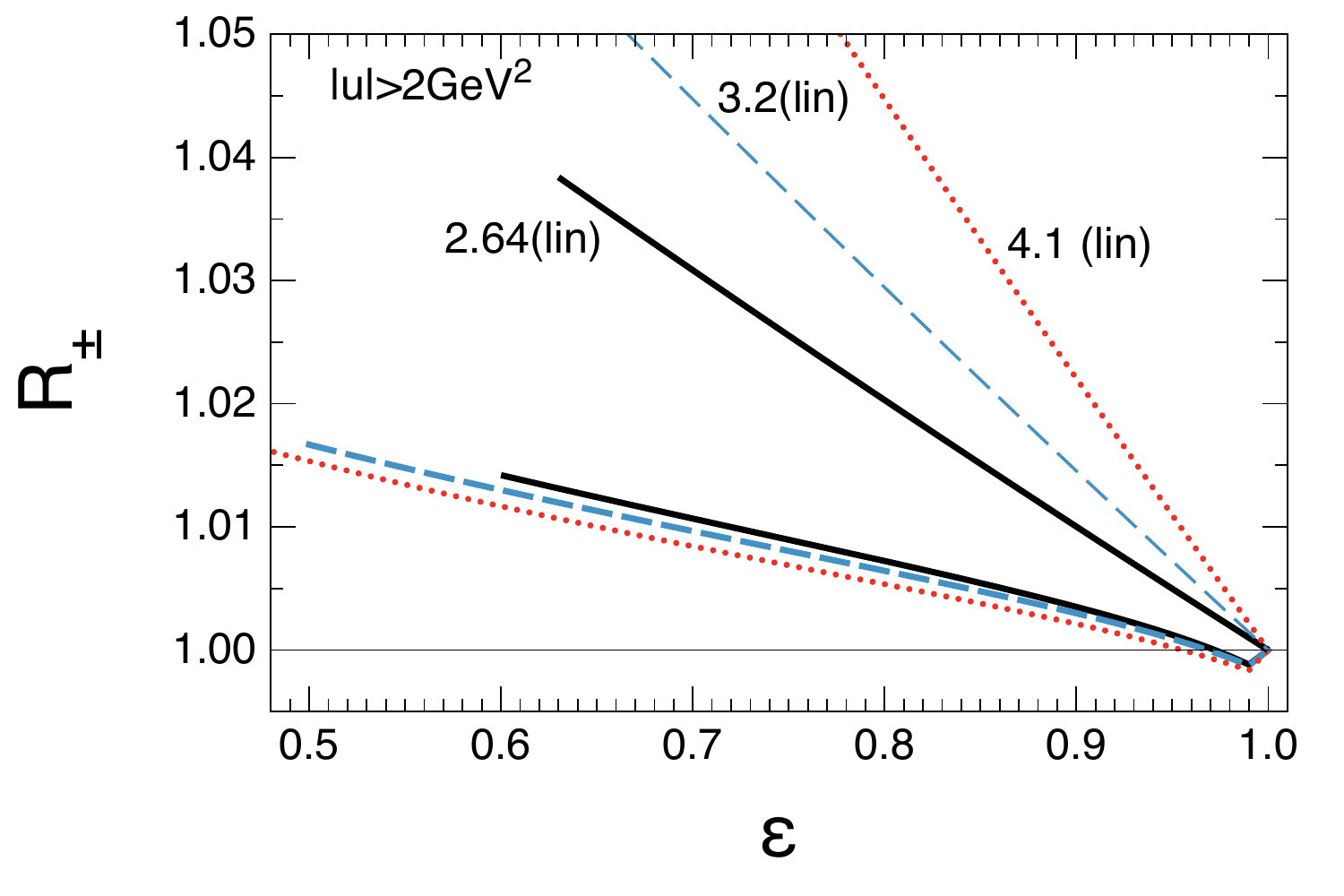}
\caption{ The ratio $R_{\pm}$ for different values of $Q^{2}$ (in GeV$^{2}$):   $Q^{2}=2.64$ (black solid curve), $Q^{2}=3.2$ (blue dashed curve) and  $Q^{2}=4.1$ ( red dotted curve). For comparison we also show the results for the linear fit obtained in \cite{Guttmann:2010au}. 
The value of $\varepsilon$ are restricted to the range  $|u|>2~$GeV$^{2}$. }
\label{Rep}%
\end{figure}

The  other good observable to access  the TPE
 corrections is to measure the ratio of the elastic $e^{\pm}p$ cross
sections
\begin{equation}
R_{\pm}=\frac{d{\sigma}_{e^{+}p}}{d{\sigma}_{e^{-}p}},
\end{equation}
If  the corresponding  reduced cross section ${\sigma}_{R}^{1\gamma}$
extracted with MT radiative corrections  \cite{Tsai:1961zz} are incomplete
then this ratio will differ from unity  due to the TPE contribution.  In our
case using Eq.(\ref{sgmRexp}) we find
\bea
R_{\pm}&\simeq&\frac{\sigma_{R}^{1\gamma}-2G_{M}\operatorname{Re}
[\delta\tilde{G}_{M}^{2 \gamma} +\varepsilon\frac{\nu}{m^{2}}\tilde{F}_{3}-G_{M}\frac12 \delta^{\text{MT}}_{2\gamma}  ]}
{\sigma_{R}^{1\gamma}+2G_{M}\operatorname{Re}[\delta\tilde{G}_{M}^{2 \gamma} +\varepsilon\frac{\nu}{m^{2}}\tilde{F}_{3}-G_{M}\frac12 \delta^{\text{MT}}_{2\gamma}  ]}
\\  
& \approx & 1-\frac{4G_{M}}{\sigma_{R}^{1\gamma}}\operatorname{Re}
[\delta\tilde{G}_{M}^{2 \gamma} +\varepsilon\frac{\nu}{m^{2}}\tilde{F}_{3}-G_{M} \frac12 \delta^{\text{MT}}_{2\gamma}  ].\label{Rpm}%
\eea
Using the values $G_{M}$ obtained from the analysis of the JLab data we obtain the ratio $R_{\pm}$
shown in Fig.\ref{Rep}  as a function
of $\varepsilon$  for three different values of
$Q^{2}=2.64,3.2$ and $4.1$GeV$^{2}$, respectively. For comparison we also show the predictions made in \cite{Guttmann:2010au}
on the basis of linear fit for the same values of momentum transfer. Because we obtained  a relative
large difference between the slopes for the reduced cross section  we also
obtain a large difference in the estimate of the ratio (\ref{Rpm}).  It is
interesting that the absolute value of the obtained TPE correction within the  factorization formalism changes
very slowly with respect to $Q^{2}$. 

Let us stress  that  the above results obtained  within the simple model of the SCET amplitudes $g_{1,3}$, see Eq.(\ref{g1fin}). 
  One cannot exclude that this rough  
model underestimates the  contribution of the amplitudes $g_{1,3}$.  Nevertheless  the above calculations clearly show that the  TPE contribution   associated with  two hard photons  cannot  describe the reduced  cross section  at large $Q^{2}$  being consistent with the  assumption about the linear behavior  
of $\sigma_{R}(\varepsilon)$ in the whole interval $0<\varepsilon<1$  at large fixed $Q^{2}$.  Therefore, in order to understand the TPE contribution better  
it is very important to obtain a  realistic estimate of the  amplitudes $g_{1,3}$. It turns out that the important information about the  
amplitude $g_{3}$  can be obtained from the analysis of  polarization observables.     

\subsection{Recoil polarization observables}

The GEp-$2\gamma$  collaboration 
 has measured the $\varepsilon$-dependence of the recoil polarizations at a
fixed value of $Q^{2}=2.64$ GeV$^{2}$ \cite{Meziane:2010xc}. These data provide additional
important constraints on the theoretical analysis  of the TPE amplitudes.

Let us to start the discussion with the longitudinal polarization $P_{l}$  which in the elastic approximation is given in 
Eq.(\ref{Pl}). Taking into account the inelastic contributions and  assuming that radiative corrections for the experimental data 
have been performed according to MT formulas we can  follow the same line as for the reduced cross section.  Introducing the new
 FFs $G_{M}$ and $G_{E}$  and expanding the reduced cross section we obtain%
\begin{align}
P_{l}/P_{l}^{Born}  & =1-\frac{2\varepsilon}{1+\frac{\varepsilon}{\tau}R^{2}}\frac{1}{G_{M}}
\left\{  
\frac{\varepsilon}{1+\varepsilon}\frac{\nu}{m^{2} }\tilde F_{3}
 +\frac{R}{\tau}
\left(  
\delta\tilde G_{E}+\frac{\nu}{m^{2}}\tilde F_{3}-G_{E}\frac12 \delta^{\text{MT}}_{2\gamma} 
\right)
\right.   
\nonumber \\  & \left. \phantom{empty space}
 -\frac{1}{\tau}R^{2}
 \left(  
 \frac{\varepsilon}{1+\varepsilon} \frac{\nu}{m^{2}}\tilde F_{3}+\delta\tilde G_{M}-G_{M}\frac12 \delta^{\text{MT}}_{2\gamma} 
 \right)  \right\}.
 \label{Pl-PlB}
\end{align}
Using Eqs.(\ref{null},\ref{fwd}) and  Eqs.(\ref{F1T2C2},\ref{def:R2}) one can easily find that in the forward
limit $\varepsilon\rightarrow 1$ this asymmetry does not vanish
\begin{equation}
P_{l}/P_{l}^{Born}(\varepsilon=1)\simeq1+\frac{1}{1+\frac1\tau R^{2}}\frac{\alpha\pi}%
{2}\frac{\mathcal{F}_{1}(Q)}{G_{M}}\simeq1+\frac{\alpha\pi}{2}\frac
{\mathcal{R}(Q)}{G_{M}}.
\end{equation}
 Taking into account  the relative smallness of $\mathcal{O}(R^{2})$-contributions and our assumption (\ref{dGEsmall}) we
can conclude that  the ratio  in Eq.(\ref{Pl-PlB}) is dominated by the first
term $\sim \tilde F_{3}$ in the brackets and can be approximated as
\begin{equation}
P_{l}/P_{l}^{Born}\simeq1-\frac{2\varepsilon^{2}}{1+\varepsilon}\frac
{\frac{\nu}{m^{2}}\tilde F_{3}}{G_{M}}.\label{Plapp}%
\end{equation}
Taking into account that $\tilde F_{3}$ is IR-finite and $\tilde F_{3}(Q^{2},\varepsilon)=0$ in the
MT-calculation we can use $P_{l}$ in order to extract 
information about the amplitude $\tilde F_{3}$. 
The kinematical factor $\varepsilon^{2}$ appearing in  Eq.(\ref{Plapp})  reduces  the possibility of such analysis only to the region
of relatively large values of $\varepsilon$ where our approximation is valid. 
\begin{figure}[h]
\centering
\includegraphics[height=2.0in]{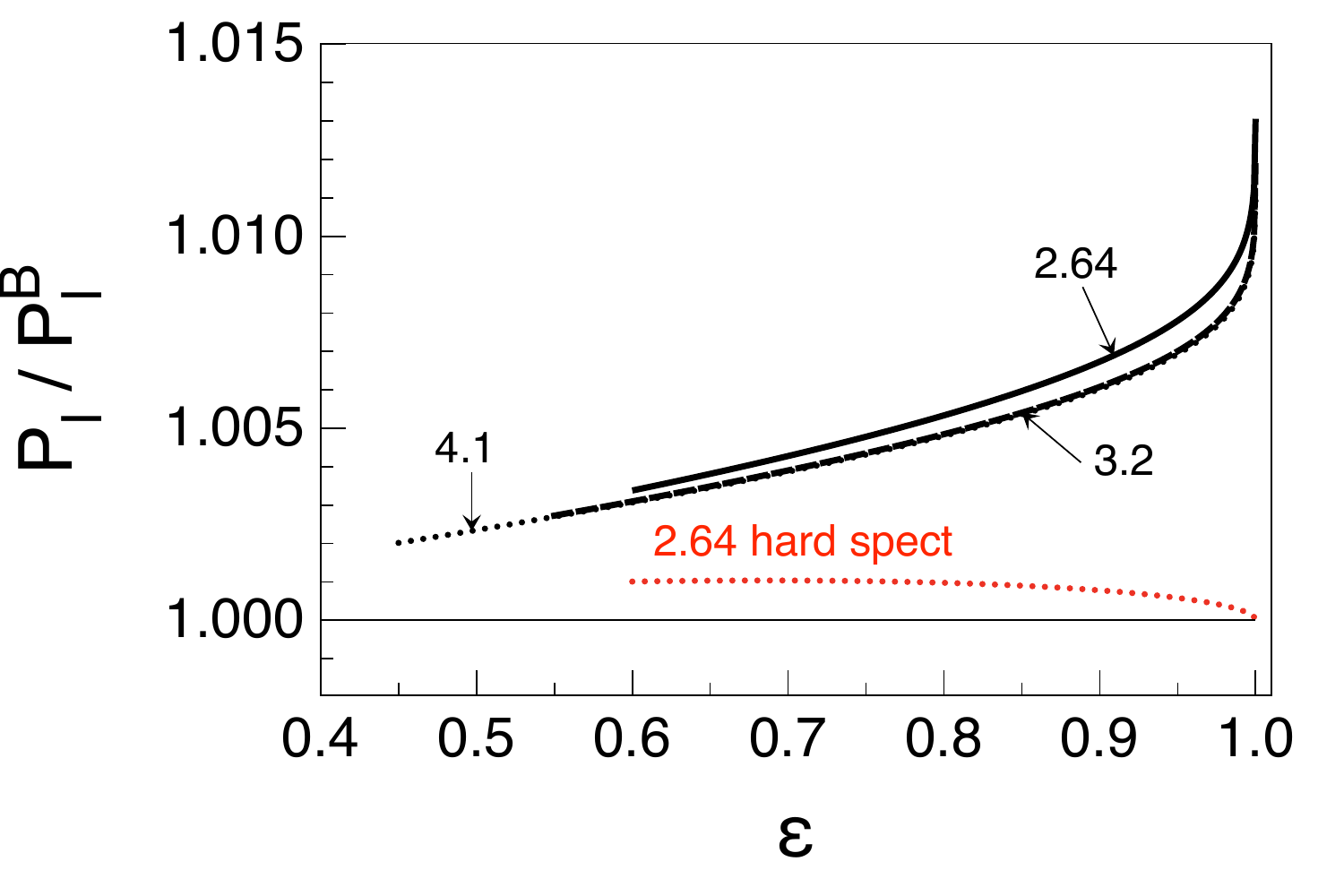}
\includegraphics[height=2.0in]{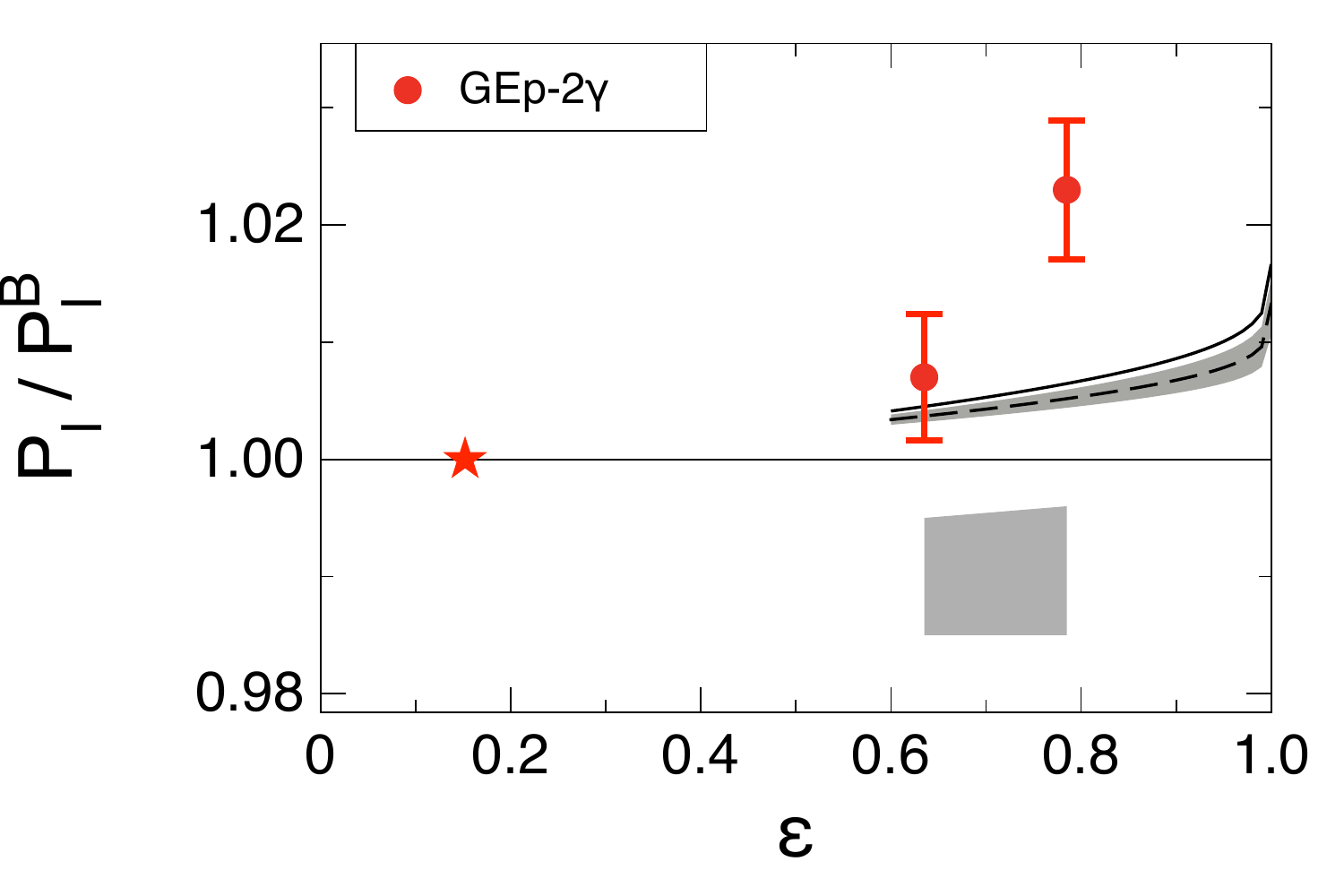}
\caption{ The ratio $P_{l}/P_{l}^{Born}$ as a function of $\varepsilon$ for different values of $Q$ (left) and comparison of our estimates 
with experimental data \cite{Meziane:2010xc} at $Q^{2}=2.64~$GeV$^{2}$.  The  shaded area around the dashed line demonstrates the ambiguity arising 
from  $\mathcal{R}$. The solid line  shows the ratio where  $\mathcal{R}$ obtained from the fit with the kinematical power corrections, as in Fig.\ref{fitF1}  }%
\label{Plratio}%
\end{figure}
In Fig.\ref{Plratio} we demonstrate our
estimates for the $P_{l}/P_{l}^{Born}$ as a function of $\varepsilon$ for
different values of $Q^{2}$. It turns out that the hard spectator contribution $\tilde F_{3}^{(h)}$
is quite small (red dotted line in left plot ) compared to the soft spectator contribution $\tilde F_{3}^{(s)}$. 
From  Fig.\ref{Plratio} we also see that the   dependence on $Q$  is rather weak.  However
this function has a steep behavior in $\varepsilon$ in the vicinity of $\varepsilon=1$. 
Such  behavior originates from the $\varepsilon$-dependence of  the hard coefficient function $C_{3}(z)$, see Eq.(\ref{C3}).  
 On the right plot in Fig.\ref{Plratio} we show a comparison of our estimate at 
$Q^{2}=2.64~$GeV$^{2}$ with the existing experimental data \cite{Meziane:2010xc}.  
For comparison  we show  the solid and dashed curves corresponding  to  the two different fits for $\mathcal{F}_{1}$: 
with and without the kinematical power corrections, respespectively.
The shaded area around the dashed curve shows $1\sigma$ error bands arising from the ambiguity in extraction of $\mathcal{R}$ from the 
WACS data.  
As we discussed in Sec.\ref{sec-wacs}  the  kinematical corrections  increase the extracted value $\mathcal{R}$ and therefore the solid curve lies a bit higher. 
The calculation  shows the qualitative trend of the data, which increase when $\varepsilon \rightarrow 1$.  However  our estimates are 
 considerably smaller than the  data point at large $\varepsilon$.  This observation can be considered as  an  indication that 
  the contribution associated with the amplitude $g_{3}$ is probably quite sizable.  In our estimate we however used approximation $g_{3}\simeq 0$. 
    If such scenario is realized  then the similar contribution to $g_{1}$  must also be large in order to describe correctly the reduced cross section.  
    
 The other important  issue  is that the TPE amplitude  $\frac{\nu}{m^{2}}\tilde F_{3}$ does not   vanish at $\varepsilon=1$ as it has been 
 assumed in phenomenological analysis in Ref.\cite{Guttmann:2010au}.  Notice that this behavior is closely associated with soft spectator
 contribution. 

Taking into account the effect of the  computed TPE corrections  in  the transverse polarization  yields 
\begin{align}
-\sqrt{\frac{\tau(1+\varepsilon)}{2\varepsilon}}\frac{P_{t}}{P_{l}}  &
=R+\frac{1}{G_{M}}
\left(  
\frac{\nu}{m^{2}}\tilde F_{3}+\delta\tilde G_{E}-G_{E}%
\delta_{2\gamma}^{\text{MT}}
\right)  \nonumber \\
& -\frac{R}{G_{M}}\left(  \frac{2\varepsilon}{1+\varepsilon}\frac{\nu}{m^{2}}\tilde F_{3}+\delta\tilde G_{M}-G_{M}\delta_{2\gamma}^{\text{MT}}\right)  .
\label{Ptfull}
\end{align}
In the limit $\varepsilon\rightarrow1$,  the TPE corrections vanish:
\begin{equation}
-\sqrt{\tau}\frac{P_{t}}{P_{l}}(\varepsilon=1)=R\equiv\frac{G_{E}}{G_{M}}.
\end{equation}
 Taking into  account  assumption (\ref{dGEsmall}) and the fact that the second term in Eq.(\ref{Ptfull}) is
proportional to $R$ we obtain the following estimate for the TPE correction 
\begin{equation}
-\sqrt{\frac{\tau(1+\varepsilon)}{2\varepsilon}}\frac{P_{t}}{P_{l}}%
=R+\frac{\alpha}{\pi}\times\mathcal{O}(R).
\end{equation}
This allows one to conclude that the expected correction can be  smaller than $1\%$.
 The  results of   GEp-$2\gamma$ collaboration  \cite{Meziane:2010xc}  are in agreement  with this conclusion.
  Within their error bars of order $1\%$  this experiment does not see any systematic TPE effect on this observable. 
 
\begin{figure}[h]
\centering
\includegraphics[height=2.0in]{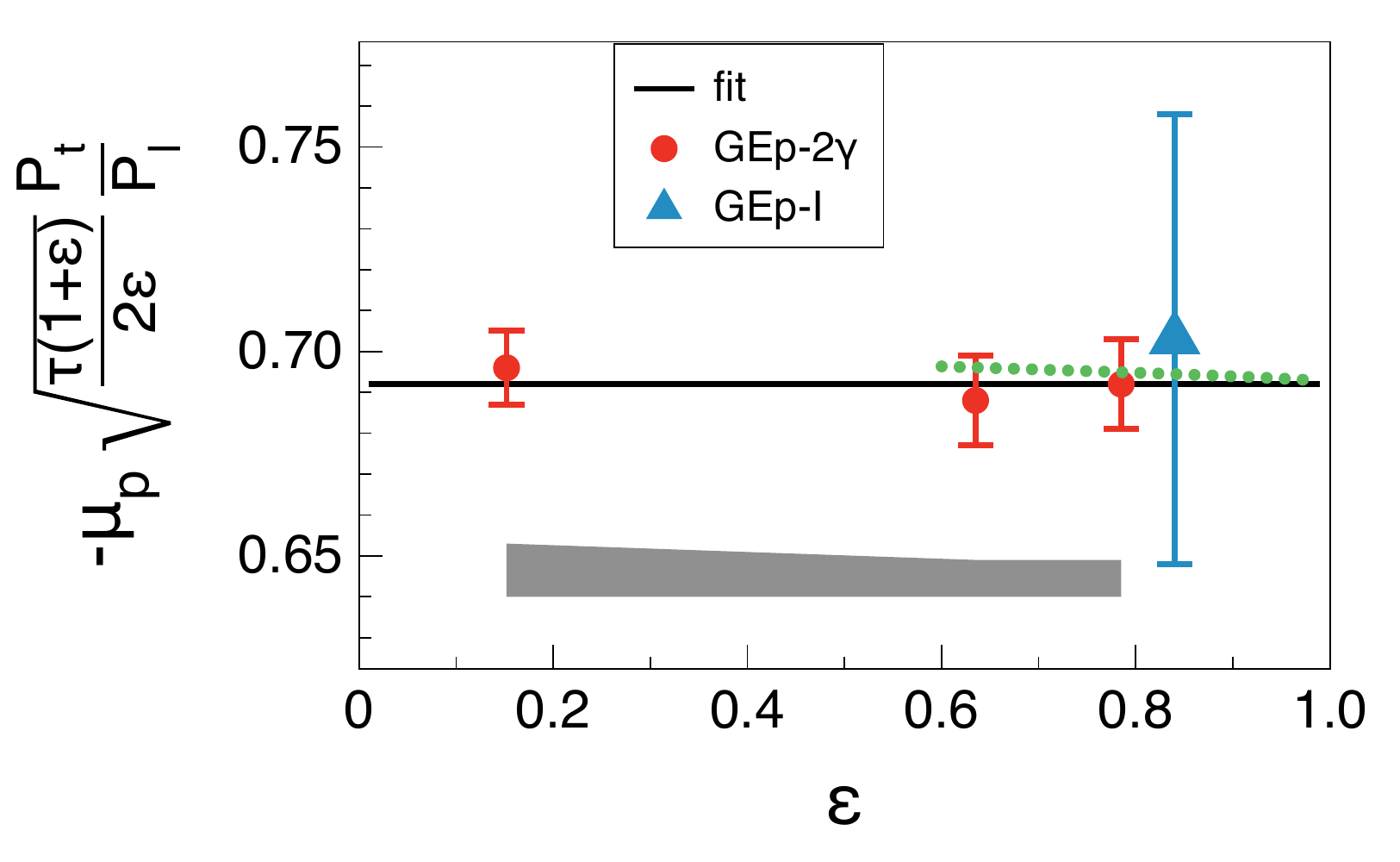}
\caption{ The ratio $-\mu_{p} \sqrt{\tau(1+\varepsilon)/2\varepsilon} P_{t}/P_{l}$ as function of $\varepsilon$  for $Q^{2}=2.5$~GeV$^{2}$. The data are from GEp-I (blue triangle)
\cite{Jones00,Punjabi:2005wq}, and GEp-$2\gamma$ (red circles)  \cite{Meziane:2010xc}. The solid line is an $\varepsilon$-independent fit, the dotted (green) line shows the effect of  the known part of the TPE-correction. }
\label{Pt-plot}
\end{figure}

For illustration  in Fig.\ref{Pt-plot} we show the effect provided by the  known TPE contribution on the $rhs$ of Eq.(\ref{Ptfull})
$  \frac{2\varepsilon}{1+\varepsilon}\frac{\nu}{m^{2}}\tilde F_{3}+\delta\tilde G_{M}-G_{M}\delta^{\text{MT}}_{2\gamma}$.
  The solid line  is an $\varepsilon$-independent fit of $R$.
The dotted  line shows the effect of the computed  part of the  TPE  correction, the curve is restricted by region $\varepsilon>0.6\ (|u|>1.83\text{GeV}^{2})$  
We can clearly see that this effect is  smaller than the experimental error bars. 

\section{Conclusions}
\label{sec-cnc}
 
 In conclusion, we have discussed the calculation of the TPE amplitudes using the QCD\   factorization\  approach\ \  in\ \  the\ \  region\ \  where\ 
 all\ \ Mandelstam\ variables are\ large\ (wide-angle scattering)\ \ $s\sim-t\sim-u\gg\Lambda^{2}$.
 The leading power behavior of these  amplitudes  are described by  the sum of the hard and soft spectator scattering contributions.  
 The hard spectator contribution at leading order accuracy in $\alpha_{s}$  was studied before in \cite{Borisyuk:2008db,Kivel:2009eg}.  
 The soft spectator contribution was  estimated before  using  the handbag  approach in \cite{Afanasev:2005mp}. 
 The  aim of  this work is to develop a more systematic  description of the latter contribution  in order to reduce
 the model dependent assumptions. 
     
  In our work we used the SCET framework and suggested a more rigorous  factorization formula 
  for the description of the soft spectator contribution.  We  carried out the leading power analysis in the hard scale  $1/Q^{2}$ for the TPE amplitudes and 
  obtained that   the soft spectator contributions to the TPE amplitudes can be presented as a sum of  two terms associated with the
  different hard configurations of the photons. 
  
   If  both photons are hard then the corresponding amplitude is described by the hard coefficient function
  and  SCET FF denoted by $\mathcal{F}_{1}$.  This function describes the soft overlap contribution between incoming and outgoing proton
  and depends on the hard-collinear scale of order $\Lambda Q$. Therefore this function depends only on the momentum transfer $Q$.  
  The $\varepsilon$-dependence is completely described by the hard coefficient function and can be computed in perturbation theory. 
   
   If one of the photons is soft then the corresponding SCET amplitudes denoted by $g_{1,3}$ are  different.  These functions describe complicated
   low-energy hard-collinear and soft dynamics.   In case of the soft spectator scattering   the soft photon is involved  in interactions
    with the hard-collinear and soft constituents of the  proton. In this case  the $\varepsilon$ dependence  arises  due to these  long distance interactions.   
    
  At  moderate values of  $Q^{2}$ when the hard-collinear scale  is not large  $Q\Lambda\leq m^{2}$ the SCET amplitudes  describe the long distance 
  interaction and cannot be computed in pQCD.   In order to estimate these quantities we used  different approaches. 
  
   The contribution of the  FF  $\mathcal{F}_{1}$  can be estimated using the universality of its definition in the SCET.  
 It turns out that  this  FF  also arises  in the factorization of the  wide-angle Compton scattering amplitude (WACS). 
 We considered this process  at leading order in $1/s$ and in the coupling  $\alpha_{s}$   and provided the  SCET factorization for the 
 soft spectator contribution  of the WACS amplitudes.  We found that the soft spectator contribution of  all three dominant amplitudes 
 are described by the same SCET FF   $\mathcal{F}_{1}$. This allows us to establish the specific relations between these independent amplitudes.
   Assuming the dominance of the soft-overlap mechanism  we 
 extracted the values of the dominant amplitudes  in the region $Q^{2}=2.5-6.5$~GeV$^{2}$ using the  JLab Hall-A  data for the WACS cross section.  
 The contribution from the  SCET FF $\mathcal{F}_{1}$ can be  related to  the physical WACS  amplitude and  this provides  
   an estimate of this part of the TPE  contribution. 
 
 A description of the amplitudes $g_{1,3}$ is more challenging because  the corresponding configuration is very specific 
 and it is difficult to find a process where they  also enter.  
  Moreover, the amplitude $g_{1}$  also includes   the  QED IR-divergent part which must cancel against inelastic terms. 
  Therefore   in order to estimate these amplitudes  we assumed that the dominant contribution might be  associated 
  with the  exchange of the very soft photon momentum.  
 Such photon interacts with the total charge of the proton and cannot resolve the charges of the constituent quarks. 
  In this case one can estimate the  $g_{1,3}$ using an effective theory with 
  nucleon degrees of freedom similar to the hadronic model  which has been used for TPE calculations at relatively small $Q^{2}$ 
 \cite{Blunden:2003sp, Blunden:2005ew, Kondratyuk:2005kk, Borisyuk:2006fh, Borisyuk:2008es, Borisyuk:2012he}. 
  We  suggested  a somewhat   similar  calculation using for  simplicity only the nucleon as  intermediate state. 
The resulting expression for $g_{1}$ provides the 
correct QED IR-behavior  and  does  not include  any large logarithms  which can be associated with the contribution of the hard region 
(in contrast to  the  MT calculation \cite{Tsai:1961zz}). 
This point is very important because the subprocesses with the hard virtualities are described by the hard coefficient functions.

   The amplitude $g_{3}$  does not have any IR-divergence  and we obtained that $g_3\simeq 0$  in the framework of the simple 
   hadronic model with the nucleon as  intermediate state. 
This allows one to conclude  that  the  dominant contribution to the amplitude $g_{3}$  may arise from  the region where the soft photon 
virtuality is of order $\Lambda^{2}$.   
In this case 
the soft photon can still   resolve the hard-collinear and soft structure of the  proton  arising from the soft spectator scattering. 
An investigation of this  dynamics is definitely beyond the simple hadronic model used here.  In our work we do not
investigate this domain and accept as a first approximation the results of the elastic hadronic calculation.

 Summarizing,   we defined  the  two dominant TPE  amplitudes in the framework of the QCD factorization approach.  We found that  the contribution  associated  
 with the two hard photons   can be computed in terms of  reasonably known nonperturbative quantities: nucleon distribution amplitudes (hard spectator contribution) and  
 SCET FF $\mathcal{F}_{1}$ (soft spectator contribution ) which  can be fixed from the WACS.   The largest ambiguity arises from the  asymmetrical configuration  
 with hard and soft photons which is described 
 in terms of the  SCET amplitudes $g_{1,3}$. The latter  can only be computed in the framework of  low-energy models.  For simplicity, we  consider a scenario 
 when these amplitudes are dominated   by a very soft photon  which already  interacts with the hadron as a point-like particle.  
 The computed   TPE corrections have been used in a  
 phenomenological analysis of  the experimental data  in the region where $|t|, |u|>2.5$~GeV$^{2}$.  
  We obtained that the corresponding TPE  corrections  are linear in $\varepsilon$ but relatively small in magnitude so that already for $Q^{2}>2.5-3$~GeV$^{2}$  
  a description of the reduced cross section   is not compatible anymore with the pure linear fit  used  for the  Rosenbluth separation in the whole  interval $0<\varepsilon<1$.   
Qualitatively this  observation coincides  with the hadronic model  calculations which  also show the similar trend for the moderate values of $Q^{2}$. 

We also  investigated   the ratio $R_{e^{+}/e^{-}}$ which is  sensitive to the same combination of the TPE amplitudes as in the reduced cross section. 
 We  obtained that our predictions with $g_{3}\simeq 0$  for the  $R_{e^{+}/e^{-}}$ at $Q^{2}=2.5-4.1$~GeV$^{2}$ are  considerably  smaller than the  estimates  
 originating from the  linear   fit of the reduced cross section, see e.g. \cite{Guttmann:2010au}.      

However  in the considered  model such a small TPE effect in $\sigma_{R}$   can also be a  consequence  of the  assumptions  about the nonperturbative amplitudes $g_{1,3}$.  
We suppose  that  the used hadronic  model can underestimate 
the contribution of these functions.  The important  information about the amplitude $g_{3}$ can be obtained from the polarization observable $P_{l}$. 
We  found that the ratio $P_{l}/P^{Born}_{l}$    is dominated   by the soft spectator contribution. However using   $g_{3}\simeq 0$  our estimates yield a value for the ratio of $0.5-1\%$  for $Q^{2}=2.5$~GeV$^{2}$ and $\varepsilon>0.7 $,   falling  short of  the data.  
 Further  measurements 
 of  $P_{l}/P^{Born}_{l}$  for   different values of  $\varepsilon$ and $Q^{2}$ with  better accuracy may  help us considerably to constrain the unknown  $g_{3}$  and  reduce the  uncertainty in our  analysis.  
 
The small effect of the TPE corrections in the transverse polarization $P_{t}$  can be explained by the  suppression of the helicity-flip amplitudes comparing to helicity conserving ones at moderate and large $Q^{2}$ values.  This assumption is realized  for the proton FFs in the smallness of the ratio  $G_{E}/G_{M}$  and we
 expect that such suppression  also holds for the  corresponding combinations of the TPE amplitudes.
 
Our work shows  that even for the wide-angle kinematics where 
 $s\sim -t\sim -u\gg \Lambda^{2}$  the  hard TPE contribution ( both photons are hard)   can not  provide the dominant  effect.  We expect that the  relatively large  
 contribution  also    arises from the region where one of the photons is soft   and associated with the nonperturbative hard-collinear and soft dynamics.  
 Such scenario is  supported  by the large TPE effect in the longitudinal polarization $P_{l}$ which is considerably larger than predictions obtained using only  the hard 
 TPE contribution.  The other indication is the relatively small  hard TPE contribution  to the reduced cross section in the large $\varepsilon$ region. 
 Further measurements of the  $\varepsilon$  dependence of the   reduced cross section or the 
 ratio $R_{\pm}$  at intermediate $Q^{2}$ values will test any non-linearities due to TPE effect.

\bigskip

\section*{ Acknowledgments}
 This work was supported by the Helmholtz Institute Mainz.

\begin{appendix}
\section{Tree level matching}
\label{tree}
In order to obtain the expressions (\ref{GMf1}) for the FFs it is enough to
consider the matrix element of the electromagnetic current. We also use
\begin{equation}
p=Q\frac{\bar{n}}{2}+\frac{m}{Q}\frac{n}{2}+\mathcal{O}(m^{2}/Q^{2}%
),~~\ p^{\prime}=\frac{m}{Q}\frac{\bar{n}}{2}+Q\frac{n}{2}+\mathcal{O}%
(m^{2}/Q^{2}).\ \
\end{equation}
Let us define%
\begin{align}
\bar{N}(p^{\prime})  &  =\bar{N}(p^{\prime})\nbn+\bar
{N}(p^{\prime})\nnb\equiv\bar{N}_{+}^{\prime}+\bar{N}%
_{-}^{\prime},\\
N(p)  &  =\nbn N(p)+\nnb N(p)\equiv N_{+}+N_{-}.
\end{align}
Then from the equations of motion for nucleon spinors one obtains%
\begin{equation}
N_{-}\simeq\frac{m}{Q}\frac{\Dslash{n}}{2}N_{+},~\bar{N}_{-}^{\prime}\simeq
\frac{m}{Q}~\bar{N}_{+}^{\prime}\frac{\Dslash {\bar{n}}}{2},\
\end{equation}
which yields %
\begin{equation}
\bar{N}(p^{\prime})\gamma^{\mu}N(p)=\bar{N}_{+}^{\prime}\gamma_{\bot}^{\mu
}N_{+}G_{M}+\frac{m}{Q}(\bar{n}^{\mu}+n^{\mu})~\bar{N}_{+}^{\prime}%
1N_{+}+\mathcal{O}(m^{2}/Q^{2})
\end{equation}%
\begin{equation}
\bar{N}(p^{\prime})1N(p)=\bar{N}_{+}^{\prime}1N_{+}+\mathcal{O}(m^{2}/Q^{2}).
\end{equation}
Therefore%
\begin{align}
\left\langle p^{\prime}\right\vert J_{e.m.}^{\mu}(0)\left\vert p\right\rangle
&  =\bar{N}(p^{\prime})\left[  \gamma^{\mu}G_{M}-\frac{P^{\mu}}{m}%
F_{2}\right]  N(p)\nonumber\\
&  \simeq\bar{N}_{+}^{\prime}\gamma_{\bot}^{\mu}N_{+}G_{M}-\frac{Q}{4m}%
(n+\bar{n})^{\mu}\left(  F_{2}-\frac{4m^{2}}{Q^{2}}G_{M}\right)  \bar{N}%
_{+}^{\prime}1N_{+}. \label{Jexp}%
\end{align}
On the other hand  the leading ordder SCET expression gives%
\begin{equation}
\left\langle p^{\prime}\right\vert J_{e.m.}^{\mu}(0)\left\vert p\right\rangle
\simeq C_{A}~\left\langle p^{\prime}\right\vert O_{+}^{\mu}(0)~\left\vert
p\right\rangle _{\text{{\footnotesize SCET}}}=\bar{N}(p^{\prime})
\nbn\gamma_{\bot}^{\mu}N(p)~C_{A}~f_{1}(Q). \label{Jscet}%
\end{equation}
Comparing (\ref{Jexp}) and (\ref{Jscet}) we obtain%
\begin{equation}
G_{M}\simeq C_{A}~f_{1}(Q),~\ F_{2}\simeq \bar F_{2} +\frac{4m^{2}}{Q^{2}}C_{A}~f_{1}(Q),\
\end{equation}
where $\bar F_{2}$ denotes the contribution from the subleading SCET operators. 

\section{Power counting analysis of the SCET operators arising in elastic $ep$-scattering }
\label{power}
In this section we consider the analysis within SCET of the different
contributions to elastic $ep$-scattering arising at the leading order. We do not pretend to provide a complete
proof of the factorization theorem but our arguments can be considered as a
first step in this direction.  

Following the standard approach,  we use a generic small parameter
$\lambda\sim\sqrt{\Lambda/Q}$. We classify the different regions following
the terminology suggested in Refs.~\cite{Hill:2002vw, Beneke:2003pa}: 
\textit{hard}$~p_{h}\sim Q(1,1,1)\footnote{Here we  imply that the light-cone momentum components are given
 as $(n\cdot p, p_{\bot},\bar n\cdot p)$},~\ $ \textit{semi-hard} $~p_{sh}\sim Q(\lambda
,\lambda,\lambda),$ \textit{hard-collinear} $p_{hc}\sim Q(1,\lambda
,\lambda^{2})$ or $p_{hc}^{\prime}\sim Q(\lambda^{2},\lambda,1),$
\textit{collinear} $p_{c}\sim Q(1,\lambda^{2},\lambda^{4})$ or $p_{c}^{\prime
}\sim Q(\lambda^{4},\lambda^{2},1)$ and \textit{soft} $p_{s}\sim Q(\lambda
^{2},\lambda^{2},\lambda^{2})$.  For the lepton momenta $k$ and $k^{\prime}$
we assume the similar scaling behavior but associated with the vectors
$\bar{v}$ and $v$, respectively. 

This implies the following  power counting  for the SCET QCD fields
\begin{eqnarray}
\text{hard-collinear sector:~}&&\xi^{hc}_{n}\sim\lambda,~\ \bar{n}\cdot
A_{hc}^{(n)}\sim1,~A_{\bot hc}^{(n)}\sim\lambda~\ ,n\cdot A_{hc}^{(n)}%
\sim\lambda^{2},
\label{hc:count} 
\\
\text{ collinear sector:~}&&\xi^{c}_{n}\sim\lambda^{2},~\ \bar{n}\cdot A_{c}%
^{(n)}\sim1,~A_{\bot c}^{(n)}\sim\lambda^{2}~\ ,n\cdot A_{c}^{(n)}\sim
\lambda^{4},
\label{col:count}
\\
\text{soft sector:~}&&
A_{s}^{\mu}\sim\lambda^{2},~\ q\sim\lambda^{3}.
\label{soft:count}
\end{eqnarray}
Similar counting rules are implied  for the QED  collinear and
soft degrees of freedom. In addition to the field relations, we also need the
counting of the energetic (collinear) hadronic state. It reads
\begin{equation}
\left\vert p_{c}\right\rangle \sim\lambda^{-2},
\end{equation}
which follows from the conventional normalization.

We first discuss the power counting of the hard-spectator contribution. The matching for the leading
order contributions  involves the six quark operator constructed only from
the collinear fields $\xi_{n}$, $\xi_{\bar{n}}$ and Wilson lines with
longitudinal collinear gluons $\bar{n}\cdot A^{(n)}$ and $n\cdot A^{(\bar{n}%
)}$ respectively. It is the product of two twist-3 \ 3-quark operators which
define the leading twist nucleon DA (\ref{DA}). When computing the hard-spectator contribution one uses 
\bea
&& \text{T}\exp\left[  i\mathcal{L}_{\text{QCD}}+i\mathcal{L}_{\text{QED}%
}\right]  \simeq
\nonumber \\ &&
\text{T}\left\{  \bar{\zeta}_{v}\gamma^{\mu}\zeta_{\bar{v}%
}~\bar{\chi}_{n}\bar{\chi}_{n}\bar{\chi}_{n}\ast\tilde{H}^{\mu}\ast\chi
_{\bar{n}}\chi_{\bar{n}}\chi_{\bar{n}}\exp\left[  i\mathcal{L}_{\text{SCET}%
}^{(n)}+i\mathcal{L}_{\text{SCET}}^{(\bar{n})}\right]  \right\}  ,
\label{lhsc}
\eea
where $\chi_{n,\bar n}$ is defined in (\ref{qjet}),  $\tilde{H}^{\mu}$ denotes the leading order hard-coefficient function in
position space, and where the asterisks denote the collinear convolutions.  
Eq.~(\ref{col:count}) then results in the power-counting for the hard-spectator amplitude~:
\begin{equation}
A_{ep}^{(h)}\sim\left\langle k^{\prime}\left\vert \bar{\zeta}_{v}\gamma^{\mu
}\zeta_{\bar{v}}\right\vert k\right\rangle \left\langle p^{\prime}\left\vert
\bar{\chi}_{n}\bar{\chi}_{n}\bar{\chi}_{n}|0\left\rangle \ast H^{\mu}%
\ast\right\langle 0|\chi_{\bar{n}}\chi_{\bar{n}}\chi_{\bar{n}}\right\vert
p\right\rangle \sim\lambda^{8}.\label{Jbot}%
\end{equation}
Thus the asymptotic behavior of the TPE amplitudes   coincides  with the scaling behavior of the nucleon
Dirac FF $F_{1}$, see e.g. \cite{Kivel:2010ns}. 

However the hard-spectator factorization is not complete because   there is a
soft-spectator contribution with the same scaling behavior. Moreover due to
the soft-collinear overlap the definition of the  hard and soft  contributions
depends on the rapidity regularization.  Therefore in order to
be consistent  we will take into account all possible soft spectator
configurations which provide the same power  $\sim\lambda^{8}$  when 
passing from the SCET-I to the SCET-II framework. 
For that purpose  we need to determine the $\lambda$ suppression factors which arise when the hard-collinear fields convert into soft 
and collinear fields through time-ordered products.

At a first step, we define the set of the relevant  SCET-I operators.   
We will construct them using gauge invariant  combinations of quark and gluon fields  which for simplicity we will call
hard-collinear or collinear jets, where  the quark jet was defined in Eq.(\ref{qjet}). The analogous gluon jet can be introduced as 
\be{gjet}
 \mathcal{A}_{\mu}^{({n})} \equiv [W_{n}^{\dagger}D_{\bot\mu}W_{n}]
\ee 
where the derivative only  acts inside the brackets.  We also assume that passing from SCET-I to SCET-II we substitute $\xi\rightarrow \xi_{c}+\xi_{hc}$ and  integrate out $\xi_{hc}$.  This implies that  the relevant SCET-I operators can be built from the collinear and hard-collinear modes.  
The SCET-I operator is relevant if the corresponding  time-ordered product  provides the overlap with the leading hard-spectator contribution in Eq.(\ref{lhsc}),
i.e.  after contraction of  all hard-collinear fields   the resulting SCET-II operators scale with $\lambda^{12}$ and have following structure 
\bea
( \bar\chi_{n}\bar\chi_{n}\bar\chi_{n} )(q \bar q [...])(\chi_{\bar n}\chi_{\bar n}\chi_{\bar n}), 
\label{str}
\eea   
where the  3-quark collinear operators are the same as in Eq.(\ref{lhsc}) and  the ellipses denote the additional soft fields.  From the  structure of Eq.(\ref{str})  it follows that  the collinear gluon fields like $\mathcal{A}_{c\bot}$, $n\cdot \mathcal{A}_{c}^{n}$ and $\bar n\cdot \mathcal{A}_{c}^{\bar n}$ can not appear   
in the SCET-I operators otherwise one obtains the operators with the subleading collinear structures.     
Keeping this in mind   we define the
following set of the SCET-I operators  (we do not specify the Dirac and color indices for simplicity). For the 2-jet operators we obtain 
\bea
\left\{  \bar{\chi}^{hc}_{n}\gamma_{\bot}\chi^{hc}_{\bar{n}},\ 
\bar{\chi}^{hc}_{\bar{n}}\gamma_{\bot}\chi^{hc}_{n}\right\}  \sim
\mathcal{O}(\lambda^{2}),
\label{2jet}
\eea
For the 3-jet operators we find (all gluon fields are hard-collinear)
\bea
&&\left\{ 
 \bar{\chi}^{hc}_{n}\gamma_{\bot}{\mathcal{A}}_{\bot}^{(n)}\chi^{hc}_{\bar{n}},~
 \bar{\chi}^{hc}_{n}\gamma_{\bot}{\mathcal{A}}_{\bot}^{(\bar{n})}\chi^{hc}_{\bar{n}},
~(n\leftrightarrow\bar{n})
 \right \} 
 \sim \mathcal{O}(\lambda^{3})
\label{3-jet3}
\\
&&
\left\{ 
\bar{\chi}^{c}_{n}\gamma_{\bot}{\mathcal{A}}_{\bot}^{(n)}\chi^{hc}_{\bar{n}},\
\bar{\chi}^{hc}_{n}\gamma_{\bot}{\mathcal{A}}_{\bot}^{(\bar{n})}\chi^{c}_{\bar{n}},\
\bar{\chi}^{hc}_{n}\gamma_{\bot}(n\cdot {\mathcal{A}}^{(n)})\chi^{hc}_{\bar{n}},\
\bar{\chi}^{hc}_{n}\gamma_{\bot}(\bar{n}\cdot{\mathcal{A}}^{(\bar{n})})\chi^{hc}_{\bar{n}},
\right. \nonumber \\
&& \left.
\phantom{empty space }
~\bar{\chi}^{hc}_{\bar{n}}\ \Gamma (n\cdot\mathcal{A}^{(n)})\chi^{hc}_{\bar{n}},
~(n\leftrightarrow\bar{n})
 \right \} 
 \sim \mathcal{O}(\lambda^{4}),
\label{3-jet4}
\\ 
 &&
\left\{ 
\bar{\chi}^{c}_{n}\gamma_{\bot}(n\cdot {\mathcal{A}}^{(n)})\chi^{hc}_{\bar{n}},\
\bar{\chi}^{hc}_{n}\gamma_{\bot}(\bar{n}\cdot{\mathcal{A}}^{(\bar{n})})\chi^{c}_{\bar{n}},\
\bar{\chi}^{c}_{\bar{n}}\ \Gamma (n\cdot\mathcal{A}^{(n)})\chi^{hc}_{\bar{n}},
\right. \nonumber \\
&& \left.
\phantom{empty space }
\bar{\chi}^{hc}_{\bar{n}}\ \Gamma (n\cdot\mathcal{A}^{(n)})\chi^{c}_{\bar{n}},
~(n\leftrightarrow\bar{n})
 \right \} 
 \sim \mathcal{O}(\lambda^{5}),
\label{3-jet5}
\eea
where we used   notation $\Gamma$ for the appropriate Dirac structures. 

The full set of  4-jet operators are split on two subsets: 4-jet quark operators and 4-jet quark-gluon operators. 
The  4-jet quark subset is described schematically as  
\bea
\left\{  
(\bar{\chi}_{n}\Gamma\bar{\chi}_{n})(\chi_{\bar{n}}\Gamma\chi_{\bar{n}}),\
(\bar{\chi}_{n}\Gamma\bar{\chi}_{\bar n})~(\chi_{\bar{n}}\Gamma\chi_{n}),\ 
(\bar{\chi}_{\bar n}\Gamma\bar{\chi}_{\bar n})~(\chi_{n}\Gamma\chi_{n})
\right\}  
\sim  \mathcal{O}(\lambda^{4})- \mathcal{O}(\lambda^{6}),
\label{4qjet}
\eea
where the  fields $\chi_{n,\bar n}$  describes   collinear or hard-collinear jets.
The subset of the 4-jet quark-gluon operators are  built from the two quark jets  and two gluon jets. This subset is quite long and we shall 
not write it here explicitly. The  corresponding operators  scale as  $\lambda^{4}$ to $\lambda^{8}$.  Schematically  these operators can be presented 
in  following way  
\bea
&& 
\bar{\chi}_{n}\gamma_{\bot} 
\left\{ 
\mathcal{A}_{\mu}^{(n)}\mathcal{A}_{\nu}^{(\bar{n})},\ 
 \mathcal{A}_{\mu}^{(n)}\mathcal{A}_{\nu}^{({n})},\ 
 \mathcal{A}_{\mu}^{(\bar n)}\mathcal{A}_{\nu}^{(\bar{n})}
\right\}
 \chi_{\bar{n}},\  (n\leftrightarrow\bar{n}),
\nonumber \\ 
&&
\bar{\chi}_{n}\Gamma
\left\{ 
\mathcal{A}_{\mu}^{(n)}\mathcal{A}_{\nu}^{(\bar{n})},\ 
\mathcal{A}_{\mu}^{(\bar{n})}\mathcal{A}_{\nu}^{(\bar{n})}
\right\}
\chi_{n},\
(n\leftrightarrow\bar{n}),  
\label{4qgjet}
\eea
where the  fields $\chi_{n}$ and $\chi_{\bar n}$  again represent   collinear or hard-collinear quantities. 

We shall not consider here the other possible higher order operators which  are required for the complete  proof of the factorization. 
 For simplicity we also skip pure gluon operators which can arise beyond the leading order approximation.  
Such consideration is quite complicated and  goes beyond the scope of this paper.  We leave this task for future a work.

\begin{figure}[h]
\centering
\includegraphics[width=5.4196in]
{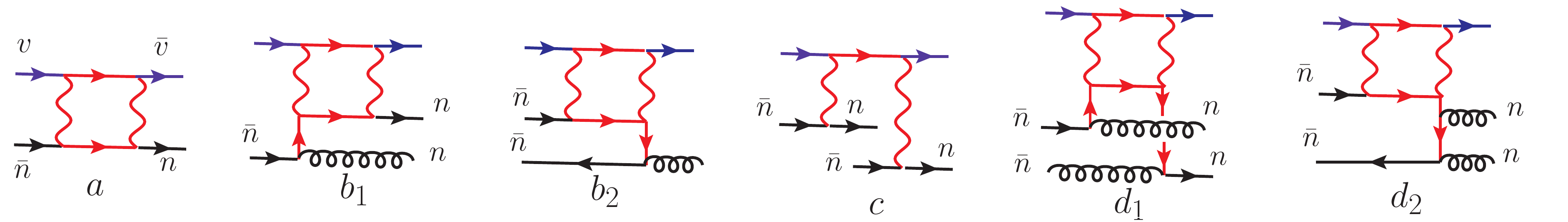}
\caption{The examples  of the TPE diagrams which can produce the different
SCET-I operators discussed in Eqs.(\ref{2jet}-\ref{4qgjet}) at leading order in the QCD running
coupling. We show the different operator families as different groups
$(a), (b_{1},b_{2}), (c), (d_{1},d_{2})$. The external collinearities are
shown by labels $n$ and $\bar{n}$. The red lines specify hard virtualities,
blue lines show the external leptons.  }%
\label{operators}%
\end{figure}

The tree level  TPE diagrams  providing the leading order coefficient functions are given
in Fig.\ref{operators}.
If only one photon is hard then the SCET factorization is
reduced to the FF case and it has been discussed in Ref.\cite{Kivel:2010ns}. 
Therefore we consider only the configurations with  two hard photons.

\begin{figure}[h]%
\centering
\includegraphics[width=1.0in]%
{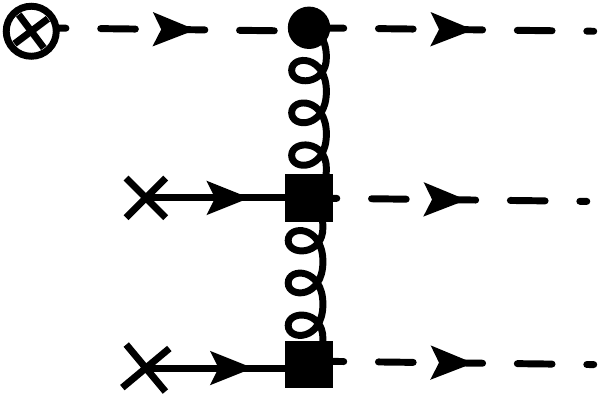}%
\caption{One of the diagrams describing the hard-collinear scattering. The
crossed circle denotes the vertex of the hard-collinear operator ($\bar{\chi
}_{n}^{hc}$), the black circle describes the vertex from $\mathcal{L}_{\xi\xi}^{(0)}$,
the black  squares denote the vertices from $\mathcal{L}_{\xi q}^{(1)}$. The
soft quark fields are shown by solid lines with crosses. }%
\label{jet-function}%
\end{figure}

Let us start the discussion with the 2-jet operators  (\ref{2jet}). Consider the first operator $\bar{\chi}^{hc}_{n}\gamma_{\bot}\chi^{hc}_{\bar{n}}$.  
The matching in this case is  similar  to the consideration for the nucleon FFs described  in Ref.\cite{Kivel:2010ns}.  
In  order to reproduce the required structure in Eq.(\ref{str}) for this case, we need the following time-ordered product 
\bea
\gamma_{\perp}\  
\int d^{4}x_{1}\int d^{4}x_{2}\int d^{4}x_{3}~\text{T}
\left(
\bar{\chi}_{n}^{hc},\mathcal{L}_{\xi q}^{(1)}(x_{1}),
\mathcal{L}_{\xi q}^{(1)} (x_{2}),\mathcal{L}_{\xi\xi}^{(0)}(x_{3})
\right)  
\nonumber \\
\times
\int d^{4} y_{1} \int d^{4} y_{2} \int d^{4} y_{3}~\text{T} 
\left(
\chi_{\bar n}^{hc},\bar{ \mathcal {L}} _{\xi q}^{(1)}(y_{1}),
\bar{\mathcal{L}}_{\xi q}^{(1)}(y_{2}),\bar{ \mathcal{L}}_{\xi\xi}^{(0)}(y_{3})
\right),  
\label{tord}
\eea
where  $\mathcal{L}$ and $\bar{\mathcal{L}}$ denote the different terms of the SCET  Largangian  associated with the  $n$ and $\bar n$ sectors, respectively. 
The explicit  expressions read \cite{BenCh} 
\bea
\mathcal{L}_{\xi\xi}^{(0)}=\bar\xi_{n}
\left(
in\cdot D+ i\Dslash D_{\bot}\frac{1}{in\cdot D}i\Dslash D_{\bot}
\right)
\frac{\Dslash{\bar n}}{2}\xi_{n},
\eea
and
\begin{equation}
\mathcal{L}_{\xi q}^{(1)}=\bar{\xi}_{n}i\hat{D}_{\bot} W_{n}q,
\label{L1scet}%
\end{equation}
which denote the leading order and
next-to-leading order $\sim\mathcal{O}(\lambda)$ SCET Lagrangians respectively.  
For a more detailed discussion of the various aspects of matching from SCET-I to SCET-II in position space formulation of SCET  we refer to 
Ref.\cite{Beneke:2003pa}. 
In order to compute the time-ordered products in each collinear sector we need the interactions terms of the following form 
\bea 
\mathcal{L}_{\xi\xi, \text{int}}^{(0)}= \bar\xi^{c}_{n}(g n\cdot A^{(n)}_{c})\xi^{hc}_{n},
\label{Vxixi1}
\\
\mathcal{L}_{\xi q, \text{int}}^{(1)}=\bar \xi^{c}g\Dslash A^{n}_{\bot hc}W^{hc}_{n} q
\label{Vqxi}
\eea
and similarly for the $\bar{\mathcal{L}}$-contributions. The calculations in both collinear sectors are very similar, therefore we consider only one of them.  
 One of the diagrams describing this subprocess is shown in   Fig.\ref{jet-function}.
The contractions of the hard-collinear fields yield:%
\begin{equation}
\int d^{4}x_{1}~\left\langle A_{hc\bot}^{\alpha}(x_{1})A_{hc\bot}^{\beta
}(x_{2})\right\rangle \sim\int d^{4}x_{2}~\left\langle (\bar{n}\cdot
A_{hc})(x_{2})(n\cdot A_{hc})(x_{3})\right\rangle \sim\lambda^{-2},\label{AA}%
\end{equation}%
\begin{equation}
\int d^{4}x_{3}\left\langle ~\bar{\xi}_{hc}(x_{1})\xi_{hc}(x_{3})\right\rangle
\sim\lambda^{-2},\label{xi-xi}%
\end{equation}
i.e. all hard-collinear contractions cost $\lambda^{-2}$, which results from
the hard-collinear propagators in momentum space. As we assume that
external hard-collinear particles are matched onto collinear ones, 
taking account of the external collinear and soft fields we obtain:%
\begin{equation}
qq\ast J_{n}\ast\underset{\text{3 coll fields}}{\underbrace{\bar{\chi}_{n}%
^{c}\bar{\chi}_{n}^{c}\bar{\chi}_{n}^{c}}}~\sim\underset{\text{2 soft
fields}}{\underbrace{\lambda^{3}\lambda^{3}}}\times\underset{\text{h-coll
contractions}}{\underbrace{\lambda^{-2}\lambda^{-2}\lambda^{-2}}\ }%
\times\underset{\text{3 coll fields}}{\underbrace{\bar{\chi}_{n}^{c}\bar{\chi
}_{n}^{c}\bar{\chi}_{n}^{c}}}~\sim\lambda^{0}~\times\bar{\chi}_{n}^{c}%
\bar{\chi}_{n}^{c}\bar{\chi}_{n}^{c},\label{T3:count}%
\end{equation}
i.e. $J_{n}\sim\lambda^{-6}$.  The same counting is valid for the second jet
function $J_{\bar{n}}$.  The  total contribution in SCET-II  now reads
\begin{equation}
A_{ep}^{(s)}\sim\underset{\lambda^{0}}{\underbrace{\left\langle k^{\prime
}\left\vert \bar{\zeta}_{v}^{c}\gamma^{\mu}\zeta_{\bar{v}}^{c}\right\vert
k\right\rangle }}\underset{\lambda^{4}}{\underbrace{\left\langle p^{\prime
}\right\vert \bar{\chi}_{n}^{c}\bar{\chi}_{n}^{c}\bar{\chi}_{n}^{c}\left\vert
0\right\rangle }}\ast\underset{\lambda^{-6}}{\underbrace{J_{n}}}%
\ast\underset{\lambda^{12}}{\underbrace{\left\langle qq\bar{q}\bar
{q}\right\rangle }}\ast\underset{\lambda^{-6}}{\underbrace{J_{\bar{n}}}}%
\ast\underset{\lambda^{4}}{\underbrace{\left\langle 0\right\vert \chi_{\bar
{n}}^{c}\chi_{\bar{n}}^{c}\chi_{\bar{n}}^{c}\left\vert p\right\rangle }}%
\sim\lambda^{8}.
\end{equation}
We obtain the same power behavior for $A_{ep}^{(s)}$ as in the case of the
hard sector mechanism in (\ref{Jbot}).  The hard-collinear jet functions
for this case and the soft-collinear overlap were studied in more detail in
Ref.~\cite{Kivel:2012mf}.  

\begin{figure}[h]%
\centering
\includegraphics[ width=1.5in]%
{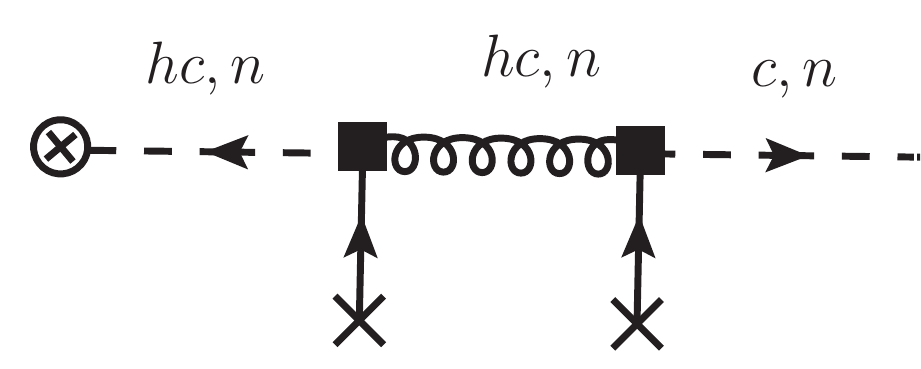}%
\caption{ Quark-antiquark transition in SCET.  The collinear and hard-collinear lines are shown by the labels.  
Notation for the vertices are the same as in Fig.\ref{jet-function}}%
\label{antiquark}%
\end{figure}

We next consider the second 2-jet operator  $\bar{\chi}^{hc}_{\bar{n}}\gamma_{\bot}\chi^{hc}_{n}$ in Eq.(\ref{2jet}).  
The difference with the previous case is that  we  have  hard-collinear antiquarks and therefore in order to 
match the structure of Eq.(\ref{str}) we  have to convert  antiquarks to quarks.  Such a transformation  in SCET-I can only be performed 
 through  two soft-collinear interactions  as shown in Fig.~\ref{antiquark}. 
 In this case  the conversion of  each hard-collinear antiquark  costs at least one power of $\lambda$  comparing to the hard-collinear quark case. 
Therefore the contribution of the ``antiquark'' operator   $\bar{\chi}^{hc}_{\bar{n}}\gamma_{\bot}\chi^{hc}_{n}$  is suppressed by a factor $\lambda^{2}$
comparing to the quark operator  $\bar{\chi}^{hc}_{n}\gamma_{\bot}\chi^{hc}_{\bar{n}}$. Therefore this operator can not provide the overlap with the 
hard-spectator contribution (\ref{lhsc}).  Nevertheless the contribution of the 2-jet ``antiquark'' operator will be suppressed only due to the
hard-collinear dynamics and therefore we suppose that  if the hard-collinear   scale is  not large $\mu_{hc}\lesssim m$ this contribution can 
provide  sizable effects.  Therefore we include it into our consideration of the soft spectator contribution.  
But the other higher order quark-gluon operators with the antiquarks which have structure  like $\bar \chi_{\bar n} ...  \chi_{n}$ or 
$\bar \chi_{\bar n}...  \chi_{\bar n}$ and  $\bar \chi_{n}...  \chi_{n}$ can be discarded  because they will be suppressed also at the level of the hard factorization.   

Neglecting the  hard-collinear antiquarks we considerably reduce the set of the 3-jet   operators which now reads
\bea
&&
 \bar{\chi}^{hc}_{n}\gamma_{\bot}
 \left\{ 
  {\mathcal{A}}_{\bot}^{(n)},\   
{\mathcal{A}}_{\bot}^{(\bar{n})}  
 \right \} \chi^{hc}_{\bar{n}}
  \sim \mathcal{O}(\lambda^{3}),~ 
\label{3-jet3-2}
\\
&&
\bar{\chi}^{c}_{n}\gamma_{\bot}{\mathcal{A}}_{\bot}^{(n)}\chi^{hc}_{\bar{n}},\
\bar{\chi}^{hc}_{n}\gamma_{\bot}{\mathcal{A}}_{\bot}^{(\bar{n})}\chi^{c}_{\bar{n}},\
\bar{\chi}^{hc}_{n}\gamma_{\bot}
 \left\{ 
(n\cdot {\mathcal{A}}^{(n)}),\
(\bar{n}\cdot{\mathcal{A}}^{(\bar{n})})
 \right \} 
 \chi^{hc}_{\bar{n}}
 \sim \mathcal{O}(\lambda^{4}),
\label{3-jet4-2}
\\ 
 &&
\left\{ 
\bar{\chi}^{c}_{n}\gamma_{\bot}(n\cdot {\mathcal{A}}^{(n)})\chi^{hc}_{\bar{n}},\
\bar{\chi}^{hc}_{n}\gamma_{\bot}(\bar{n}\cdot{\mathcal{A}}^{(\bar{n})})\chi^{c}_{\bar{n}},\
 \right \} 
 \sim \mathcal{O}(\lambda^{5}),
\label{3-jet5-2}
\eea
Constructing the time-ordered product for these operators we need to consider the new 2-jet operator vertex  which consists of  
quark and gluon jets.  In the case of the hard-collinear quark jet we can proceed as before in Eq.(\ref{tord}).  
Therefore  in the case of  quark-gluon vertex we need the time-product which is matched onto the  soft-collinear operator 
with the same structure  $(\bar \xi^{c}\bar \xi^{c}\bar\xi^{c}) (qq)$.   One  can consider the same  T-product, for instance
\bea
\text{T}
\left(
\bar{\chi}^{hc}_{n} {\mathcal{A}}_{\bot}^{(n)},\mathcal{L}_{\xi q}^{(1)},
\mathcal{L}_{\xi q}^{(1)} ,\mathcal{L}_{\xi\xi}^{(0)}\right),
\label{T3jet-hc}
\eea
where the integrations are not shown for simplicity.  If the quark jet is collinear then the leading order Lagrangian 
is not relevant and to the same order  we obtain 
\bea
\text{T}
\left(
\bar{\chi}^{c}_{n} {\mathcal{A}}_{\bot}^{(n)},\mathcal{L}_{\xi q}^{(1)},
\mathcal{L}_{\xi q}^{(1)} \right)\ \text{or }  \text{T}
\left(
\bar{\chi}^{c}_{n} {\mathcal{A}}_{\bot}^{(n)},\mathcal{L}_{\xi q}^{(2)},
\mathcal{L}_{\xi q}^{(1)} \right).
\label{T3jet-c}
\eea
However  using the  SCET counting rules  we obtain that all these time-ordered products are suppressed by a factor $\lambda^{2}$ comparing to the 1-jet case  
considered in Eq.(\ref{T3:count}).  For  the T-products in Eqs.(\ref{T3jet-c})  it follows from the observation that the number of the hard-collinear propagators (contractions) is  smaller but the external configuration is the same and this produces the extra $\lambda^{2}$ factor.  If we  use  the leading order vertex 
 (\ref{Vxixi1})  to compute  the T-product  in  (\ref{T3jet-hc})   then 
we  obtain three transverse  fields  $A^{(n)}_{\bot}$  and therefore one of them can not be contracted and must be collinear,  which not match the required  structure  (\ref{str}).  In order to avoid this situation  instead  of (\ref{Vxixi1})  we can take the term with   $\Dslash{A}^{(n)}_{hc\bot}$ and a transverse derivative 
\bea 
\mathcal{L}_{\xi\xi, \text{int}}^{(0)}\simeq \bar\xi^{c}_{n}g  \Dslash{A}^{(n)}_{hc\bot}\frac{1}{in\partial}\Dslash{\partial}_{\bot}\xi^{hc}_{n},
\label{Vxixi1}
\eea
The  transverse derivative  in this case yields  the  factor $\lambda$ and the T-product scales with $\lambda^{7}$.  
 However the  real suppression is stronger.   At the tree level the transverse momenta of all particles
(including the hard-collinear modes) are soft $p_{\perp}\sim\Lambda$ and the transverse derivative 
yields $\lambda^{2}$ factor.  We expect that in case of the hard-collinear loops the additional  $\lambda$ factor  also arises, see
for instance the discussion in Ref.\cite{Beneke:2003pa}.   

The consideration  of the  vertices with the longitudinal projection $(n\cdot {\mathcal{A}}^{(n)})$ is quite similar and shows that 
these contributions  are also suppressed by a factor 
$\lambda^{2}$ compared to the 1-jet contribution in (\ref{T3:count}).  Therefore we conclude that  3-jet operators  can not provide the  
operator with structure  (\ref{str})   which scales with $\lambda^{12}$. Therefore  3-jet operators can be  neglected as subleading contributions.   

\begin{figure}[h]%
\centering
\includegraphics[ width=3.5 in]%
{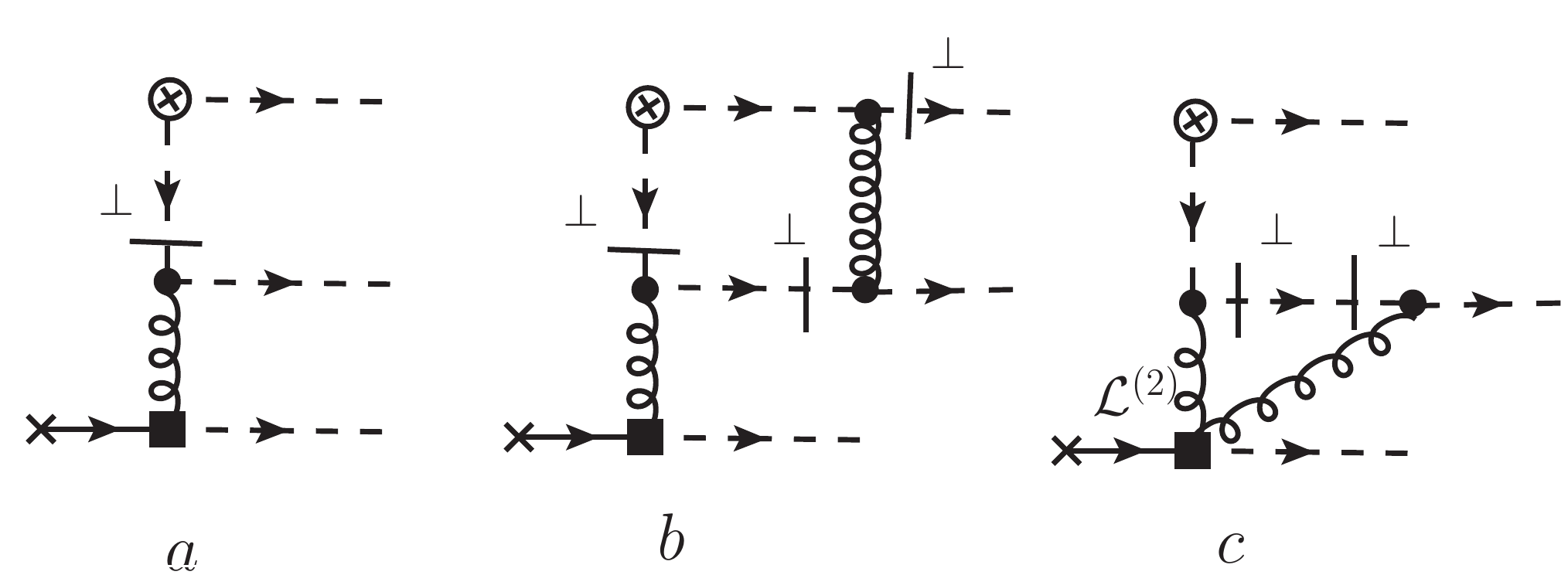}%
\caption{An example of a diagram originating in the calculation of different  T-products  associated with 4-jet quark operators.
 The notation for the lines and vertices is the same as in  Fig.\ref{jet-function}. The short lines crossing the hard-collinear quark propagators denote
 the transverse derivatives. }%
\label{2jet-diagrams}%
\end{figure}

The similar consideration   shows that the different 4-jet quark-gluon contributions are  also suppressed. 
But this analysis is rather lengthy  and we shall not consider it here. 
Let us  discuss  only the four-jet operators listed in Eq.(\ref{4qjet}).  Neglecting the antiquark operators we obtain only one structure
\bea
\left\{  
(\bar{\chi}_{n}\Gamma\bar{\chi}_{n})(\chi_{\bar{n}}\Gamma\chi_{\bar{n}})
\right\} .
\label{4qjet}
\eea
 We can conclude that the minimal
configuration which can match the required structure (\ref{str}) is  described by an operator with the two soft fields $q\bar q$ (or one soft spectator quark).  
The operators in the both collinear sectors are the same hence we consider only one of them. The required time-ordered product reads
\begin{equation}
\int d^{4}x_{1}\int d^{4}x_{2}~\text{T}\left\{  
(\bar{\chi}_{n}^{c}\Gamma\bar{\chi}_{n}^{hc}),\mathcal{L}_{\xi q}^{(1)}(x_{1}),\mathcal{L}
_{\xi\xi}^{(0)}(x_{2})\right\} , 
\label{J2n}%
\end{equation}
where we consider the collinear and hard-collinear jets for clarity.  
This choice implies that we use the interaction vertices as in (\ref{Vqxi}) and (\ref{Vxixi1}).  An example of
a  diagram  described by this T-product   is shown in Fig.~\ref{2jet-diagrams}$a$.
Taking into account that in tree level diagrams $p_{\perp}\sim\Lambda$ which yields $\lambda^{2}$ 
 we obtain that the T-product of Eq.(\ref{J2n}) scales with $\lambda^{7}$.  This is suppressed 
by a factor $\lambda$ comparing to the T-product of single quark jet  in (\ref{T3:count}).  The hard-collinear loops  have the same scaling behavior because the number of   transverse derivatives is always odd, see for example the diagram in Fig.~{\ref{2jet-diagrams}}$b$.  Computing such loops one must have an even number of the transverse derivatives in the integrand  otherwise the loop integral is trivial.  Then the one  derivative must always act on an external field which scales as 
$\lambda^{2}$ as it is shown in Fig.~{\ref{2jet-diagrams}}$b$ giving $\lambda^{2}$. Therefore the loop corrections also scale as $\lambda^{7}$.    

An additional  possibility is to
insert the higher order vertex which is generated by the $\mathcal{ O}(\lambda^{2})$ Lagrangian $\mathcal {L}^{(2)}$
\bea
\mathcal {L}_{int}^{(2)}\sim \xi^{c}\Dslash {A}^{hc}_{\bot}\frac{1}{i n\cdot\partial}\Dslash {A}^{hc}_{\bot}q
\eea
In this case one can use the even number of the vertices generated by the leading order Lagrangian  $\mathcal {L}^{(0)}_{\xi\xi}$. The example of the diagram generated by the corresponding time-ordered product is shown in Fig.~\ref{2jet-diagrams}$c$. The  corresponding contribution scales again as $\lambda^{7}$.  
Therefore we conclude that 4-jet quark operators provide the $\mathcal{O}(\lambda^{2})$ correction to the leading order contribution. We also take into account that these operators have larger dimension than the 2-jet operators  and are therefore stronger suppressed  by a hard scale after the hard factorization.  Therefore we also discard  these contributions.   

In order to provide  the complete proof that only the 2-jet operator is relevant we must  also  consider some other  higher order operators which can potentially provide the structure of Eq.(\ref{str}). However such consideration  goes beyond the subject of this publication  and we will  not present it here.  

\section{Calculation of the one-loop diagrams in Fig.\ref{g1-renorm}}
\label{calculation}
The combination of the matrix elements which define the SCET amplitude $g_{1}^{q}$
reads
\begin{equation}
\left\langle p^{\prime}\right\vert T\{O_{+}^{\mu}Y_{v}^{\dag}(0)S_{\bar{v}%
}(0)\}~\left\vert p\right\rangle _{\text{{\footnotesize SCET}}}+
\left\langle\bar{p}^{\prime}\right\vert T\{O_{+}^{\mu}Y_{v}^{\dag}(0)S_{\bar{v}%
}(0)\}~\left\vert \bar{p}\right\rangle _{\text{{\footnotesize SCET}}}%
=\frac{\alpha}{\pi}g_{1}^{q}(z,Q,\mu_{F}), \label{def:g1q}%
\end{equation}
where the second term $\left\langle\bar{p}^{\prime}\right\vert \dots \left\vert \bar{p}\right\rangle$ denotes the antiquark contribution.
We consider only calculation of the fist matrix element on lhs (\ref{def:g1q}%
). The explicit expressions for the diagrams in Fig.\ref{g1-renorm} can be
written as follows%
\begin{equation}
D_{1a}+D_{1b}=\frac{ie_{q}^{2}e^{2}\bar{\mu}^{2\varepsilon}}{(2\pi)^{D}}%
~\bar{\xi}_{n}\gamma^{\mu}\xi_{\bar{n}}\int\frac{d^{D}l}{\left[  l^{2}%
-\lambda^{2}+i\varepsilon\right]  }\frac{1}{\left[  (l\bar{n})+i\varepsilon
\right]  }\left\{  \frac{(v\bar{n})}{\left[  (vl)+i\varepsilon\right]  }%
+\frac{(\bar{v}\bar{n})}{\left[  -\left(  \bar{v}l\right)  +i\varepsilon
\right]  }\right\}  , \label{D1ab}%
\end{equation}%
\begin{equation}
D_{2a}+D_{2b}=\frac{ie_{q}^{2}e^{2}\bar{\mu}^{2\varepsilon}}{(2\pi)^{D}}%
~\bar{\xi}_{n}\gamma^{\mu}\xi_{\bar{n}}\int\frac{d^{D}l}{\left[  l^{2}%
-\lambda^{2}+i\varepsilon\right]  }\frac{1}{\left[  (ln)+i\varepsilon\right]
}\left\{  \frac{(\bar{v}n)}{\left[  (\bar{v}l)+i\varepsilon\right]  }%
+\frac{(v\bar{n})}{\left[  -\left(  vl\right)  +i\varepsilon\right]
}\right\}  . \label{D2ab}%
\end{equation}
Here we assume that the renormalization scale $\bar{\mu}^{2}$ is defined in
MS-scheme%
\begin{equation}
\bar{\mu}^{2}\equiv\mu_{F}^{2}e^{-\psi(1)-\ln4\pi}.
\end{equation}
One can easily see that each integral $D_{ik}$ represents the evaluation of
the $T$-product of the WLs associated with the couple of the light-cone
vectors: $v$ and $n$, $v$ and $\bar{n}$ and so on. In fact, soft photons can
be decoupled from the hard-collinear quarks performing the similar
redefinition of the quark fields as we did for the lepton fields in
Eq.(\ref{spdec}). Then one obtains that the diagrams in Fig.\ref{g1-renorm}
represents  the  calculation of the $T$-product of the light-like WLs.

Computation of the integrals in Eqs. (\ref{D1ab}) and (\ref{D2ab}) are very
similar therefore it is enough to consider only one combination.\ Each
integral $D_{ij}$ are not well defined even in dimensional regularization and
therefore one needs to introduce some additional regularization. Consider the
integral in (\ref{D1ab}) and let us introduce the following regularized
expressions%
\begin{equation}
D_{1a,b}=\frac{ie_{q}^{2}e^{2}\bar{\mu}^{2\varepsilon}}{(2\pi)^{D}}~\bar{\xi
}_{n}\gamma^{\mu}\xi_{\bar{n}}~J_{1a,b}~,
\end{equation}
with
\begin{equation}
~J_{1a}=\int\frac{d^{D}l}{\left[  l^{2}-\lambda^{2}+i\varepsilon\right]
}\frac{1}{\left[  (l\bar{n})-\tau_{-}+i\varepsilon\right]  }\frac{(\bar{v}%
\bar{n})}{\left[  -\left(  \bar{v}l\right)  -\tau_{+}+i\varepsilon\right]  },
\end{equation}%
\begin{equation}
J_{1b}=\int\frac{d^{D}l}{\left[  l^{2}-\lambda^{2}+i\varepsilon\right]  }%
\frac{1}{\left[  (l\bar{n})-\tau_{-}+i\varepsilon\right]  }\frac{(v\bar{n}%
)}{\left[  (vl)-\tau_{+}+i\varepsilon\right]  },
\end{equation}
where $\tau_{\pm}$ are additional regulators required in the intermediate
calculations. With these regulators we can compute the integrals separately.
Consider $J_{1b}$, using light-cone vectors $\bar{n}$ and $\bar{v}$ as basic
light-cone vectors:
\begin{equation}
l=(l\bar{n})\frac{v}{(v\bar{n})}+(l\bar{v})\frac{\bar{n}}{(\bar{v}\bar{n}%
)}+l_{\perp}\equiv l_{-}\frac{v}{(v\bar{n})}+l_{+}\frac{\bar{n}}{(\bar{v}%
\bar{n})}+l_{\perp},
\end{equation}
we rewrite the integrals in terms of these Sudakov variables%
\begin{equation}
~J_{1b}=\int dl_{+}dl_{-}dl_{\bot}\frac{1}{\left[  zl_{+}l_{-}-l_{\bot}%
^{2}-\lambda^{2}\right]  }\frac{1}{\left[  l_{-}-\tau_{-}\right]  }\frac
{1}{\left[  l_{+}-\tau_{+}\right]  },
\end{equation}
where we used that
\begin{equation}
d^{D}l=\frac{1}{(v\bar{n})}dl_{+}dl_{-}d^{D-2}l_{\bot},~(v\bar{n}%
)=\frac{2(k^{\prime}p)}{Q^{2}}=\frac{-u}{Q^{2}}=\frac{\bar{z}}{z}.
\end{equation}

The obtained integral can be easily computed using residues and then
integrating over transverse momenta:
\begin{equation}
~J_{1b}=(-2\pi i)\int_{0}^{\infty}dl_{-}dl_{\bot}~\frac{1}{\left[  zl_{-}%
\tau_{+}+\lambda^{2}+l_{\bot}^{2}\right]  }~\frac{1}{\left[  l_{-}+\tau
_{-}\right]  }%
\end{equation}%
\begin{equation}
=(-2\pi i)\Gamma(\varepsilon)\int_{0}^{\infty}dl_{-}~\frac{1}{\left[
zl_{-}\tau_{+}+\lambda^{2}\right]  ^{\varepsilon}}~\frac{1}{\left[  l_{-}%
+\tau_{-}\right]  }.
\end{equation}%
\begin{equation}
=(-2\pi i)\frac{\Gamma(\varepsilon)}{\lambda^{2\varepsilon}}\left\{  \frac
{1}{\varepsilon}-\ln\left[  \frac{z\tau_{+}\tau_{-}}{\bar{z}\lambda^{2}%
}\right]  -\psi(1+\varepsilon)+\psi(1)\right\}  +\mathcal{O}(\tau_{+}\tau
_{-}).
\end{equation}
The second integral can be computed in the same way, we need only to take into
account that%
\begin{equation}
(\bar{v}\bar{n})=\frac{2(kp)}{Q^{2}}=\frac{s}{Q^{2}}=\frac{1}{z}.
\end{equation}%
\begin{equation}
J_{1a}=2\pi i\frac{\Gamma(\varepsilon)}{\lambda^{2\varepsilon}}\left\{
\frac{1}{\varepsilon}-\ln\left[  \frac{-z\tau_{+}\tau_{-}}{\lambda^{2}%
}\right]  -\psi(1+\varepsilon)+\psi(1)\right\}  +\mathcal{O}(\tau_{+}\tau
_{-}).
\end{equation}
Therefore the sum reads%
\begin{equation}
~J_{1a}+J_{1b}=2\pi i\frac{\Gamma(\varepsilon)}{\lambda^{2\varepsilon}%
}\left\{  \ln\left[  \frac{z\tau_{+}\tau_{-}}{\bar{z}\lambda^{2}}\right]
-\ln\left[  \frac{z\tau_{+}\tau_{-}}{\bar{z}\lambda^{2}}\right]  \right\}
\end{equation}%
\begin{equation}
=-2\pi i\frac{\Gamma(\varepsilon)}{\lambda^{2\varepsilon}}\ln\bar{z},
\end{equation}
where we keep only the real part. Then%
\begin{equation}
D_{1a}+D_{1b}=\frac{e_{q}^{2}e^{2}}{8\pi^{2}}~\bar{\xi}_{n}\gamma^{\mu}%
\xi_{\bar{n}}\left\{  \frac{1}{\varepsilon}\ln\bar{z}+\ln\bar{z}\ln\frac
{\mu_{F}^{2}}{\lambda^{2}}\right\}  .
\end{equation}
The second sum can be computed similarly and the sum of the all terms reads%
\begin{equation}
D_{1a}+D_{1b}+D_{2a}+D_{2b}=\frac{\alpha}{\pi}~\bar{\xi}_{n}\gamma^{\mu}%
\xi_{\bar{n}}~e_{q}^{2}\left\{  \frac{1}{\varepsilon}\ln\bar{z}+\ln\bar{z}%
\ln\frac{\mu_{F}^{2}}{\lambda^{2}}\right\}  .
\end{equation}
The antiquark can be computed in the same way. The UV-pole $1/\varepsilon$ is
removed by the renormalization and the final expression for the SCET ampltude
$g_{1}^{q}$ reads
\begin{equation}
g^{q}_{1}(z,Q,\mu_{F})=\ln\bar{z}\ln\frac{\mu_{F}^{2}}{\lambda^{2}}\mathcal{F}%
_{1}^{q}.
\end{equation}

\section{Cancellation of  the contribution from the  hard-collinear regions in the box diagrams  }
\label{cancellation}
In order to see the cancellation of  contributions with the hard-collinear photons  we 
suggest to compute  the  diagrams in Fig.\ref{box-quarks}  using the   strategy  of regions \cite{Beneke:1997zp, Smirnov:2002pj}. 
 Using the IR-regularization with off-shell external momenta we obtain%
\begin{equation}
D_{1,2}= -i \frac{e^{4}e_{q}^{2}}{(2\pi)^{D}}~J_{1,2},
\end{equation}
where subscript $1,2$ denotes the box and the crossed box diagram in Fig.\ref{box-quarks}, respectively.  
Corresponding integrals read 
\begin{equation}
~J_{1}=\int dl~\frac{\bar{u}_{\bar v}\gamma^{\nu}(\Dslash k-\Dslash l)\gamma^{\mu}u_{v}}{\left[
l^{2}-2(lk)-k_{\bot}^{2}\right]  }~\frac{1}{\left[  l^{2}\right]  \left[
(l+p-p^{\prime})^{2}\right]  }\frac{\bar{\xi}_{n}\gamma^{\nu}\left(  \Dslash{l}+\Dslash{p}\right)  \gamma^{\mu}\xi_{\bar{n}}}{\left[  l^{2}+2(lp)-p_{\bot
}^{2}\right]  },
\end{equation}%
\begin{equation}
J_{2}=\int dl~\frac{\bar{u}_{\bar v}\gamma^{\nu}(\Dslash k^{\prime}+\Dslash l)\gamma^{\mu}u_{v}}{\left[
l^{2}+2(lk^{\prime})-k_{\bot}^{2}\right]  }~\frac{1}{\left[  l^{2}\right]
\left[  (p^{\prime}-p-l)^{2}\right]  }\frac{\bar{\xi}_{n}\gamma^{\nu}\left(
\Dslash{l}+\Dslash{p}\right)  \gamma^{\mu}\xi_{\bar{n}}}{\left[  l^{2}%
+2(lp)-p_{\bot}^{2}\right]  }.
\end{equation}
where we assume that  poles in the brackets $[...]$ are always defined with $+i\varepsilon$ prescription.  

In the collinear to $p$ region we obtain
\begin{equation}
J_{1cp}=\bar{u}_{\bar v}\gamma^{\nu}u_{v}~\bar{\xi}_{n}\gamma^{\nu}\xi_{\bar{n}}\int
dl~\frac{~2k^{\mu}}{\left[  -2(lk)-k_{\bot}^{2}\right]  }~\frac{1}{\left[
-2p^{\prime}(l+p)\right]  }~\left\{  \frac{2(l+p)^{\mu}}{\left[  l^{2}\right]
\left[  l^{2}+2(lp)-p_{\bot}^{2}\right]  }\right\}  ,
\end{equation}
and%
\begin{equation}
J_{2cp}=\bar{u}_{\bar v}\gamma^{\nu}u_{v}~\bar{\xi}_{n}\gamma^{\nu}\xi_{\bar{n}}\int
dl~\frac{~2k^{\prime\mu}}{\left[  2(lk^{\prime})-k_{\bot}^{2}\right]  }%
~\frac{1}{\left[  -2p^{\prime}(l+p)\right]  }~\left\{  \frac{2(l+p)^{\mu}%
}{\left[  l^{2}\right]  \left[  l^{2}+2(lp)-p_{\bot}^{2}\right]  }\right\}  .
\end{equation}
Therefore%
\begin{align}
J_{1cp}+J_{2cp}  &  =\bar{u}_{\bar v}\gamma^{\nu}u_{v}~\bar{\xi}_{n}\gamma^{\nu}\xi
_{\bar{n}}\int dl~~\frac{1}{\left[  -2p^{\prime}(l+p)\right]  }~\left\{
\frac{2(l+p)^{\mu}}{\left[  l^{2}\right]  \left[  l^{2}+2(lp)-p_{\bot}%
^{2}\right]  }\right\} 
 \nonumber \\ & 
\times \left(  \frac{~2k^{\mu}}{\left[  -2(lk)-k_{\bot}^{2}\right]  }%
+\frac{~2k^{\prime\mu}}{\left[  2(lk^{\prime})-k_{\bot}^{2}\right]  }\right)
\label{cpsum}
.
\end{align}
Consider expression in $(...).$ 
\begin{equation}
(...)\simeq\frac{~k_{-}n^{\mu}}{\left[  -l_{+}k_{-}-k_{\bot}^{2}\right]
}+\frac{~k_{-}^{^{\prime}}n^{\mu}}{\left[  l_{+}k_{-}^{\prime}-k_{\bot}%
^{2}\right]  }=\frac{~n^{\mu}}{\left[  -l_{+}-k_{\bot}^{2}/k_{-}\right]
}+\frac{~n^{\mu}}{\left[  l_{+}-k_{\bot}^{2}/k_{-}^{\prime}\right]  }.
\end{equation}
Notice that the IR-regulator $k_{\bot}^{2}$ introduces the small difference. 
Consider the soft limit $l_{\mu}\sim \Lambda$ in the collinear integrals in Eqs.(\ref{cpsum}). 
We obtain
\begin{align}
\left(  J_{1cp}+J_{2cp}\right)  _{s}  &  =\bar{u}_{\bar v}\gamma^{\nu}u_{v}~\bar{\xi}%
_{n}\gamma^{\nu}\xi_{\bar{n}}\int dl~~\frac{1}{q^{2}}~\left\{  \frac{2p^{\mu}%
}{\left[  l^{2}\right]  \left[  2(lp)-p_{\bot}^{2}\right]  }\right\}
\nonumber  \\
&
~\ \ \ \ \ \ \ \ \ \ \ \ \ \ \ \ \ \ \ \ \ \ 
\times \left(
\frac{~n^{\mu}}{\left[  -l_{+}-k_{\bot}^{2}/k_{-}\right]  }+\frac{~n^{\mu}%
}{\left[  l_{+}-k_{\bot}^{2}/k_{-}^{\prime}\right]  }\right) .
\end{align}
This expression coincides with the soft limit taken in the original integrals $J_{1,2}$. Therefore in order to obtain the 
collinear contribution we must subtract the overlapping soft part, see e.g. \cite{Collins:1999dz,Manohar:2006nz}.   This yields
\begin{equation}
\left(  J_{1cp}+J_{2cp}\right)  -\left(  J_{1cp}+J_{2cp}\right)  _{s}=
\bar{u}_{\bar v}\gamma^{\nu}u_{v}~\bar{\xi}_{n}\gamma^{\nu}\xi_{\bar{n}}~J_{cp},
\end{equation}%
with
\begin{align}
J_{cp}  &  =\int dl~\left(  ~\frac{1}{\left[  -2p^{\prime}(l+p)\right]
}~\frac{2(l+p)^{\mu}}{\left[  l^{2}\right]  \left[  l^{2}+2(lp)-p_{\bot}%
^{2}\right]  }-\frac{1}{q^{2}}\frac{2p^{\mu}}{\left[  l^{2}\right]  \left[
2(lp)-p_{\bot}^{2}\right]  }\right)
\nonumber \\
&  ~\ \ \ \ \ \ \ \ \ \ \ \ \ \ \ \ \ \ \ \ \ \ \times\left[  \frac{~n^{\mu}%
}{\left(  -l_{+}-k_{\bot}^{2}/k_{-}\right]  }+\frac{~n^{\mu}}{\left[
l_{+}-k_{\bot}^{2}/k_{-}^{\prime}\right)  }\right]  .
\end{align}
The integral $J_{cp}$ is not singular in the region of  small $l_{+}$  and we can put $k_{\bot}^{2}=0$  that yileds
\begin{equation}
\left(  \frac{~n^{\mu}}{\left[  -l_{+}\right]  }+\frac{~n^{\mu}%
}{\left[  l_{+}\right]  }\right)  =-2\pi i~\delta(l_{+})n^{\mu}.
\end{equation}
Hence
\begin{equation}
J_{cp}=\left(  -4\right)  \pi i\int dl~\frac{\delta(l_{+})}{\left[
l^{2}\right]  }\left(  ~\frac{1}{\left[  -p_{-}^{\prime}(l_{+}+p_{+})\right]
}~\frac{(l_{+}+p_{+})}{\left[  l^{2}+2(lp)-p_{\bot}^{2}\right]  }-\frac
{1}{q^{2}}\frac{p_{+}}{\left[  2(lp)-p_{\bot}^{2}\right]  }\right)
\end{equation}%
\begin{equation}
=\frac{-4\pi i}{q^{2}}p_{+}\int dl_{-}dl_{\bot}~\frac{1}{\left[  -l_{\bot}%
^{2}\right]  }\left(  ~~\frac{1}{\left[  -l_{\bot}^{2}+l_{-}p_{+}-p_{\bot}%
^{2}\right]  }-\frac{1}{\left[  l_{-}p_{+}-p_{\bot}^{2}\right]  }\right)
\end{equation}%
\begin{equation}
=\frac{4\pi i}{q^{2}}p_{+}\int dl_{-}dl_{\bot}~\frac{1}{\left[  -l_{\bot}%
^{2}+l_{-}p_{+}-p_{\bot}^{2}\right]  \left[  l_{-}p_{+}-p_{\bot}^{2}\right]
}.
\end{equation}
This integral is trivial 
because the poles in $l_{-}$ lie on the same complex semi-plane. Thus we demonstrated
that after soft subtraction the collinear to $p$ integral is zero. The same
manipulations also can be done for the other collinear regions.

\section{Compensation of the QED IR-divergencies in the elastic cross section }
\label{compensation}
 The dependence from the IR-regulator $\lambda$ in amplitudes $\delta\tilde G_{M,E}^{2 \gamma}$ arises only in
the SCET amplitude  $g_{1}$.  Using Eqs.(\ref{sigmR}), (\ref{dGM}) and
(\ref{g1fin}) we obtain%
\begin{equation}
\delta\tilde G^{2 \gamma}_{M}=\frac{\alpha}{\pi}g_{1}(z,\mu_{0})+~...=\frac{\alpha}{\pi}G_{M}\frac
{\lambda^{2}}{\mu_{0}^{2}}\ln\left\vert \frac{\tilde{s}}{\tilde{u}}\right\vert
+~...~,
\end{equation}
where dots denote the $\lambda$-independent terms.  
Therefore for the corresponding difference we obtain 
\bea
\operatorname{Re}[\delta\tilde{G}_{M}^{2 \gamma} -G_{M}\frac12 \delta^{\text{tpe,MT}}_{2\gamma}]
&=&\frac{\alpha}{\pi}G_{M}\ln\frac{\lambda^{2}}{\mu_{0}^{2}}\ln\left\vert
\frac{\tilde{s}}{\tilde{u}}\right\vert -\frac{\alpha}{\pi}G_{M}\ln
\frac{\lambda^{2}}{\tilde{s}}\ln\left\vert \frac{\tilde{s}}{\tilde{u}%
}\right\vert +~... 
\\  
&=& \frac{\alpha}{\pi}G_{M}\ln\frac{\tilde{s}}{\mu_{0}^{2}}\ln\left\vert
\frac{\tilde{s}}{\tilde{u}}\right\vert +~...~.
\eea

The expressions  (\ref{GMf1})  and (\ref{dF2s}) for  the  FF $F_{2}$ and amplitude $\delta\tilde F^{(s)}_{2}$    are incomplete because we neglected  the contributions with the SCET subleading operators.
Therefore we can consider  cancellation of the IR-regulator $\lambda$  in the combination  $\delta\tilde G_{E}-G_{E}\delta_{2\gamma}$  in Eq.(\ref{sgmRexp})  
only between the corresponding kinematical power corrections.  
 Using Eqs.(\ref{dF2s}) yields%
\bea
\delta\tilde G^{(s)}_{E}   & = &\delta\tilde G_{M}-(1+\tau)\delta\tilde F^{(s)}_{2}
\\
  &\simeq &  -\frac{\alpha}{\pi}\frac{1}{\tau}\left\{  ~g_{1}(z,\mu_{0})+C_{M}(z,\mu
_{0})\mathcal{F}_{1}+\frac{\nu}{s}C_{3}(z)\mathcal{F}_{1}\right\}  -\frac{\nu
}{s}C_{3}(z)\mathcal{F}_{1},
\eea
Using  Eqs.(\ref{sigmR}) and (\ref{g1fin}) we obtain
\begin{equation}
\delta\tilde  G^{(s)}_{E}\simeq-\frac
{\alpha}{\pi}\frac{1}{\tau}\left\{  ~g_{1}(z,\mu_{0})+C_{M}(z,\mu
_{0})\mathcal{F}_{1}+\frac{\nu}{s}C_{3}(z)\mathcal{F}_{1}\right\}
\end{equation}%
\begin{equation}
=-\frac{\alpha}{\pi}\frac{1}{\tau}~g_{1}(z,\mu_{0})+~...=-\frac{\alpha}{\pi
}\frac{1}{\tau}G_{M}(Q)\ln\frac{\lambda^{2}}{\mu_{0}^{2}}\ln\left\vert
\frac{\tilde{s}}{\tilde{u}}\right\vert +...~.
\end{equation}

Then taking into account (\ref{GMf1}) where we also neglected  the
 subleading SCET operators we obtain%
\begin{align}
G_{E}  &  =G_{M}-(1+\tau)F_{2}\simeq G_{M}-(1+\tau)\frac{1}{\tau}G_{M} =-\frac{1}{\tau}G_{M}.
\end{align}
Taking into account  Eq.(\ref{dlt2g}) yileds
\bea
\operatorname{Re}[\delta\tilde G^{(s)}_{E}-\frac{\alpha}{\pi}G_{E}~\frac12 \delta^{\text{tpe,MT}}_{2\gamma}]
&\simeq &
-\frac{\alpha}{\pi}\frac{1}{\tau}G_{M}\ln\frac{\lambda^{2}}{\mu_{0}^{2}} 
\ln\left\vert \frac{\tilde{s}}{\tilde{u}}\right\vert +\frac{\alpha}{\pi} \frac{1}{\tau}G_{M}  \ln\frac{\lambda^{2}}{\tilde{s}}
\ln\left\vert \frac{\tilde{s}}{\tilde{u}}\right\vert +~...
\nonumber \\
&=&-\frac{\alpha}{\pi}\frac{1}{\tau}G_{M}\ln\frac{\tilde{s}}{\mu_{0}^{2}}
\ln\left\vert \frac{\tilde{s}}{\tilde{u}}\right\vert +~...~.
\eea
From  this calculation we can clarify why we keep unexpanded the argument of
the logarithm in Eq.(\ref{g1fin}): we need this otherwise   the cancellation of the IR-mass
$\lambda^{2}$ will be incomplete because the expression for $\delta_{2\gamma}$
in Eq.(\ref{dlt2g}) includes such power suppressed contributions. 

\end{appendix}

\end{document}